%% file: paper.tex
\documentclass{aa}
\usepackage{textcomp}
\usepackage{graphicx} %
\usepackage{amsmath} %
\usepackage{amssymb} %
\usepackage{bm} %
\usepackage{upgreek} %
\usepackage{IEEEtrantools} %
\usepackage{multirow}
\usepackage[colorlinks=true,allcolors=blue,urlcolor=blue]{hyperref}
\usepackage{txfonts}
\usepackage[normalem]{ulem} 
\usepackage[dvipsnames]{xcolor}

\newcommand{\eqn}[1]{\text{Eq.~\ref{#1}}}
\newcommand{\sect}[1]{\text{Sect.~\ref{#1}}}
\newcommand{\fig}[1]{\text{Fig.~\ref{#1}}}
\newcommand{\tab}[1]{\text{Table~\ref{#1}}}

\newcommand{\multitd}{\texttt{Multi3D}}
\newcommand{\balder}{\texttt{Balder}}
\newcommand{\blue}{\texttt{Blue}}
\newcommand{\marcs}{\texttt{MARCS}}
\newcommand{\sme}{\texttt{SME}}
\newcommand{\pysme}{\texttt{PySME}}

\newcommand{\kms}{\mathrm{km\,s^{-1}}}
\newcommand{\teff}{T_{\mathrm{eff}}}
\newcommand{\lgg}{\log{g}}
\newcommand{\lggu}{\log{g / \mathrm{cm\,s^{-2}}}}

\newcommand{\vbroad}{\varv_{\text{broad}}}
\newcommand{\lggf}{\log{gf}}
\newcommand{\feh}{\mathrm{\left[Fe/H\right]}}

\newcommand{\xfe}[1]{\mathrm{\left[#1/Fe\right]}}
\newcommand{\xh}[1]{\mathrm{\left[#1/H\right]}}
\newcommand{\lgeps}[1]{\log{\epsilon_{\mathrm{#1}}}}
\newcommand{\sh}{S_{\mathrm{H}}}
\newcommand{\dex}{\mathrm{dex}}

\newcommand{\nm}{\mathrm{nm}}
\newcommand{\K}{\mathrm{K}}

\newcommand{\vmic}{\xi_{\text{mic}}}

\newcommand{\level}[1]{n_{#1}}
\newcommand{\levellte}[1]{n^{*}_{#1}}
\newcommand{\dc}[1]{\beta_{#1}}
\newcommand{\rcoef}[2]{R_{#1#2}}
\newcommand{\ccoef}[2]{C_{#1#2}}

\newcommand{\colsmall}{CornflowerBlue}
\newcommand{\colbig}{SeaGreen}
\begin{document} 

\title{The GALAH Survey: Non-LTE departure coefficients for large
spectroscopic surveys\thanks{Grids of departure coefficients can
be found online \citep{grid_nlte}
or by contacting the lead author directly.}}
\titlerunning{Non-LTE departure coefficients for large
spectroscopic surveys}
\author{A.~M.~Amarsi\inst{\ref{uu1}} 
\and
K.~Lind\inst{\ref{su},\ref{mpia}} 
\and
Y.~Osorio\inst{\ref{iac},\ref{ull}} 
\and
T.~Nordlander\inst{\ref{rsaa},\ref{astro3d}} 
\and
M.~Bergemann\inst{\ref{mpia}} 
\and
H.~Reggiani\inst{\ref{jhu}} 
\and
E.~X.~Wang\inst{\ref{rsaa},\ref{astro3d}} 
\and
S.~Buder\inst{\ref{rsaa},\ref{astro3d}} 
\and
M.~Asplund\inst{\ref{rsaa},\ref{astro3d}} 
\and
P.~S.~Barklem\inst{\ref{uu1}} 
\and
A.~Wehrhahn\inst{\ref{uu2}} 
\and
\'A.~Sk\'ulad\'ottir\inst{\ref{florence},\ref{arcetri}} 
\and
C.~Kobayashi\inst{\ref{astro3d},\ref{hertfordshire}} 
\and
A.~I.~Karakas\inst{\ref{astro3d},\ref{monash}} 
\and
X.~D.~Gao\inst{\ref{mpia}} 
\and
J.~Bland-Hawthorn\inst{\ref{astro3d},\ref{sifa}} 
\and
G.~M.~De~Silva\inst{\ref{astro3d},\ref{aao}} 
\and
J.~Kos\inst{\ref{slovenia}} 
\and
G.~F.~Lewis\inst{\ref{sifa}} 
\and
S.~L.~Martell\inst{\ref{astro3d},\ref{unsw}} 
\and
S.~Sharma\inst{\ref{astro3d},\ref{sifa}}
\and
J.~D.~Simpson\inst{\ref{unsw}} 
\and
D.~B.~Zucker\inst{\ref{macq1},\ref{macq2}} 
\and
K.~\v{C}otar\inst{\ref{slovenia}}
\and
J.~Horner\inst{\ref{usq}}
\and
the GALAH Collaboration}
\institute{\label{uu1} Theoretical Astrophysics, 
Department of Physics and Astronomy,
Uppsala University, Box 516, SE-751 20 Uppsala, Sweden\\
\email{anish.amarsi@physics.uu.se} 
\and 
\label{su} Department of Astronomy, Stockholm
University, AlbaNova University Centre, SE-106 91 Stockholm, Sweden 
\and
\label{mpia} Max Planck Institute for
Astronomy, Königstuhl 17, 69117 Heidelberg, Germany 
\and
\label{iac}
Instituto de Astrof\'isica de Canarias, E-38205 La Laguna, Tenerife, Spain
\and
\label{ull} 
Departamento de Astrof\'isica, Universidad de La Laguna (ULL), E-38206 La Laguna, Tenerife, Spain
\and
\label{rsaa} Research School of Astronomy and Astrophysics, Australian
National University, Canberra, ACT 2611, Australia 
\and
\label{astro3d} ARC Centre of
Excellence for All Sky Astrophysics in 3 Dimensions (ASTRO 3D), Australia 
\and 
\label{jhu}
Department of Physics and Astronomy, Johns Hopkins University,
3400 N Charles St., Baltimore, MD 21218, USA
\and
\label{uu2} Observational Astrophysics, 
Department of Physics and Astronomy,
Uppsala University, Box 516, SE-751 20 Uppsala, Sweden
\and
\label{florence} Dipartimento di Fisica e Astronomia, Universit\'a 
degli Studi di Firenze, Via G. Sansone 1, 50019 Sesto Fiorentino, Italy 
\and
\label{arcetri} INAF/Osservatorio Astrofisico di Arcetri,
Largo E. Fermi 5, 50125 Firenze, Italy
\and
\label{hertfordshire} 
Centre for Astrophysics Research, 
Department of Physics, Astronomy and Mathematics,
University of Hertfordshire, Hatfield, AL10
9AB, UK
\and
\label{monash}
School of Physics \& Astronomy, Monash University, Clayton VIC 3800, Australia
\and
\label{sifa} Sydney Institute for Astronomy, 
School of Physics, A28, The University of Sydney, NSW 2006, Australia
\and
\label{aao} Australian Astronomical Optics, 
Macquarie University, 105 Delhi Rd, North Ryde, 211, Australia
\and
\label{slovenia} Faculty of Mathematics and Physics, 
University of Ljubljana, Jadranska 19, 1000 Ljubljana, Slovenia
\and
\label{unsw} 
School of Physics, UNSW, Sydney, NSW 2052, Australia
\and
\label{macq1} 
Department of Physics and Astronomy, 
Macquarie University, Sydney, NSW 2109, Australia
\and
\label{macq2} Macquarie University Research Centre for Astronomy, 
Astrophysics \& Astrophotonics, Sydney, NSW 2109, Australia
\and
\label{usq}
Centre for Astrophysics, University of Southern Queensland, 
Toowoomba, QLD 4350, Australia}

\abstract{Massive sets of stellar spectroscopic observations
are rapidly becoming available and these can be used
to determine the chemical composition and evolution of the Galaxy
with unprecedented precision.
One of the major challenges in this endeavour
involves constructing realistic models of stellar spectra with which to 
reliably determine stellar abundances.
At present, large stellar surveys commonly use simplified models
that assume that the stellar atmospheres are approximately in
local thermodynamic equilibrium (LTE).
To test and ultimately relax this assumption, we 
have performed non-LTE calculations
for $13$ different elements (H, Li, C, N, O,
Na, Mg, Al, Si, K, Ca, Mn, and Ba),
using recent model atoms that have physically-motivated 
descriptions for the inelastic collisions with neutral hydrogen,
across a grid of $3756$ 1D \marcs{} model atmospheres 
that spans $3000\leq\teff/\K\leq8000$,
$-0.5\leq\lggu\leq5.5$, and $-5\leq\feh\leq1$.
We present the grids of departure coefficients
that have been implemented into the 
GALAH DR3 analysis pipeline in
order to complement the extant non-LTE grid for iron.
We also present a detailed line-by-line 
re-analysis of $50126$ stars from GALAH DR3.
We found that relaxing LTE can change the abundances
by between $-0.7\,\dex$ and $+0.2\,\dex$ for different lines and stars.
Taking departures from LTE into account can reduce the dispersion in 
the $\xfe{A}$ versus $\feh$ plane by up to $0.1\,\dex$, and 
it can remove 
spurious differences between the dwarfs and giants by up to $0.2\,\dex$.
The resulting abundance slopes can thus be qualitatively different
in non-LTE, possibly with important implications
for the chemical evolution of our Galaxy.
The grids of departure coefficients are publicly available
and can be implemented into LTE pipelines 
to make the most of observational data sets
from large spectroscopic surveys.}


\date{Received 15 June 2020 /
    Accepted 20 August 2020}

\maketitle
\section{Introduction}
\label{introduction}

Stellar astronomy has entered a new era
characterised by extremely large surveys of stars and their spectra.
Massive studies of stellar parameters and elemental abundances,
based on medium- or high-resolution spectra
of around $10^{5}$ stars,
are close to completion or have recently finished, including
RAVE (\citealt{2020AJ....160...83S}; $R\sim7500$),
Gaia-ESO (\citealt{2013Msngr.154...47R}; $R\sim20000$),
APOGEE (\citealt{2020ApJS..249....3A}; $R\sim22500$), 
and 
LAMOST (\citealt{2020arXiv200507210L};
$R\sim7500$ in the medium-resolution
setting).
The near future will see a jump in the number statistics.
The planned WEAVE \citep{2016SPIE.9908E..1GD}
and 4MOST \citep{2019Msngr.175....3D} surveys
in the northern and southern hemispheres, respectively,
should obtain high-resolution spectra for around $10^{7}$ stars,
while the third \textit{Gaia} data release \citep{2018A&A...616A...1G}
will include medium-resolution spectra of around $10^{8}$
stars \citep{2016A&A...585A..93R}.
The ongoing Galactic Archaeology with HERMES
(GALAH) survey falls somewhere in between these two groups,
with the goal to observe $10^{6}$ stars 
at a relatively high resolution of $R\sim28000$
\citep{2015MNRAS.449.2604D}.

Theoretical stellar spectra are used by all of these surveys
in order to obtain stellar parameters and
the elemental abundances.
This is usually achieved by 
comparing the theoretical spectra against observations
directly \citep[][]{2016AJ....151..144G}.
Alternatively, they can also be used to train data-driven approaches
\citep{2015ApJ...808...16N,2019ApJ...879...69T}
by directly using
the theoretical spectra \citep{2016A&A...585A..93R}
or gradient spectra \citep{2019ApJS..245...34X},
or by first inferring precise stellar parameters and elemental abundances
of a smaller sample of stars, which are then used
as a training set \citep{2018MNRAS.478.4513B}.

The accuracy of these surveys depends on the
reliability of their theoretical stellar spectra.
This is particularly relevant for the determination of
stellar elemental abundances.
While extra constraints can be placed on
effective temperatures, $\teff$, and surface gravities, $\lgg$,
via interferometry, photometry, astrometry, and asteroseismology
\citep{2012ApJ...757...99S,2018MNRAS.475L..81K,2020A&A...640A..25K},
elemental abundances, in contrast, are most directly probed
through absorption and emission lines in the stellar spectrum.
It follows that deficiencies in the theoretical spectra 
have a direct impact on the reliability of the 
elemental abundance determinations.

A potential pitfall in classical spectroscopic analyses
is the assumption that the stellar atmospheric matter 
satisfies local thermodynamic equilibrium 
(LTE; \citealt{1973ARA&amp;A..11..187M}).
When calculating synthetic stellar spectra
via radiative transfer post-processing of
pre-computed model atmospheres (which themselves
either assume strict LTE or adopt coherent isotropic continuum
scattering; \citealt{2008A&amp;A...486..951G}),
this simplifying assumption allows one to 
describe the populations of the different 
energy levels of a given absorbing species using Saha-Boltzmann statistics.
This enables an immediate, analytical solution for
the populations $\level{i}$ and $\level{j}$ of the lower and
upper levels $i$ and $j$ for any given spectral line $i\leftrightarrow j$.

In reality, the radiation field is non-Planckian
in stellar photospheres.
Consequently, interactions between light and matter cause
the latter to depart from LTE.
A more general description of the level populations is given
by the equations of statistical equilibrium
\citep{2003rtsa.book.....R}:
\phantomsection\begin{IEEEeqnarray}{rCl}
\label{eq:stateq}
    \level{i} 
    \displaystyle\sum\limits_{j} 
    \left[\rcoef{i}{j} + \ccoef{i}{j}\right]
    &=&
    \displaystyle\sum\limits_{j} 
    \level{j} \left[\rcoef{j}{i} + \ccoef{j}{i}\right]\,.
\end{IEEEeqnarray}
Or in other words, the net rate 
out of a level $i$ is set by
the balance between all the outwards and inwards
radiative ($R$) and collisional ($C$) transitions.
\eqn{eq:stateq} is satisfied trivially in the stellar interior
by virtue of the principle of detailed balance for LTE populations
$\levellte{i}$:
\phantomsection\begin{IEEEeqnarray}{rCl}
\label{eq:detailedbalancerad}
    \levellte{i}\,\rcoef{i}{j} &\equiv& \levellte{j}\,\rcoef{j}{i}\,, \\
\label{eq:detailedbalancecol}
    \levellte{i}\,\ccoef{i}{j} &\equiv& \levellte{j}\,\ccoef{j}{i}\,.
\end{IEEEeqnarray}
However in the stellar photosphere,
\eqn{eq:detailedbalancerad} no longer holds;
nevertheless, since 
the particles have Maxwell-Boltzmann distributions
to a good approximation \citep{2014tsa..book.....H},
\eqn{eq:detailedbalancecol} remains true 
for LTE populations $\levellte{i}$.
Thus, the matter can only satisfy LTE, $\level{i}=\levellte{i}$,
if the collisional rates dominate over the radiative
rates; in general, this is not the case.

Solving for the statistical equilibrium is much more demanding
than simply assuming LTE.
\eqn{eq:stateq} indicates that all radiative and collisional
transitions can affect the absorber populations
and thus the spectral line.
As well as being a much larger system of equations to solve,
\eqn{eq:stateq} must be iterated with a solution for
the radiation field \citep{1992A&amp;A...262..209R},
and convergence problems are often encountered.
Moreover, non-LTE calculations require comprehensive sets of 
energy levels, and radiative and collisional
transition probabilities;
the final result is only as reliable as this input atomic data
\citep{2016A&amp;ARv..24....9B}.
Therefore care is needed to calculate these data accurately,
and then to consistently merge different data sources
into what are referred to as model atoms.
In contrast, when modelling a spectral line in LTE,
the populations are fixed and
one only needs to have the partition functions describing that
particular chemical species and the parameters of that particular line.

One way to improve the accuracy of classical spectroscopic analyses
without significantly increasing their cost
is to apply pre-computed non-LTE solutions to them.
A common approach is to use pre-computed absolute or differential abundance
corrections $\Delta_{l}$ 
for a given spectral line $l$\footnote{The 
absolute abundance of
    element $\mathrm{A}$ is defined as
$\lgeps{A}\equiv\log_{10}\left(N_{\mathrm{A}}/N_{\mathrm{H}}\right)+12$,
where $N_{\mathrm{A}}$ and $N_{\mathrm{H}}$ are
the number of nuclei of element $A$ and of hydrogen.
Abundance ratios differential to the Sun are defined as
$\xh{A}\equiv\lgeps{A}-\lgeps{A}_{\odot}$
and $\xfe{A}\equiv\xh{A}-\feh$.}:
\phantomsection\begin{IEEEeqnarray}{rCl}
\label{eq:abcor}
    \lgeps{A}^{\text{Non-LTE}}_{l} &=& 
    \lgeps{A}^{\text{LTE}}_{l} + \Delta_{l}^{\text{abs.}}\,, \\
\label{eq:abcor2}
    \xh{A}^{\text{Non-LTE}}_{l} &=& 
    \xh{A}^{\text{LTE}}_{l} + \Delta_{l}^{\text{diff.}}\,.
\end{IEEEeqnarray}
Vast grids of line-by-line abundance corrections
for many different atomic species
already exist in the literature
\citep[e.g.][]{2012A&A...546A..90B,
2012MNRAS.427...50L,2015A&A...581A..70K,
2016AstL...42..606M,2016A&A...586A.120O,
2019A&A...630A.104A},
that can readily be adopted and applied by
the stellar spectroscopy community.

The line-by-line abundance corrections approach
(\eqn{eq:abcor}) can become prohibitively complicated
when a large number of spectral lines need to be studied
simultaneously.
It is simpler to instead apply non-LTE corrections
to the level populations, because the number of relevant levels 
that would need to be considered roughly scales
with the square root of the corresponding number of relevant spectral lines.
This can be accomplished using
pre-computed grids of non-LTE departure coefficients $\dc{i}$,
for a given energy level $i$:
\phantomsection\begin{IEEEeqnarray}{rCl}
\label{eq:depcoef}
    \dc{i} &\equiv& \frac{\level{i}}{\levellte{i}}\,.
\end{IEEEeqnarray}
This approach requires some extra initial effort, because
LTE spectrum synthesis codes need to be modified to 
read and manipulate the grids of departure coefficients.
Nevertheless, grids of departure coefficients are desirable for
large spectroscopic surveys, which are typically based on
full-spectrum analyses.

Here, we present publicly-available
grids of departure coefficients for $13$ different elements:
H, Li, C, N, O, Na, Mg, Al, Si, K, Ca, Mn, and Ba.
For many of these elements, this is the most extensive
set of non-LTE calculations: they 
cover $3756$ 1D model atmospheres that span
the HR diagram from M-dwarfs, up the main sequence past the turn-off,
to the tip of the red giant branch, with $-5\leq\feh\leq1$.
This is also the first time 
that consistently computed grids of departure coefficients for
multiple elements have been released in the literature.

In \sect{method} we describe the
non-LTE radiative transfer calculations,
and the implementation of the
resulting grids of departure coefficients 
into the GALAH analysis pipeline.
In \sect{results} we present a line-by-line
re-analysis of $55159$ spectra corresponding
to $50126$ FGK-type field stars from GALAH DR3,
and in \sect{discussion}
use these results to discuss the impact of the departure
coefficients in practice. 
In \sect{gce} we briefly discuss the implication
of our non-LTE abundance analysis in the context
of Galactic chemical evolution, and
in \sect{conclusion} we summarise and 
make some concluding remarks about the outlook 
of quantitative non-LTE stellar spectroscopy.

\section{Calculation of departure coefficients}
\label{method}

\begin{figure*}
    \begin{center}
        \includegraphics[scale=0.21]{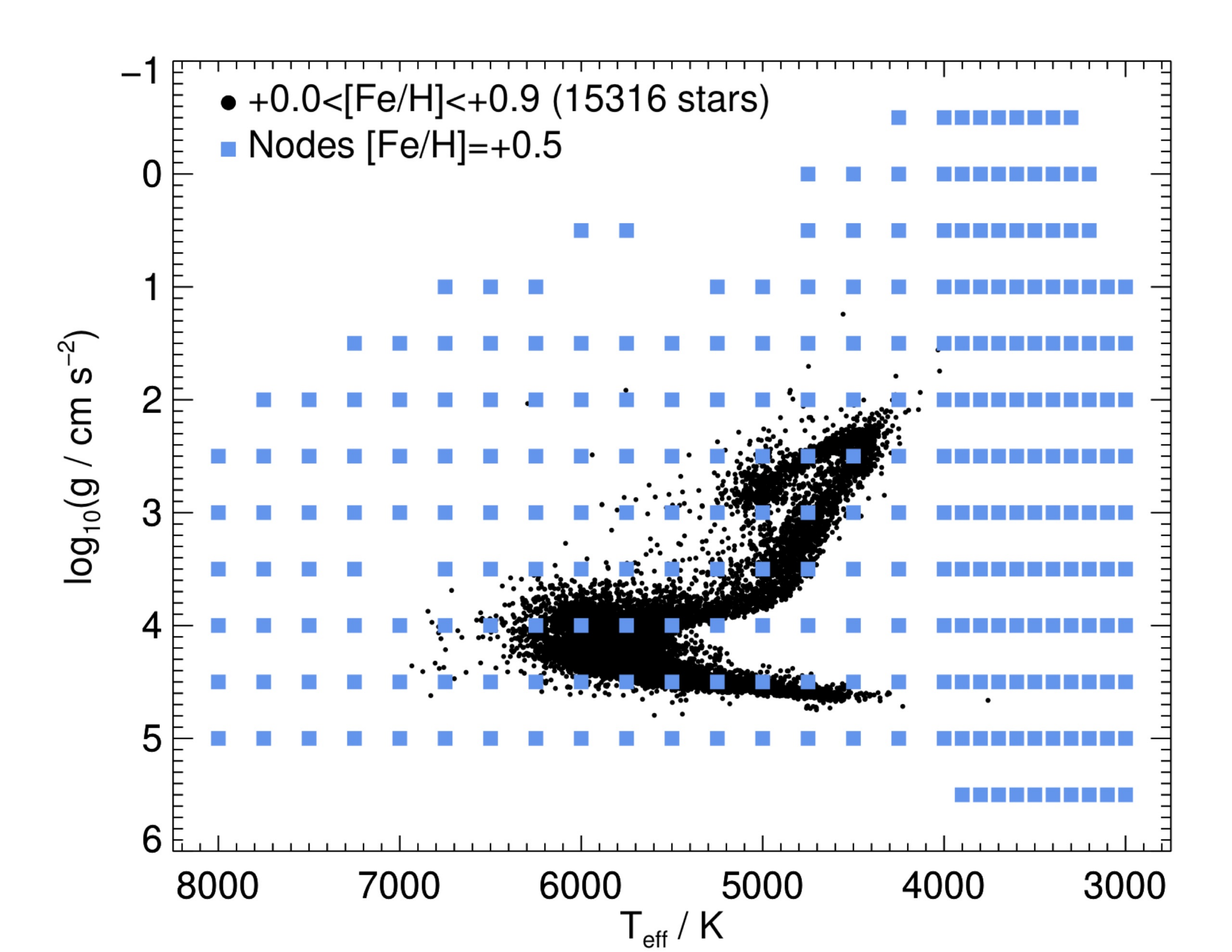}
        \includegraphics[scale=0.21]{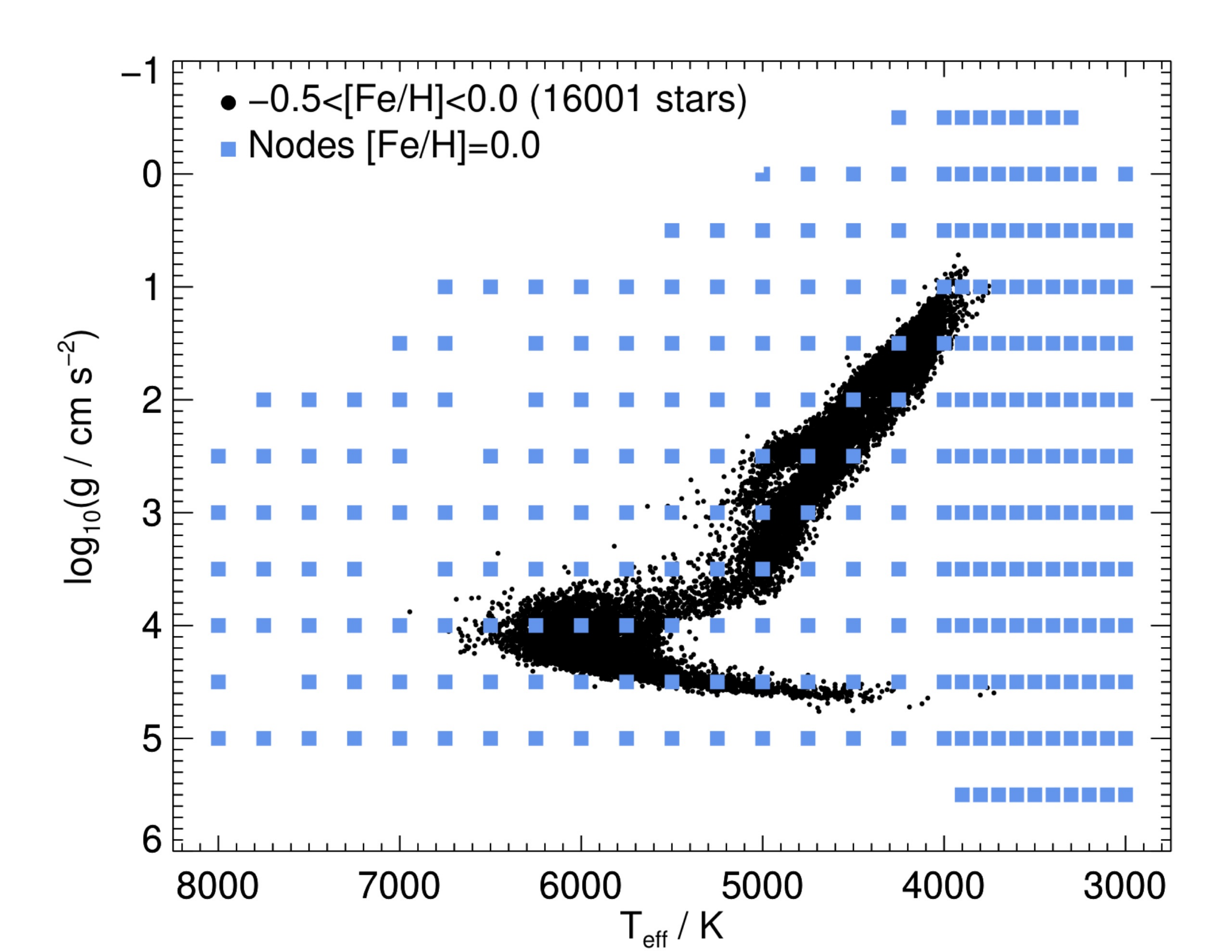}
        \includegraphics[scale=0.21]{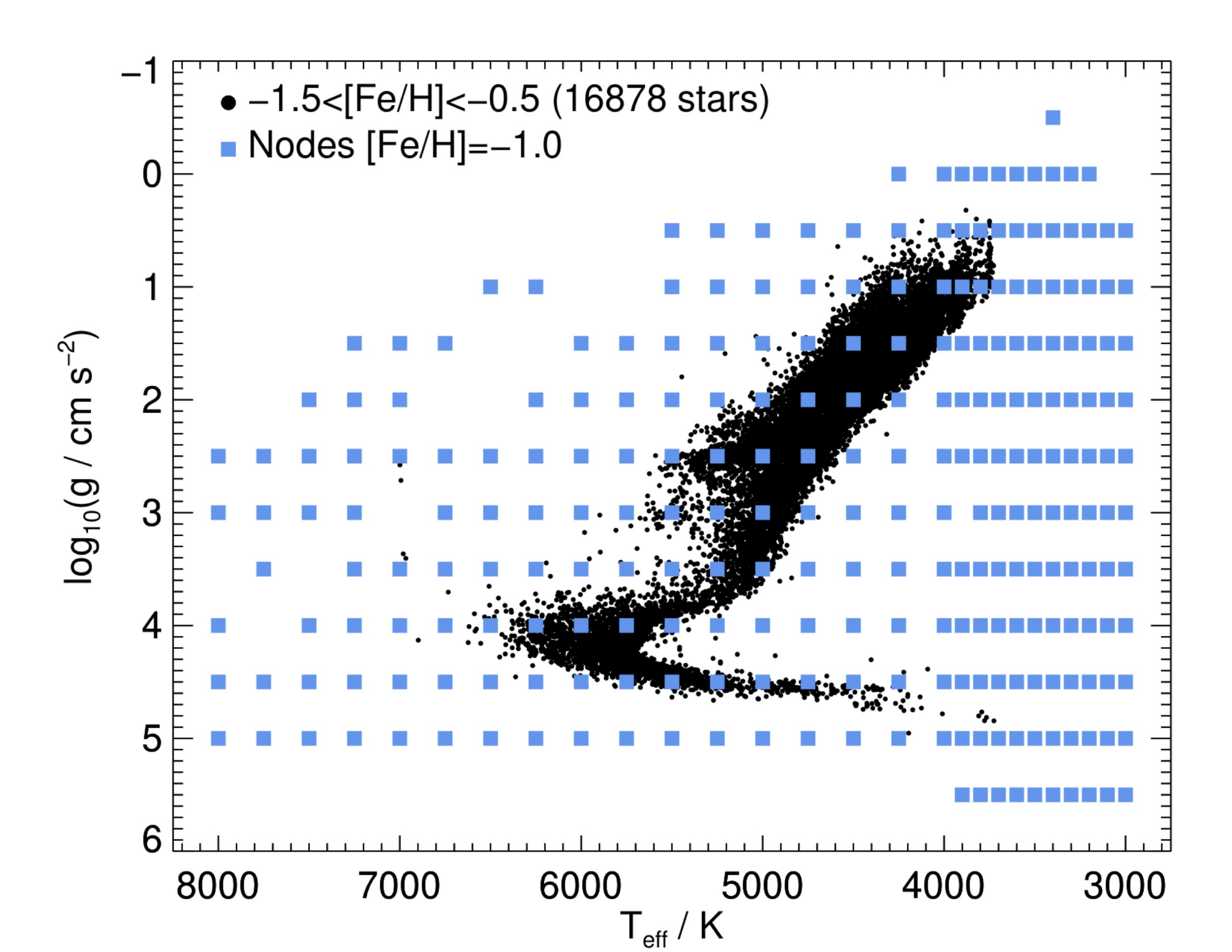}
        \includegraphics[scale=0.21]{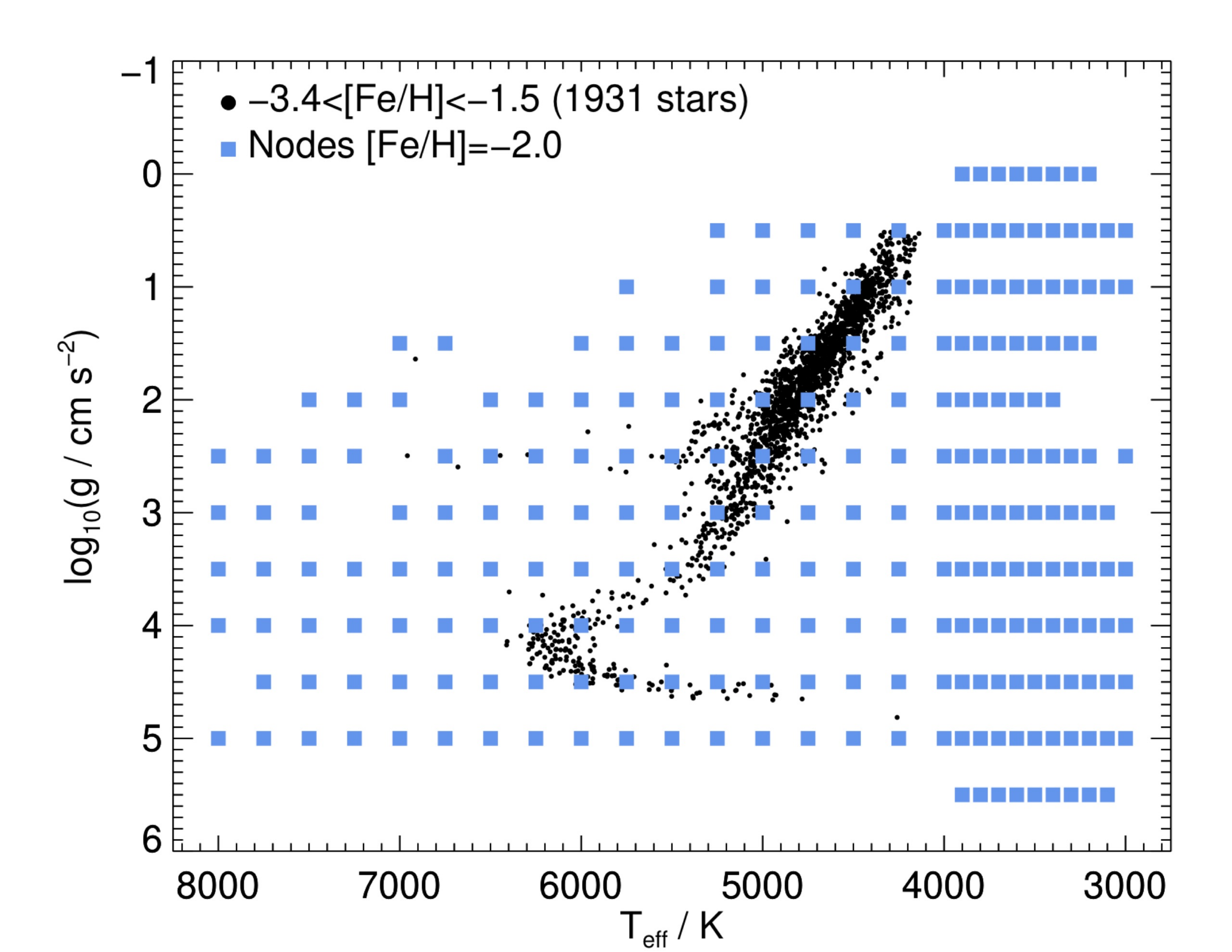}
        \caption{Grid nodes (surface gravities and effective temperatures)
            at which the non-LTE radiative transfer
            calculations were carried out.
            The missing nodes are where
            standard \marcs{} model atmospheres do not 
            presently exist, owing to convergence issues 
            (see Sect. 5.4 of \citealt{2008A&amp;A...486..951G}).
            Also shown in black are the 
            parameters of the $50126$ stars drawn 
            from GALAH DR3, studied in this work.}
        \label{fig:grid}
    \end{center}
\end{figure*}

\begin{figure}
    \begin{center}
        \includegraphics[scale=0.21]{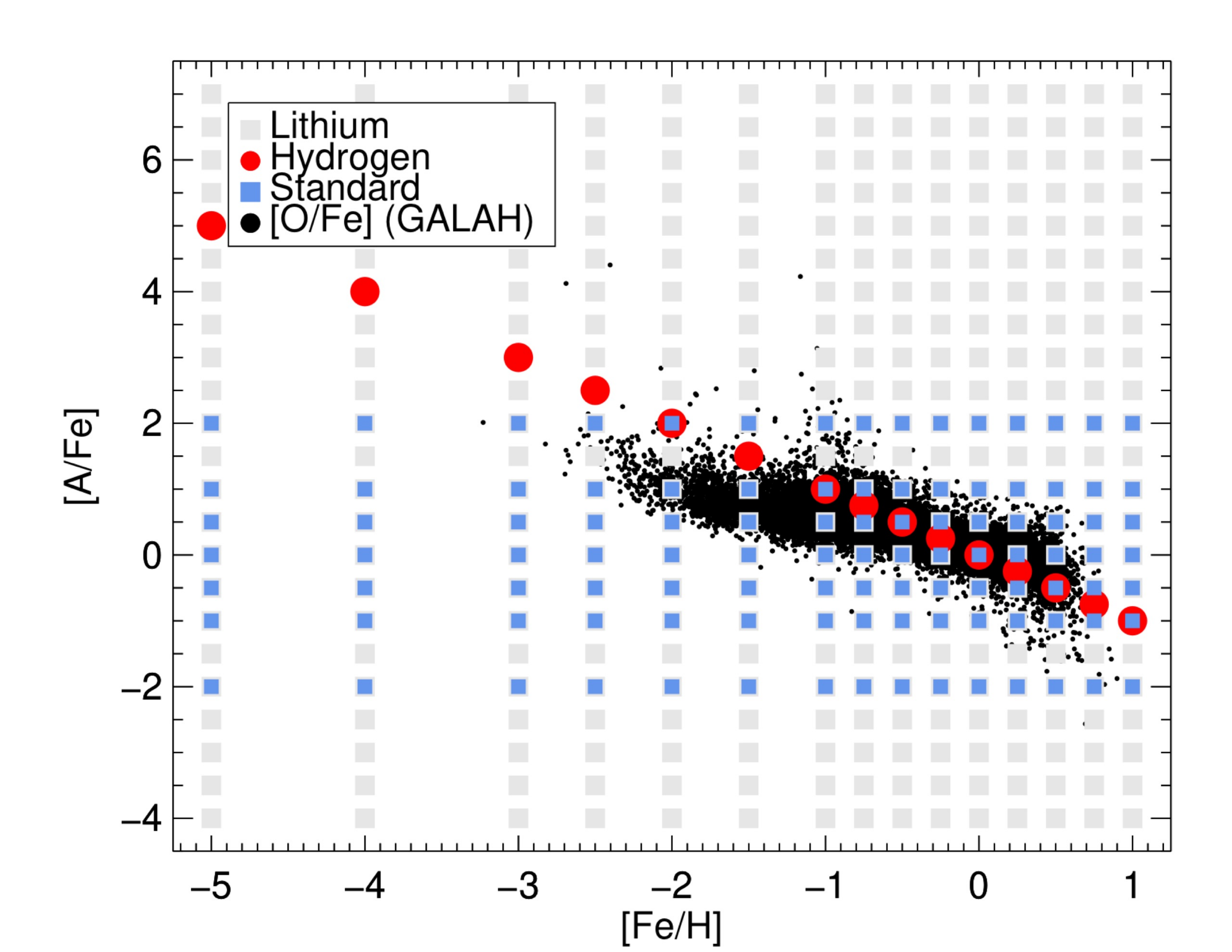}
        \caption{Grid nodes (abundance ratios and metallicities)
            at which the
            non-LTE radiative transfer calculations were carried out.
            The calculations for hydrogen and lithium
            are specially noted.
            Also shown in black are $\xfe{O}$ for the sample
            of stars discussed in the present study.}
        \label{fig:abgrid}
    \end{center}
\end{figure}

\input{atoms.tex}

\subsection{Model atoms}
\label{methodatom}

We list the $13$ elements considered in this work
and give the sizes of the corresponding model atoms
in \tab{tab:atoms}.
Full details about the physics included in each of the models
can be found in the original papers, references to which are also
provided in \tab{tab:atoms}.
Here we provide a very brief overview of the models.

The adopted model atoms were all constructed recently,
the oldest being that for sodium \citep{2011A&amp;A...528A.103L}.
Minor changes were made to some of the 
models compared to what was presented in the original papers,
in order to reduce the computational cost.
In particular, for carbon, nitrogen, magnesium, and calcium,
all fine structure levels were collapsed.
This has previously been verified
for carbon \citep{2019A&A...624A.111A} and 
oxygen \citep{2018A&A...616A..89A}
to have only a small impact on
the predicted departures from LTE:
typically of the order $0.01\,\dex$ in
terms of abundances.

For completeness, the model atom for iron is also listed 
in \tab{tab:atoms}.  Iron is treated in non-LTE 
in GALAH DR2 \citep{2018MNRAS.478.4513B} and
DR3 (Buder et al. in prep.). However the departure coefficients
originate from an older set of calculations
described in \citet{2016MNRAS.463.1518A}. 
They are not presented and discussed in the present study.

Inelastic collisions with neutral hydrogen have historically
been one of the largest sources of uncertainty in non-LTE models
\citep{2005ARA&amp;A..43..481A}.
Until recently, if such processes were not neglected completely,
the classical Drawin recipe would usually be adopted.
This recipe is based on the Thomson cross-section
for ionisation by electron collisions  \citep{thomson1912xlii},
as modified for ionisation by atomic collisions by
\citet{1969ZPhy..225..483D,1968ZPhy..211..404D},
and later extended by \citet{1984A&amp;A...130..319S} and
\citet{1993PhST...47..186L} to also cover excitation.
However, the Drawin recipe does not reflect the actual
physics of the low-energy collisions occurring in stellar atmospheres
\citep{2011A&amp;A...530A..94B,2016A&amp;ARv..24....9B,
2017A&A...606A.147B}.
To attempt to correct for this, usually a single fudge factor
$\sh$ is calibrated and applied to the Drawin rate coefficients
\citep{2004A&amp;A...423.1109A,2015A&amp;A...583A..57S}.
This approach is not guaranteed to improve the reliability
of the models: apart from
introducing more free parameters into the spectroscopic
analysis, this does not take into account that the
errors in the Drawin recipe vary depending on the stellar parameters,
and are transition-dependent. 
The Drawin recipe is also unable
to describe charge transfer processes
($\mathrm{A+H}\leftrightarrow\mathrm{A^{+}+H^{-}}$),
which are of astrophysical importance
\citep{2011A&amp;A...530A..94B}.

To avoid the inherent uncertainties of the Drawin recipe,
all of the adopted model atoms use alternative,
physically-motivated descriptions for the inelastic collisions
with neutral hydrogen.
For transitions involving low- and intermediate-lying levels of
lithium \citep{2003PhRvA..68f2703B,2003A&amp;A...409L...1B},
sodium \citep{2010PhRvA..81c2706B,2010A&amp;A...519A..20B},
and magnesium \citep{2012PhRvA..85c2704B,2012A&amp;A...541A..80B},
cross-sections based on full quantum chemistry calculations were used.
For the other species, calculations based on
asymptotic methods were used
\citep{2013PhRvA..88e2704B,2016PhRvA..93d2705B}.
In addition, with the exception of barium
\citep{2020A&A...634A..55G},
transitions involving Rydberg levels
were described using the free electron method
\citep{1985JPhB...18L.167K,kaulakys1986free,1991JPhB...24L.127K}.

\subsection{Model atmospheres}
\label{methodatmosphere}

The departure coefficients were calculated 
for a grid of 1D \marcs{} model atmospheres
\citep{2008A&amp;A...486..951G},
available from the online repository\footnote{\url{marcs.astro.uu.se}}.
The models specify the gas temperature, density, 
and other thermodynamic quantities, on $56$ depth points. 
In this work, these quantities were not re-sampled and interpolated onto
a new depth scale.
This means that the departure coefficients presented here
can readily be adopted and used consistently with
the standard \marcs{} models.

We illustrate the extent of the grid of 
model atmospheres in \fig{fig:grid} and \fig{fig:abgrid}.
The $3756$ different models are labelled by $\teff$, $\lgg$, and $\feh$,
and have the standard \marcs{} chemical compositions:
namely, they adopt the
solar chemical compositions of \citet{2007coma.book..105G},
scaled with $\feh$, and with an enhancement to $\upalpha$ elements
of $+0.1$, $+0.2$, $+0.3$, and $+0.4$ for
$\feh=-0.25$, $-0.5$, $-0.75$, and $\feh\leq-1.0$ respectively.
Plane-parallel models with $\vmic=1.0\,\kms$
were adopted for dwarfs 
($\lggu>3.5$ in \fig{fig:grid}),
and spherically-symmetric models with
$\vmic=2.0\,\kms$ were adopted for 
giants ($\lggu\leq3.5$
in \fig{fig:grid}).

\subsection{Non-LTE radiative transfer}
\label{methodbalder}

The departure coefficients were calculated using 
the MPI-parallelised non-LTE radiative transfer 
code \balder{} \citep{2018A&A...615A.139A},
which is
our modified version of \multitd{} \citep{2009ASPC..415...87L_short}.
The code takes as input 
a model atom (\sect{methodatom})
and a model atmosphere (\sect{methodatmosphere}).
\balder{} uses the algorithm 
described in Sect.~2.4 of \citet{1992A&amp;A...262..209R}
to solve the equations of statistical equilibrium
\eqn{eq:stateq}, together with the radiative transfer
equation on short characteristics
\citep{2013A&amp;A...549A.126I}; the system
of equations are closed by enforcing population conservation.
Acceleration of convergence was achieved using
the generalised conjugate residual algorithm
(GCR; see the Appendix of \citealt{1990MWRv..118.1551K}, and 
\citealt[][]{saad2003iterative}).

The code \blue{} \citep{2016MNRAS.463.1518A}
was used within \balder{} to
determine the equation-of-state (assumed to be in LTE)
and the background opacities 
for around $10^{6}$ lines from atomic and ionic species
and $10^{8}$ lines from molecular species.
The bound-free (continuous) opacities were calculated on-the-fly, 
whilst the bound-bound (line) opacities were pre-computed
for different temperature-density-metallicity combinations,
and interpolated at runtime.  
Care was taken not to double-count opacities already included
in the model atoms. Partition functions
and dissociation constants were drawn
from \citet{2016A&amp;A...588A..96B},
and details of which background bound-bound and bound-free
transitions were included can be found in
Sect.~2.1.2 of \citealt{2016MNRAS.463.1518A}.

The non-LTE radiative transfer calculations were performed
independently for different elements,
for a variety of elemental abundances,
as we illustrate in \fig{fig:abgrid}.
Generally, for a given element $\mathrm{A}$, 
and for a given model atmosphere with chemical composition
labelled by $\feh$, calculations were performed for 
different abundances $-2\leq\xfe{A}\leq+2$, 
keeping the abundances of all other elements fixed 
(employing the trace element approximation),
but with the equation-of-state recomputed each time.
There are some exceptions, however:
for hydrogen, the calculations 
were kept strictly 
consistent with the composition of the model atmosphere;
and for lithium, calculations were performed for $-4\leq\xfe{Li}\leq+7$.

The departure coefficients are very weakly sensitive to 
the input microturbulence parameter.
A depth-independent microturbulence was adopted,
the value of which was chosen to be 
consistent with that of the model atmosphere 
(\sect{methodatmosphere}): namely,
$\vmic=1.0\,\kms$ for dwarfs ($\lggu>3.5$ in \fig{fig:grid}),
and $\vmic=2.0\,\kms$ for giants ($\lggu\leq3.5$ in \fig{fig:grid}).
Although the microturbulence is fixed to these values 
for the calculation of the grid of departure coefficients,
the microturbulence is varied in GALAH DR3 
and in the re-analysis we present in \sect{results},
using the empirical formula
described in \citet{2018MNRAS.478.4513B}.

The non-LTE calculations were assumed to have converged
once the monochromatic emergent intensities 
changed by less than $0.01\%$ between successive iterations.
Out of the total number of individual non-LTE runs ($383036$),
only a small number of them
($2089$) did not reach this convergence criterion.
Various types of convergence problems affected
different runs, depending on the element
and on the region of the parameter space.
Given the scope of this project these
problems could not be addressed individually.
Nevertheless, we found that typically the departure coefficients
of the runs that did not formally converge looked physically reasonable,
after comparing against those 
from converged runs having similar stellar parameters.
We include them in the data set, but cannot preclude
the possibility of some unphysical behaviour
in some regions of the parameter space.

\subsection{Background scattering}
\label{methodscattering}

Owing to limited computational resources at the time,
the results presented here and adopted into GALAH DR3
are based on an initial set of non-LTE calculations wherein 
all background species are assumed to strictly satisfy LTE.
In other words, the background source function is equal to
the Planck function:
\phantomsection\begin{IEEEeqnarray}{rCl}
\label{eq:background}
    S^{\text{background}}_{\nu}&=&B_{\nu}\,.
\end{IEEEeqnarray}
Separate calculations were later
carried out wherein scattering 
for lines from background atomic and ionic species 
were included, assuming it to be isotropic and coherent:
\phantomsection\begin{IEEEeqnarray}{rCl}
\label{eq:background}
    S^{\text{background}}_{\nu}&=&
    \epsilon_{\nu} B_{\nu} + (1-\epsilon_{\nu}) J_{\nu}\,.
\end{IEEEeqnarray}
Here, $J_{\nu}$ is the
mean radiation field, and $\epsilon_{\nu}$ is
the photon destruction probability.
For background lines, the latter quantity were estimated using
Eq.~3.98 of \citet{2003rtsa.book.....R}, which is valid for
two level atoms:
\phantomsection\begin{IEEEeqnarray}{rCl}
    \label{eq:destruction}
    \epsilon_{\nu;ji}&=&\frac{C_{ji}} {C_{ji}+A_{ji}+B_{ji} B_{\nu}}\,.
\end{IEEEeqnarray}
When estimating
the photon destruction probabilities,
electron collisions were assumed to dominate,
and these rates were estimated
using Eq.~22 of \citet{1962ApJ...136..906V}.
In addition, Thomson scattering of free electrons
and Rayleigh scattering of atomic hydrogen
in the red wing of the Lyman series
were also included, the latter
following \citet{2004MNRAS.347..802L}.

The two sets of calculations allow 
us to quantify the impact of background scattering on
GALAH DR3 abundances.  In general, the effects on
most of the GALAH lines are small, across the entire parameter space.
The effects on the \ion{Li}{I}, \ion{O}{I}, 
\ion{Na}{I}, \ion{Al}{I}, \ion{Si}{I},
and \ion{K}{I} lines in \tab{tab:linelist} are 
at most $0.01\,\dex$, for dwarfs and giants.
The effects are more severe for the
\ion{C}{I}, \ion{Mg}{I}, \ion{Ca}{I}, and \ion{Mn}{I} lines.
For these species, 
the effects can be important for giants ($\lggu\leq3.5$),
towards lower metallicities. Our tests imply that the
non-LTE GALAH results for carbon 
may be slightly overestimated, 
and the non-LTE GALAH results for magnesium, calcium,
and manganese abundances may be slightly underestimated,
in metal-poor giants by $0.01$ to $0.05\,\dex$; and in extreme cases
(mainly in the parameter space around $\teff\approx5000\,\K$,
$\lggu\approx2$, and $\feh\lesssim-2$)
by up to $0.1\,\dex$.
Nevertheless, metal-poor giants are only a small fraction of the GALAH 
sample (\fig{fig:grid}), 
hence treating the background in LTE is well justified
for the vast majority of stars,
including for the 
\ion{C}{I}, \ion{Mg}{I}, \ion{Ca}{I}, and \ion{Mn}{I} lines 
considered in this study.

\subsection{Implementation of departure coefficients into
LTE spectrum synthesis codes}
\label{methodimplementation}

The grids of departure coefficients can be found online
\citet{grid_nlte}
or by contacting the lead author directly.
These departure coefficients include
background scattering as described in \sect{methodscattering};
the departure coefficients that do not
include background scattering, which were used in
GALAH DR3, are also available upon request.
The departure coefficients, defined by 
\eqn{eq:depcoef}, are given for different energy levels $i$ of 
the different elements $\mathrm{A}$ in \tab{tab:atoms},
at each depth point $z$ in the different model atmospheres 
labelled by $\teff$, $\lgg$, and $\feh$ in
\sect{methodatmosphere}:
\phantomsection\begin{IEEEeqnarray}{rCl}
\label{eq:depcoefgrid}
    \dc{i} &\equiv& 
    \dc{i\left(\mathrm{A}\right)}\left(\xfe{A},\teff,\lgg,\feh,z\right)\,.
\end{IEEEeqnarray}
The levels in each of the model atoms were matched to 
the species, electron configurations, spectroscopic terms, and 
total angular momentum quantum numbers $J$ listed in the
NIST Atomic Spectra Database version 5.7 \citep{kramida2012nist}.
If $J$ was not resolved in the model atom,
the fine structure levels were assigned identical departure coefficients.
Levels in NIST that are missing in the model atoms were assigned
departure coefficients corresponding to those of the levels nearest
in energy, within the same spin system,
with zero-order extrapolation for most ions
(the exceptions being the ionisation
stages that do not extend to Rydberg states in the model atoms, 
including \ion{Ba}{I} and \ion{Mn}{II}).

In addition to ASCII format, grids are provided that are compatible 
with the spectrum synthesis code
Spectroscopy Made Easy (\sme{}; \citealt{1996A&amp;AS..118..595V}),
and the python version \pysme{}\footnote{\url{https://github.com/AWehrhahn/SME}}
(Wehrhahn et al. in prep.).
Details about the way in which the departure coefficients
are used to generate non-LTE synthetic spectra can be found in
Sect.~3 of \citet{2017A&amp;A...597A..16P}
and in the online documentation\footnote{\url{https://pysme-astro.readthedocs.io}}.
Given a line list, \sme{} or \pysme{} looks 
for the departure coefficients of the lower and upper levels
of all spectral lines by matching the species,
electron configurations, spectroscopic terms, and 
total angular momentum quantum numbers $J$ 
specified in the grids. It is therefore important
that the user ensures the format of these labels in their line list  
is consistent with the format used in the 
departure coefficient grids.

The grids of departure coefficients
(without background scattering)
have been implemented into the GALAH analysis pipeline,
that is based on \sme{} as described in \citet{2018MNRAS.478.4513B}.
The grids have been used for determining 
elemental abundances in GALAH DR3 (Buder et al. in prep.).
We have also used them in the line-by-line re-analysis of 
GALAH DR3 stars, as we discuss in \sect{results} below.

\section{Re-analysis of GALAH DR3 stars} 
\label{results}

\input{linelist.tex}

\begin{figure}
    \begin{center}
        \includegraphics[scale=0.31]{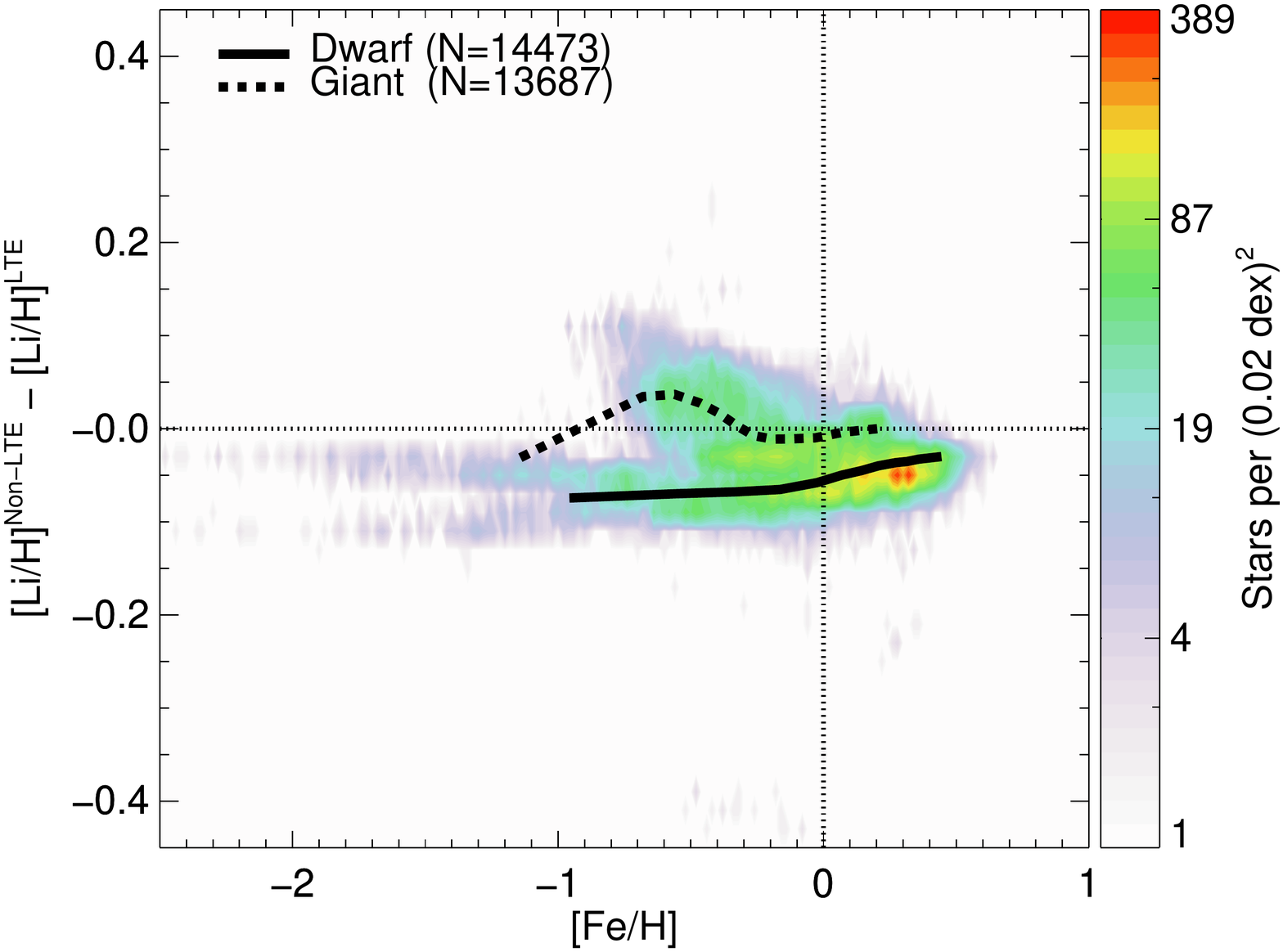}
        \caption{Differential abundance corrections for lithium.
        Binned data for dwarfs ($\lggu>3.5$) and
        giants ($\lggu\leq3.5$) are overplotted.}
        \label{fig:abcorli}
    \end{center}
\end{figure}

\begin{figure}
    \begin{center}
        \includegraphics[scale=0.31]{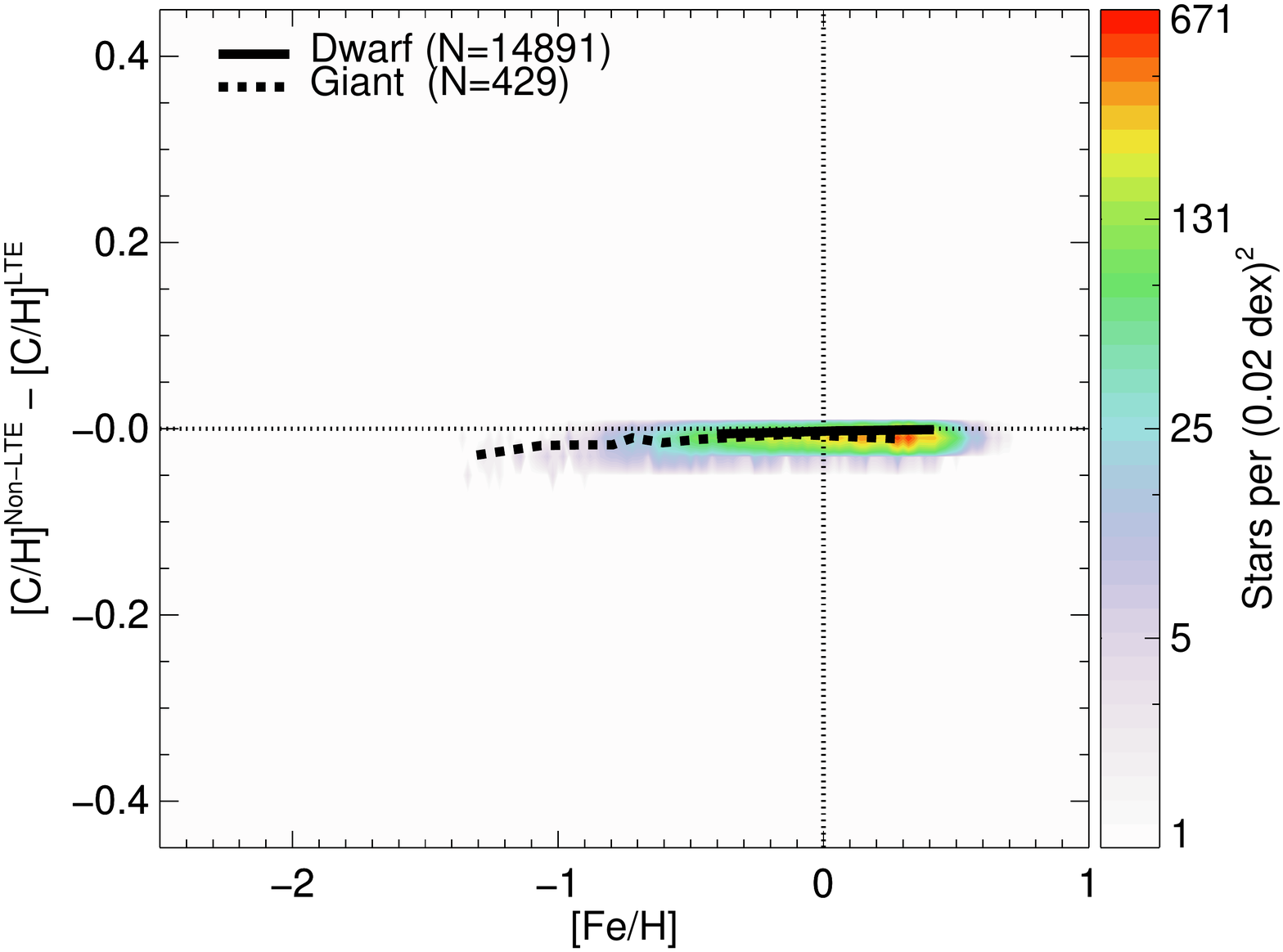}
        \caption{Differential abundance corrections for carbon.
        Binned data for dwarfs ($\lggu>3.5$) and
        giants ($\lggu\leq3.5$) are overplotted.}
        \label{fig:abcorc}
    \end{center}
\end{figure}

\begin{figure}
    \begin{center}
        \includegraphics[scale=0.31]{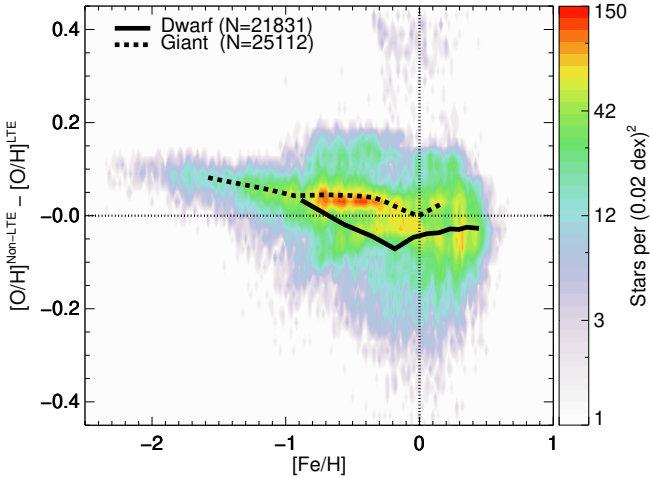}
        \caption{Differential abundance corrections for oxygen.
        Binned data for dwarfs ($\lggu>3.5$) and
        giants ($\lggu\leq3.5$) are overplotted.}
        \label{fig:abcoro}
    \end{center}
\end{figure}

\begin{figure}
    \begin{center}
        \includegraphics[scale=0.31]{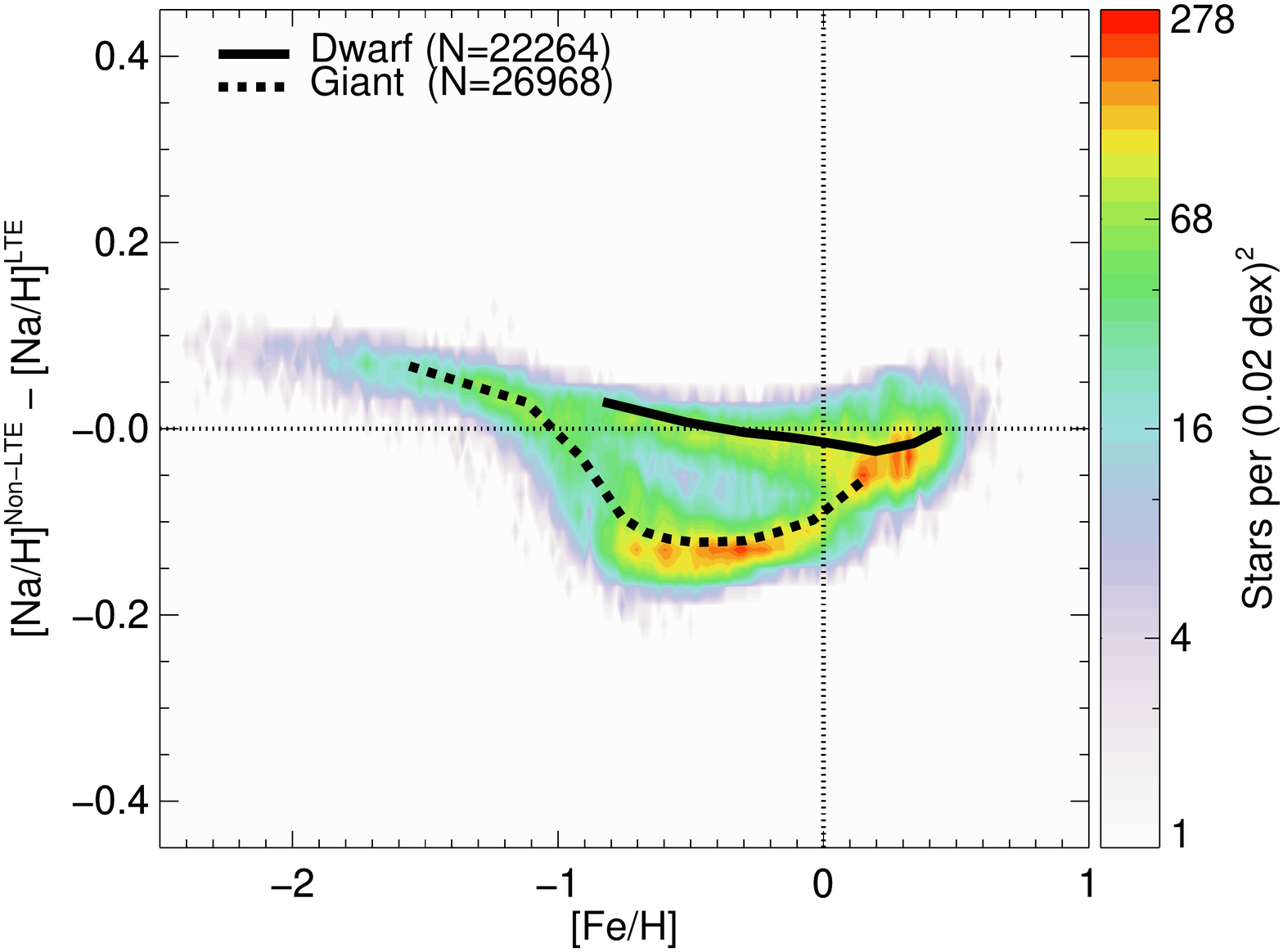}
        \caption{Differential abundance corrections for sodium.
        Binned data for dwarfs ($\lggu>3.5$) and
        giants ($\lggu\leq3.5$) are overplotted.}
        \label{fig:abcorna}
    \end{center}
\end{figure}

\begin{figure}
    \begin{center}
        \includegraphics[scale=0.31]{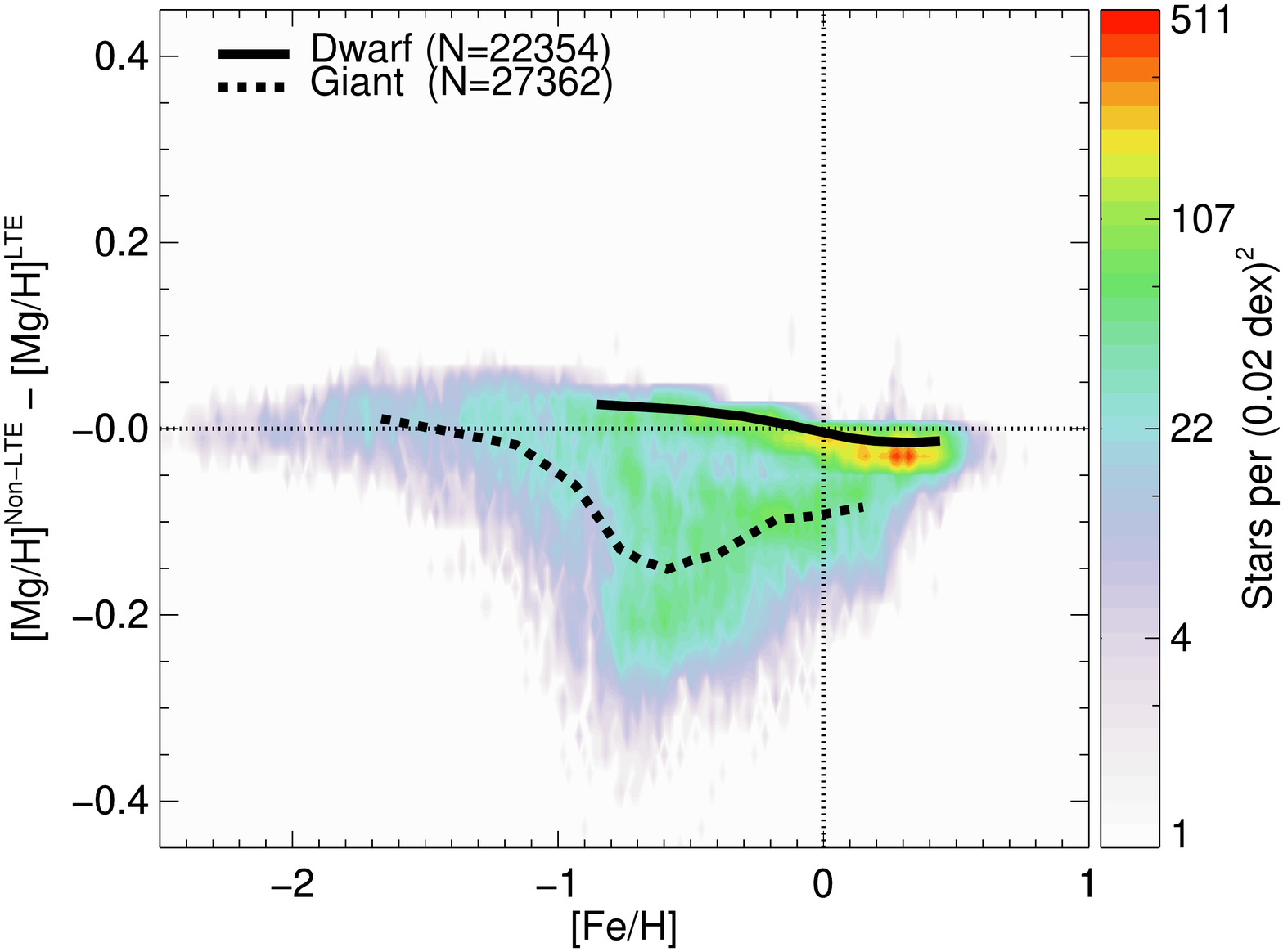}
        \caption{Differential abundance corrections for magnesium.
        Binned data for dwarfs ($\lggu>3.5$) and
        giants ($\lggu\leq3.5$) are overplotted.}
        \label{fig:abcormg}
    \end{center}
\end{figure}

\begin{figure}
    \begin{center}
        \includegraphics[scale=0.31]{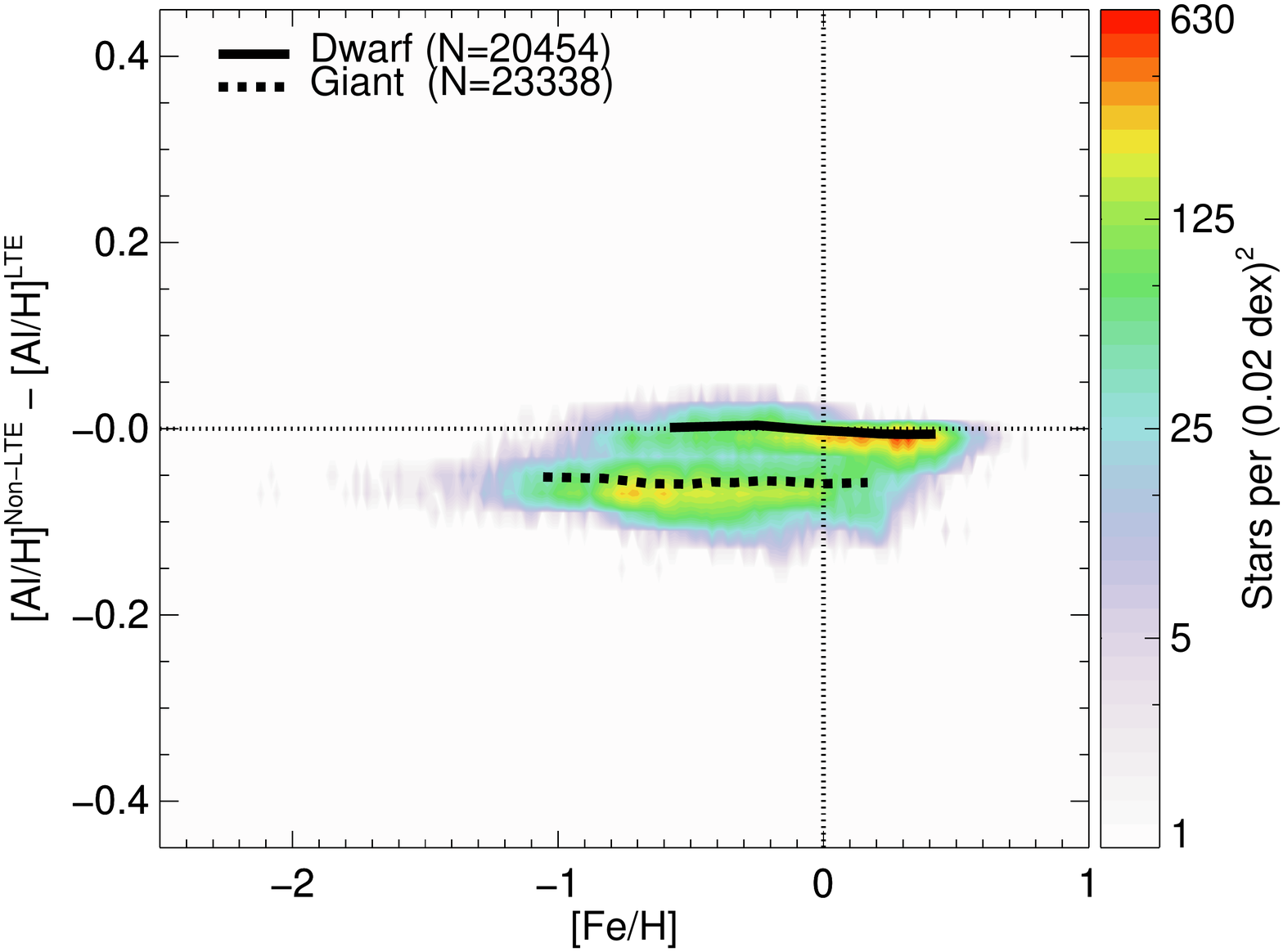}
        \caption{Differential abundance corrections for aluminium.
        Binned data for dwarfs ($\lggu>3.5$) and
        giants ($\lggu\leq3.5$) are overplotted.}
        \label{fig:abcoral}
    \end{center}
\end{figure}

\begin{figure}
    \begin{center}
        \includegraphics[scale=0.31]{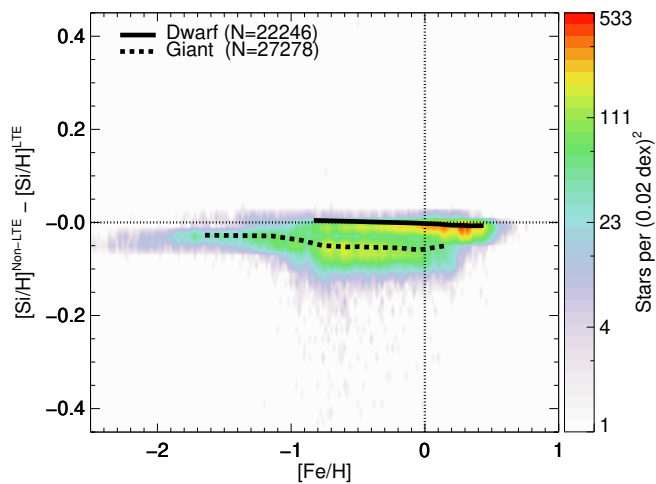}
        \caption{Differential abundance corrections for silicon.
        Binned data for dwarfs ($\lggu>3.5$) and
        giants ($\lggu\leq3.5$) are overplotted.}
        \label{fig:abcorsi}
    \end{center}
\end{figure}

\begin{figure}
    \begin{center}
        \includegraphics[scale=0.31]{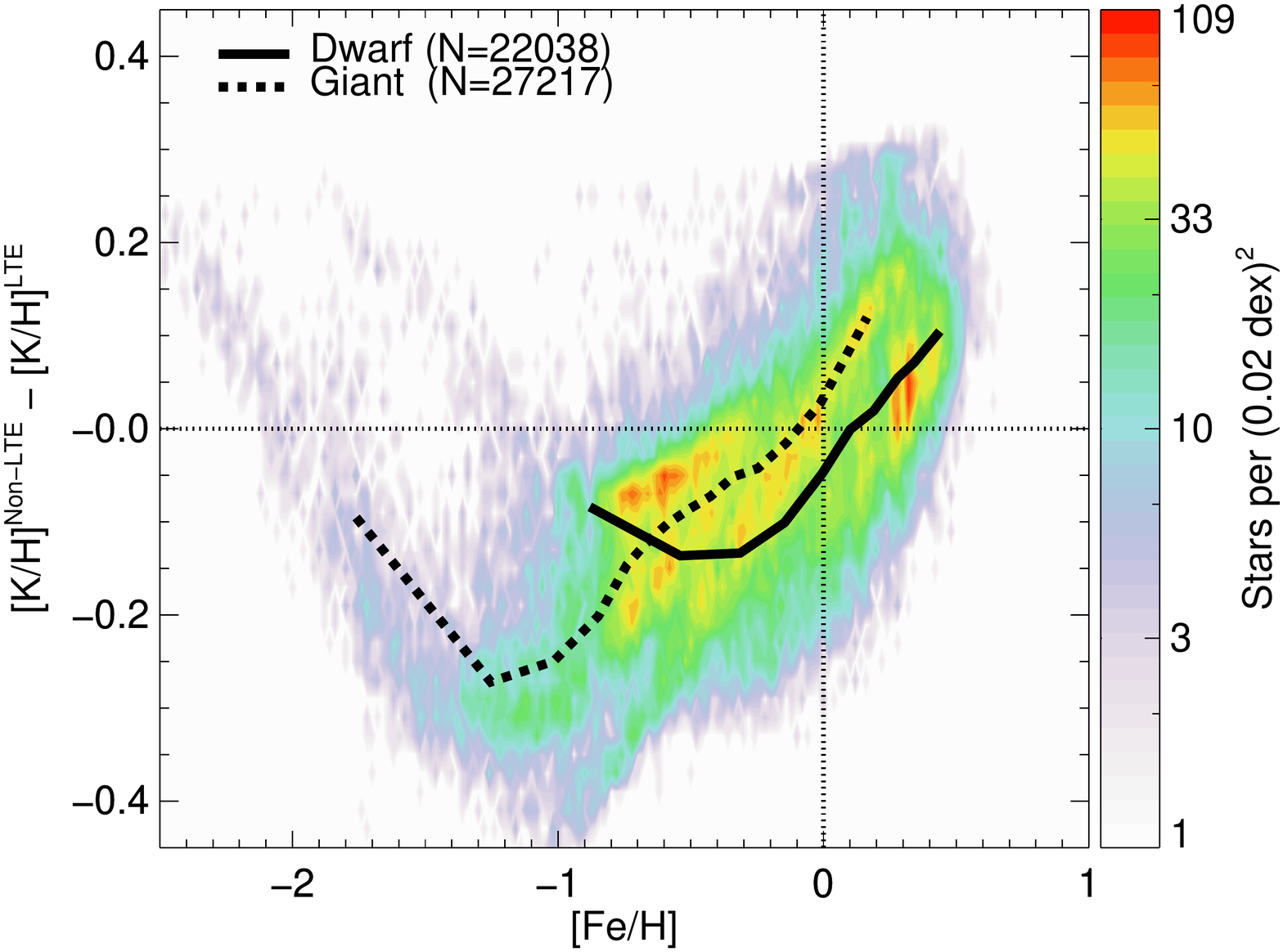}
        \caption{Differential abundance corrections for potassium.
        Binned data for dwarfs ($\lggu>3.5$) and
        giants ($\lggu\leq3.5$) are overplotted.}
        \label{fig:abcork}
    \end{center}
\end{figure}

\begin{figure}
    \begin{center}
        \includegraphics[scale=0.31]{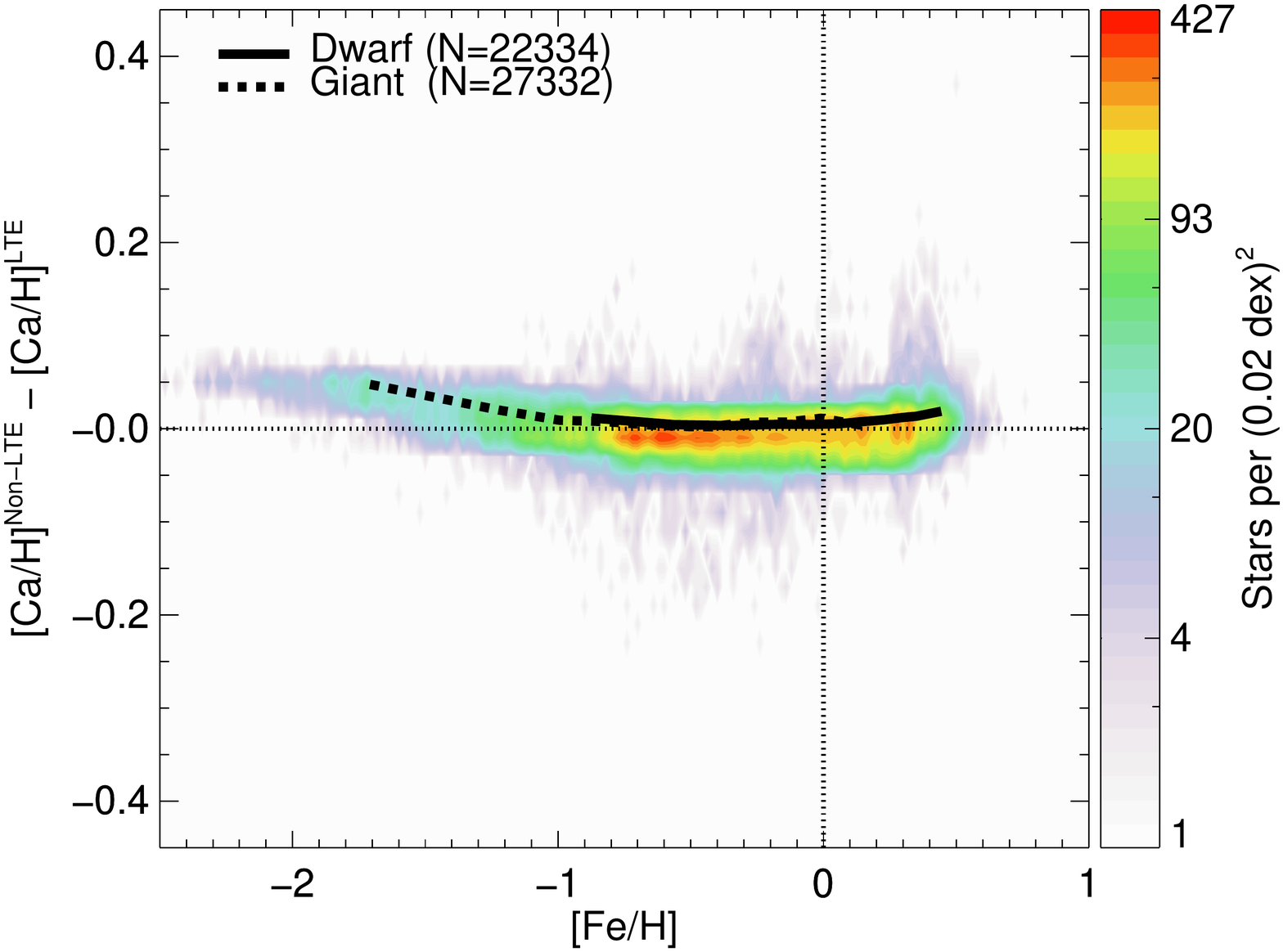}
        \caption{Differential abundance corrections for calcium.
        Binned data for dwarfs ($\lggu>3.5$) and
        giants ($\lggu\leq3.5$) are overplotted.}
        \label{fig:abcorca}
    \end{center}
\end{figure}

\begin{figure}
    \begin{center}
        \includegraphics[scale=0.31]{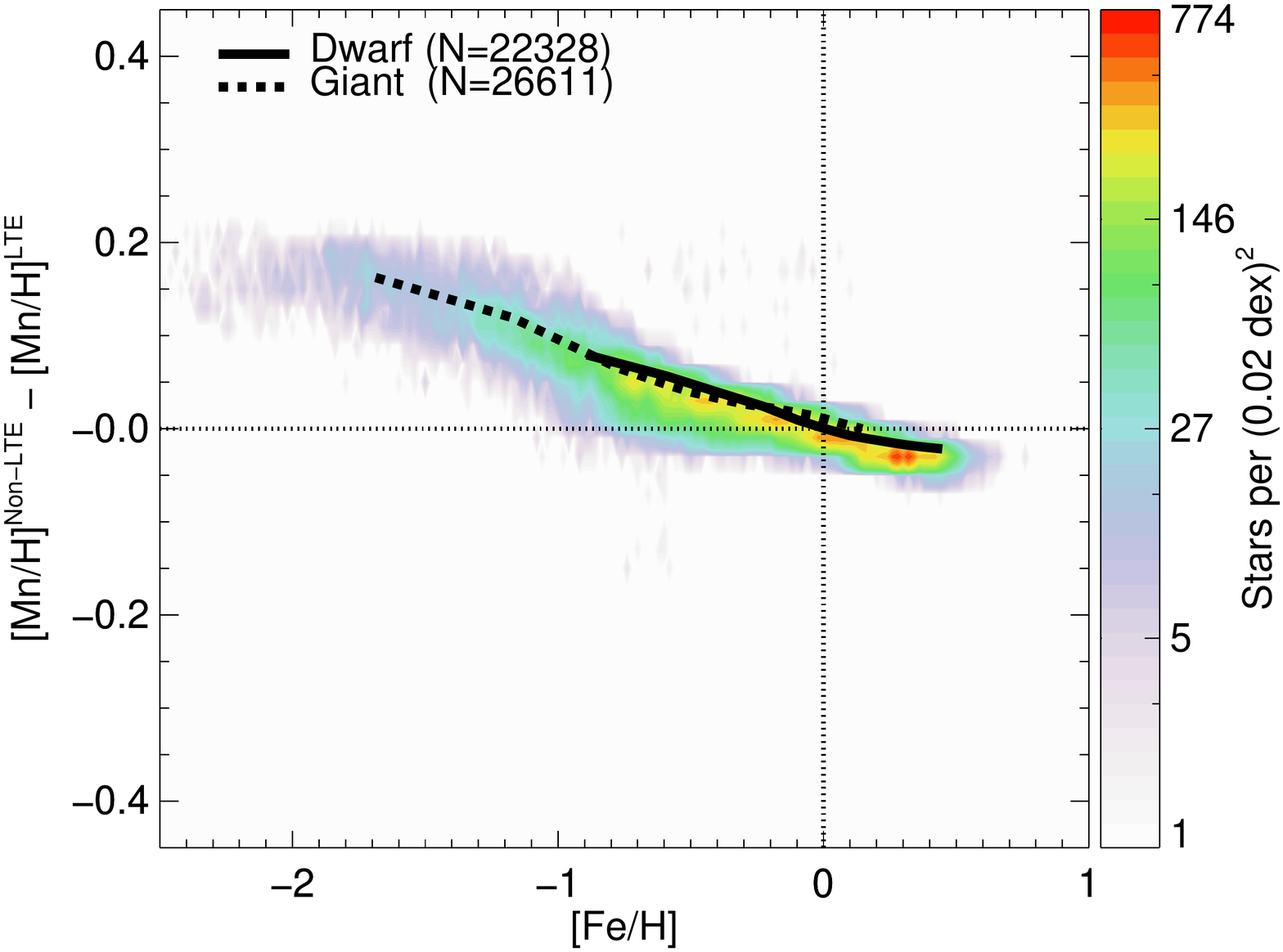}
        \caption{Differential abundance corrections for manganese.
        Binned data for dwarfs ($\lggu>3.5$) and
        giants ($\lggu\leq3.5$) are overplotted.}
        \label{fig:abcormn}
    \end{center}
\end{figure}

\begin{figure}
    \begin{center}
        \includegraphics[scale=0.31]{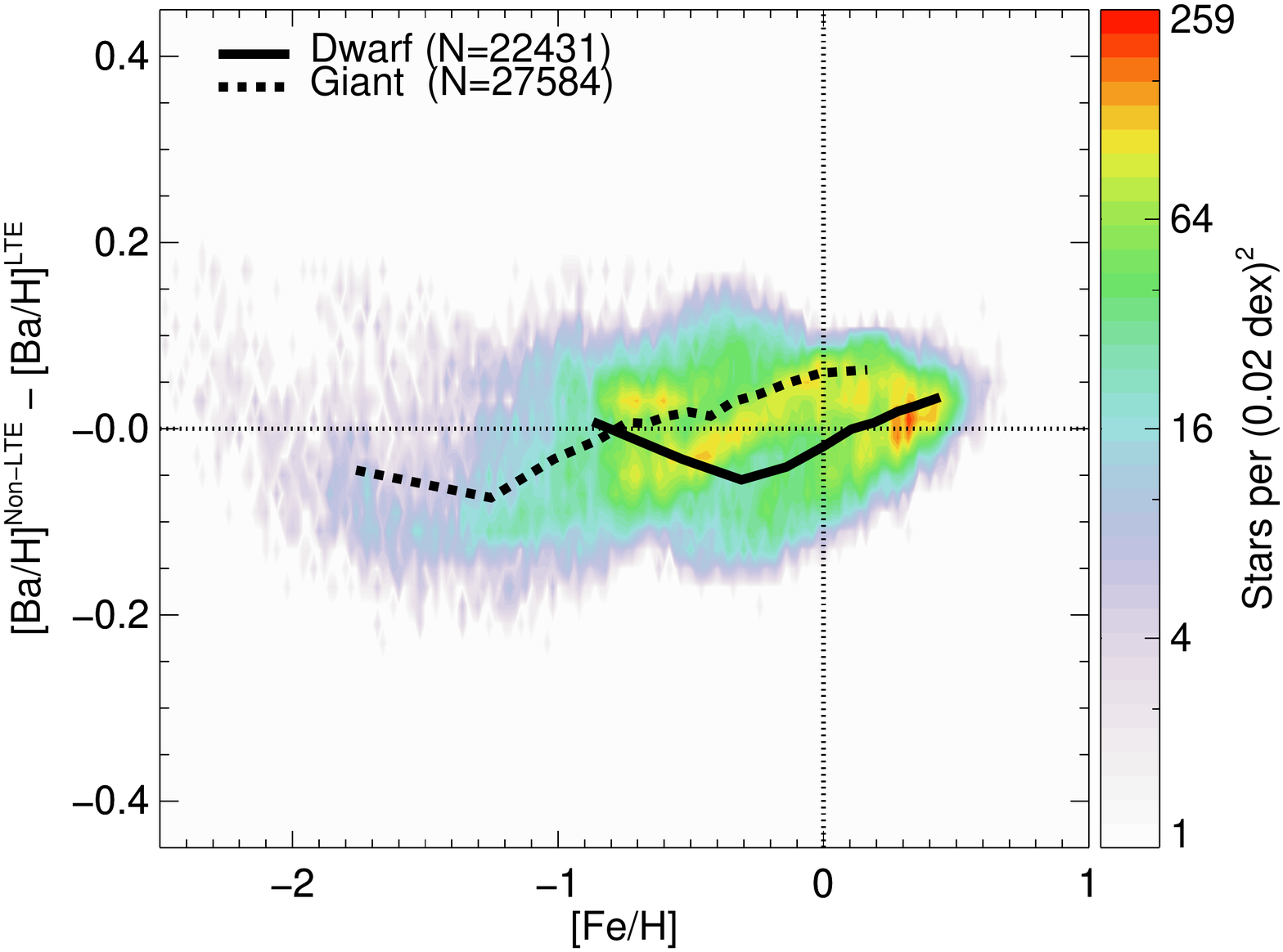}
        \caption{Differential abundance corrections for barium.
        Binned data for dwarfs ($\lggu>3.5$) and
        giants ($\lggu\leq3.5$) are overplotted.}
        \label{fig:abcorba}
    \end{center}
\end{figure}

\begin{figure*}
    \begin{center}
        \includegraphics[scale=0.31]{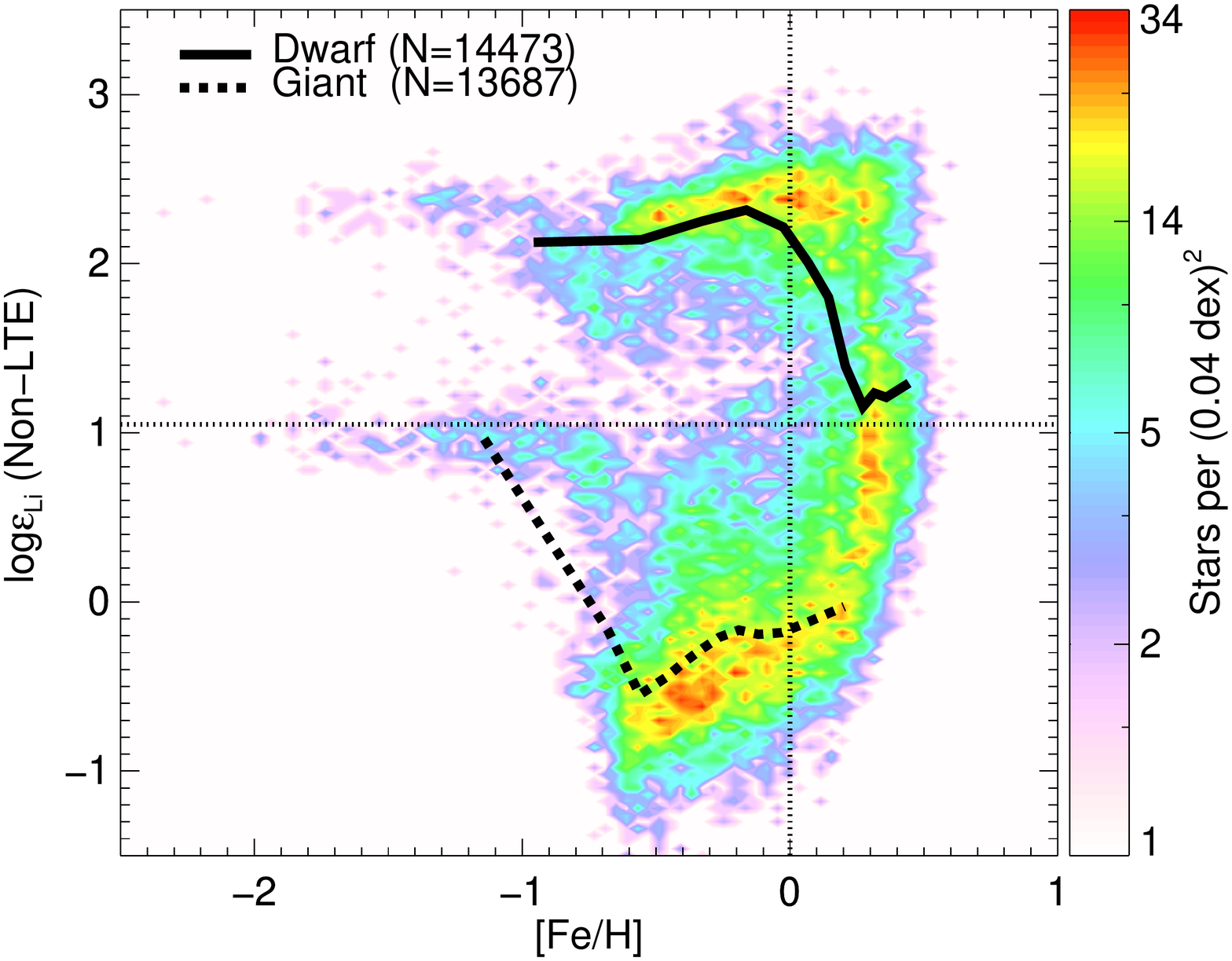}\includegraphics[scale=0.31]{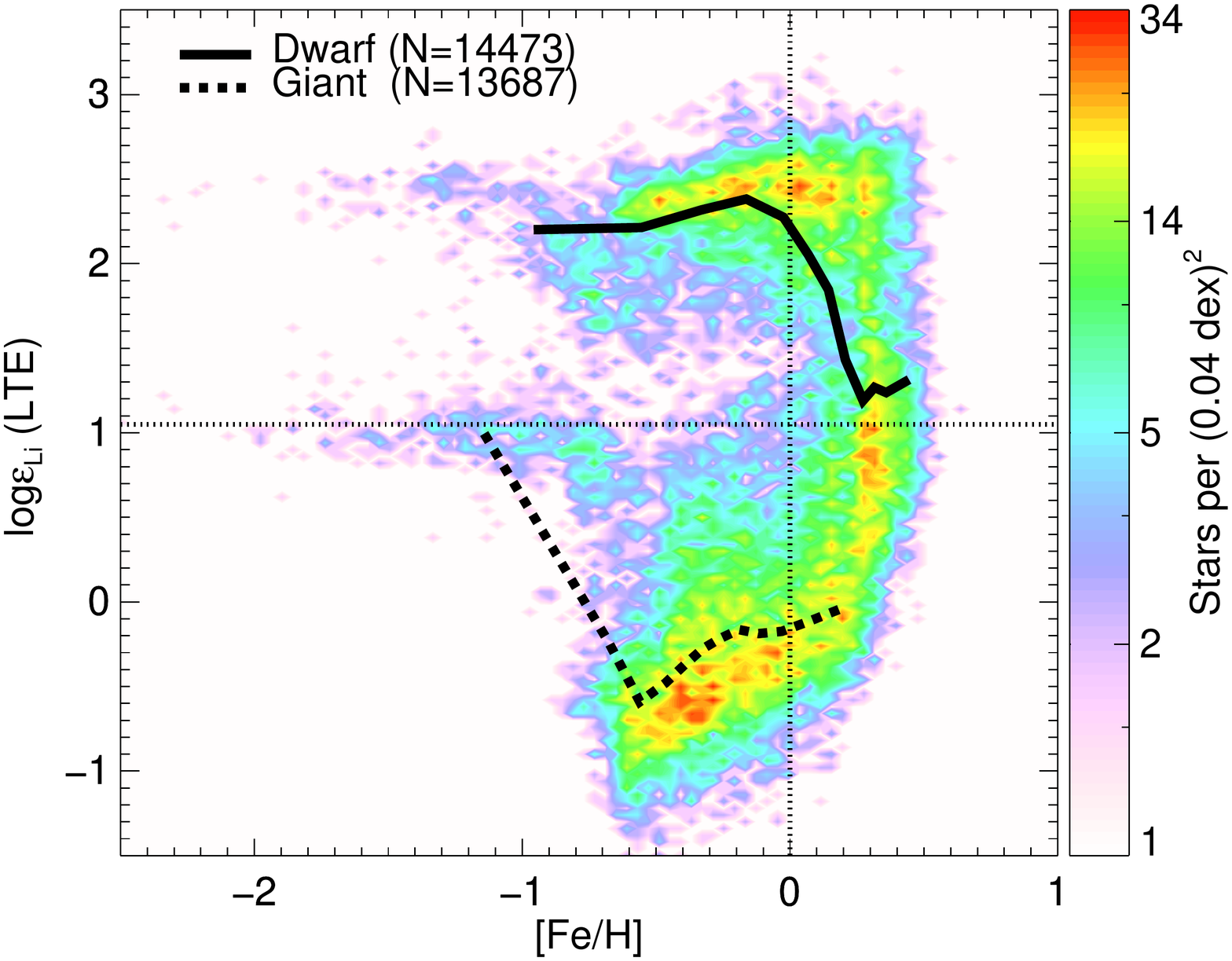}
        \caption{Non-LTE (left) and LTE (right) lithium abundances,
        with non-LTE $\feh$ adopted from GALAH DR3 in both cases.
        Overplotted are binned data for dwarfs ($\lggu>3.5$) and
        giants ($\lggu\leq3.5$).}
        \label{fig:li}
    \end{center}
\end{figure*}

\begin{figure*}
    \begin{center}
        \includegraphics[scale=0.31]{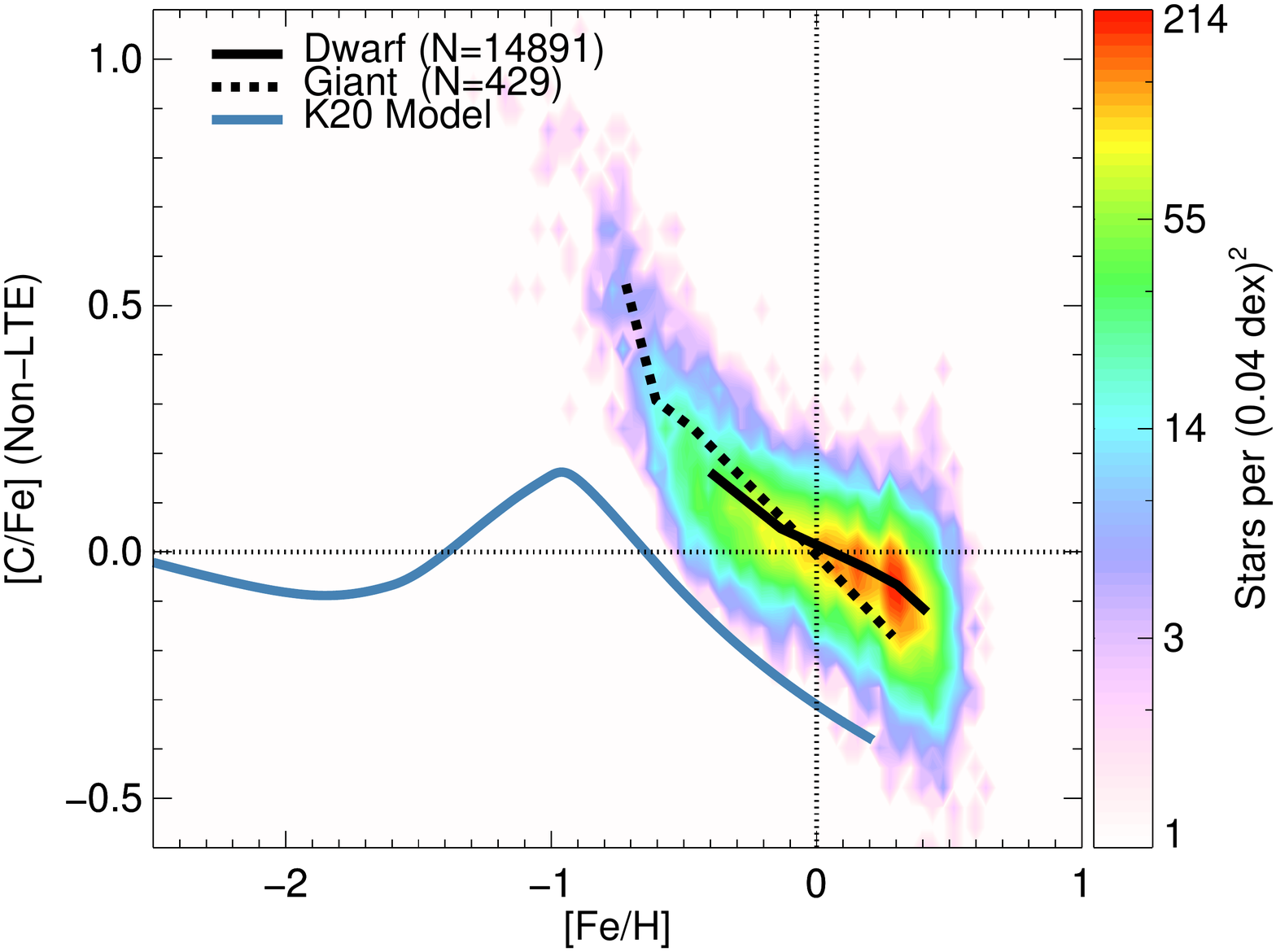}\includegraphics[scale=0.31]{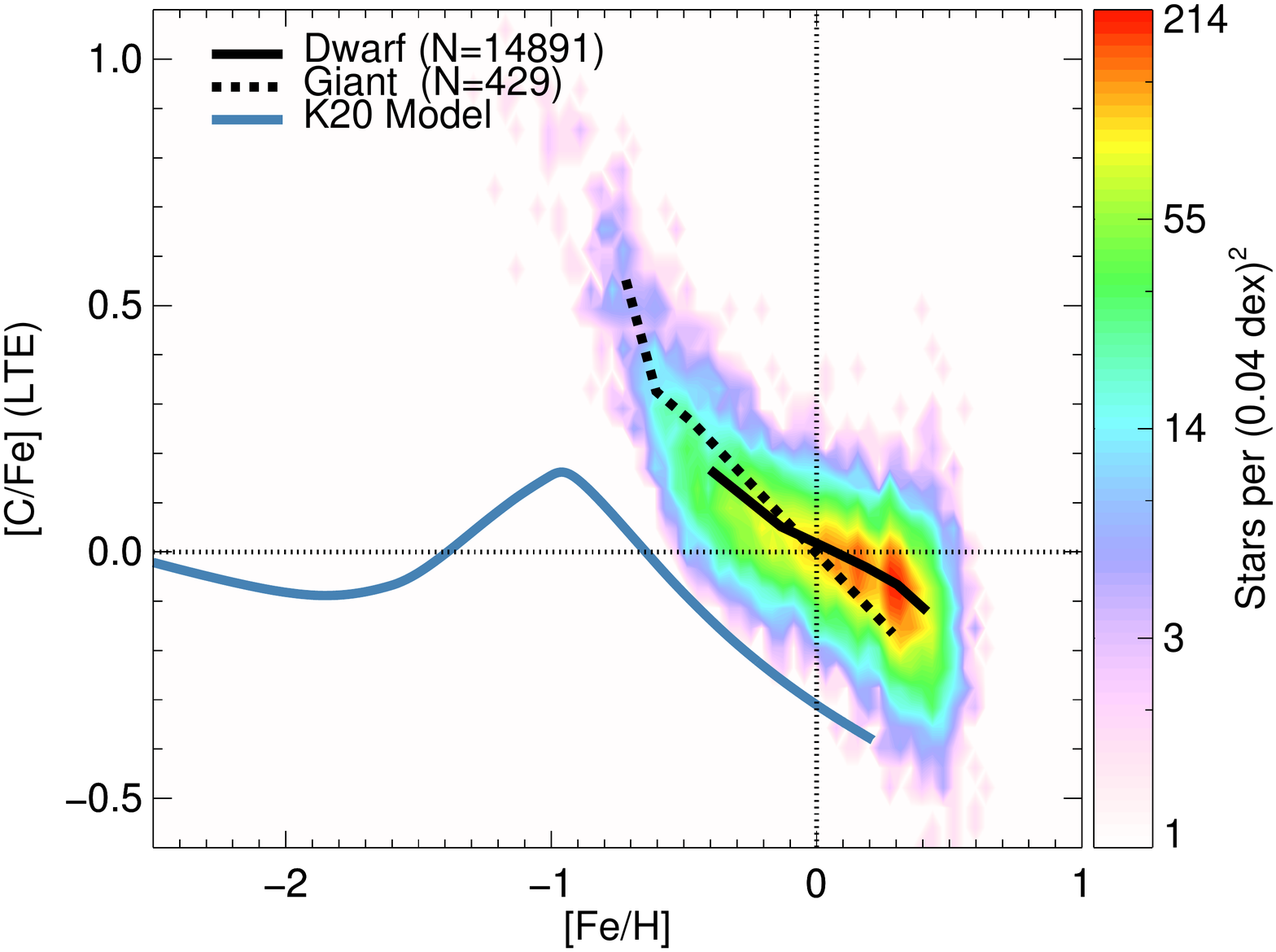}
        \caption{Non-LTE (left) and LTE (right) carbon abundances,
        with non-LTE $\feh$ adopted from GALAH DR3 in both cases.
        Overplotted are binned data for dwarfs ($\lggu>3.5$) and
        giants ($\lggu\leq3.5$), and the GCE model of K20.} 
        \label{fig:c}
    \end{center}
\end{figure*}

\begin{figure*}
    \begin{center}
        \includegraphics[scale=0.31]{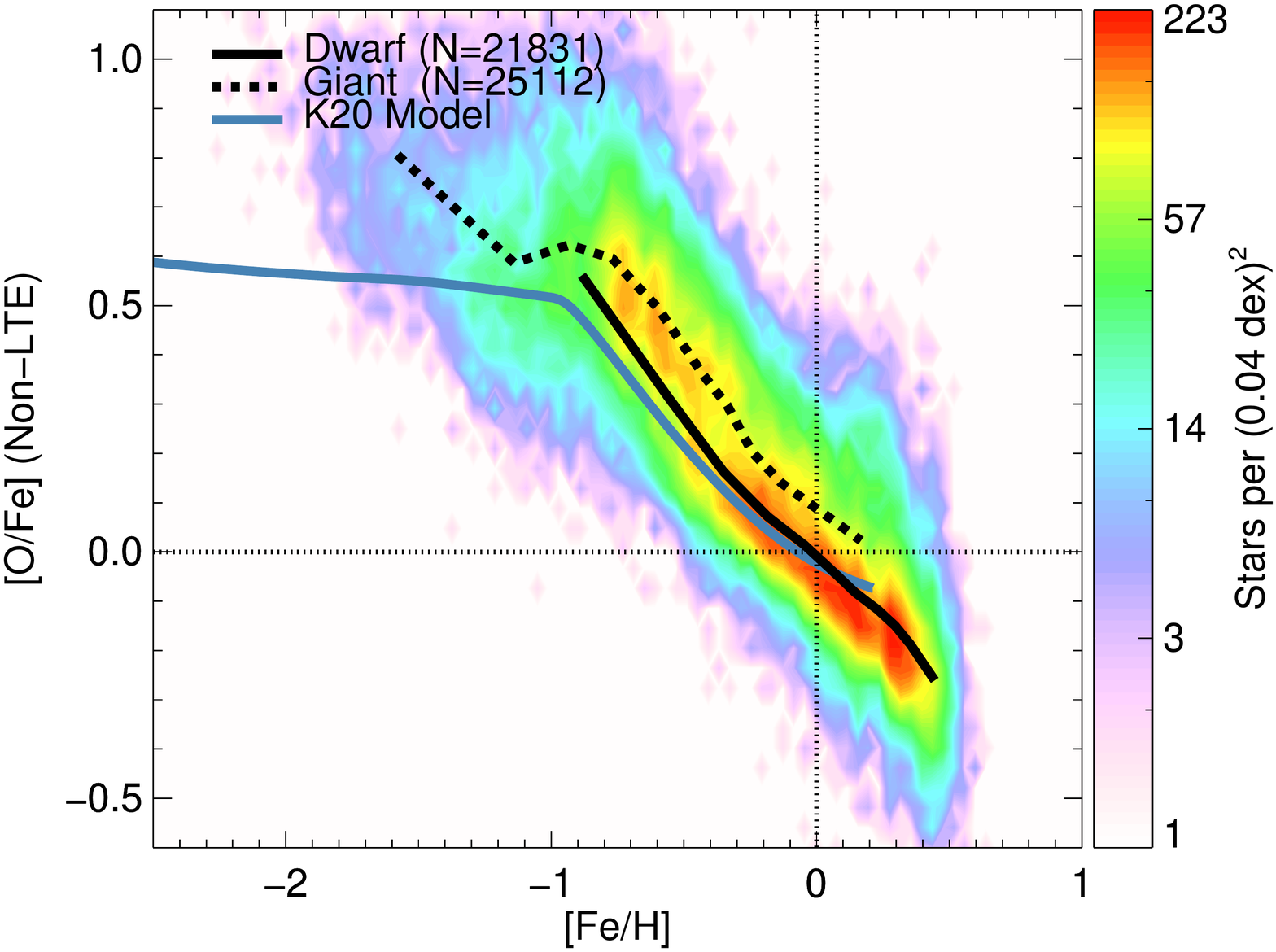}\includegraphics[scale=0.31]{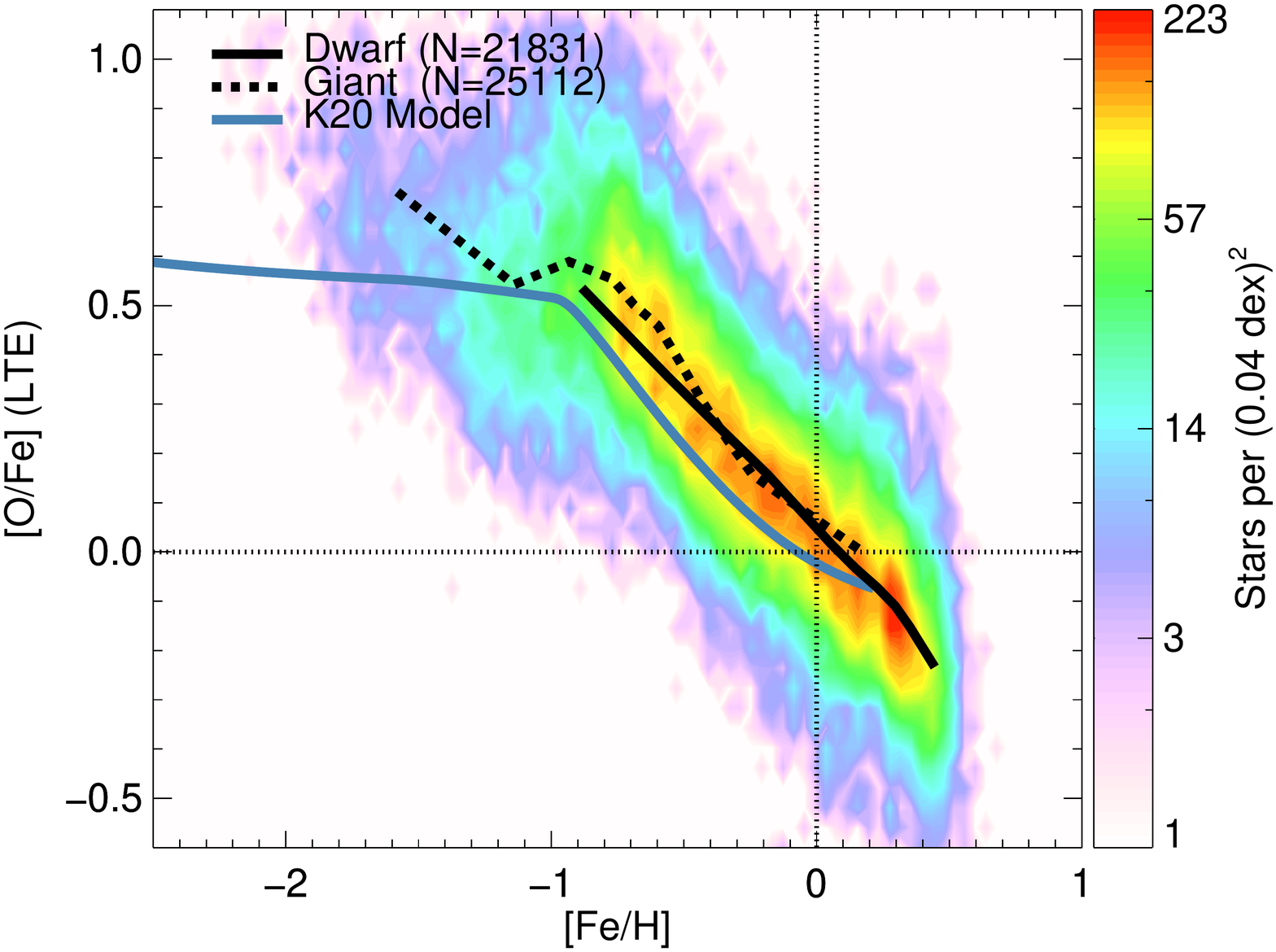}
        \caption{Non-LTE (left) and LTE (right) oxygen abundances,
        with non-LTE $\feh$ adopted from GALAH DR3 in both cases.
        Overplotted are binned data for dwarfs ($\lggu>3.5$) and
        giants ($\lggu\leq3.5$), and the GCE model of K20.} 
        \label{fig:o}
    \end{center}
\end{figure*}

\begin{figure*}
    \begin{center}
        \includegraphics[scale=0.31]{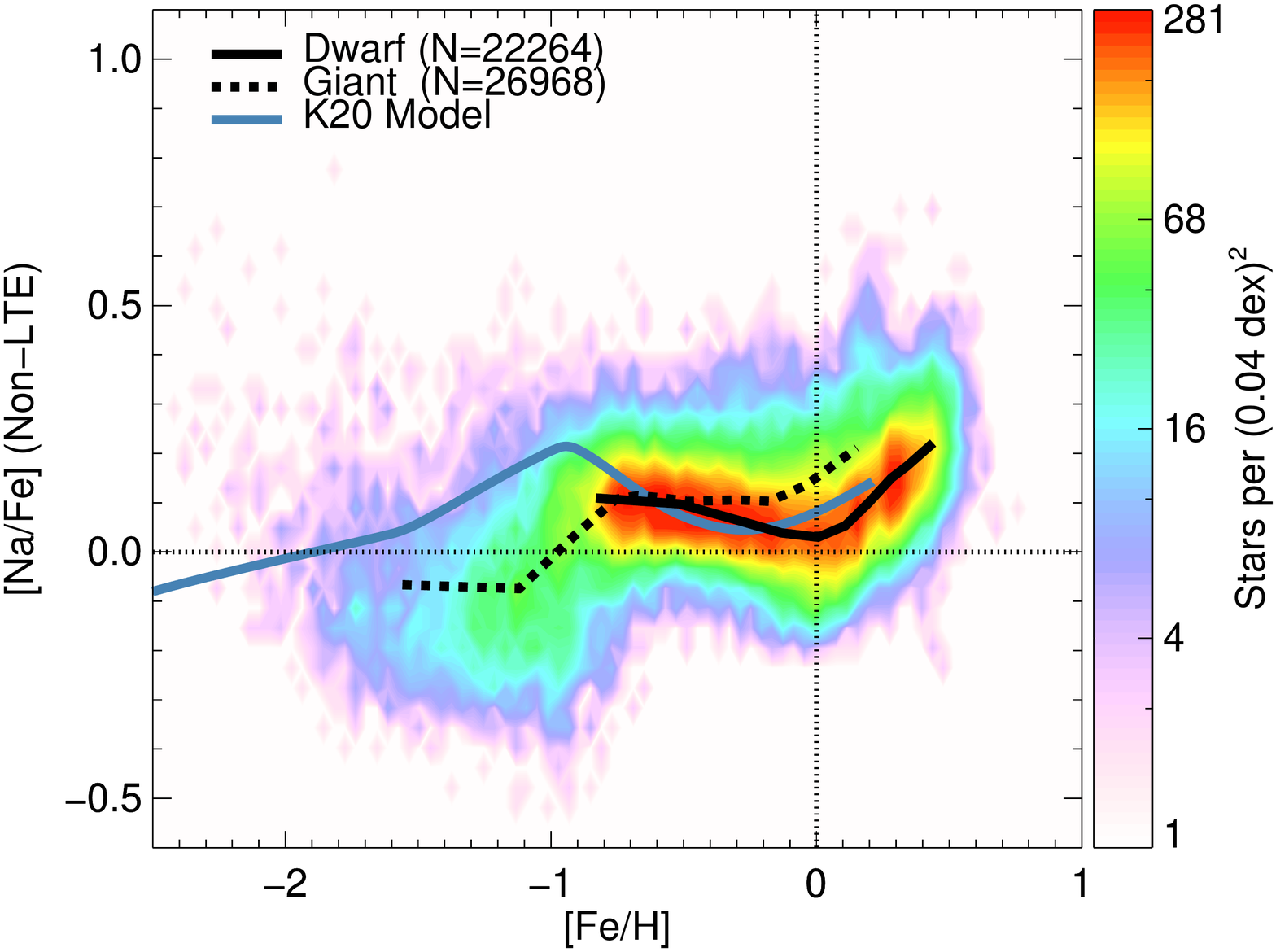}\includegraphics[scale=0.31]{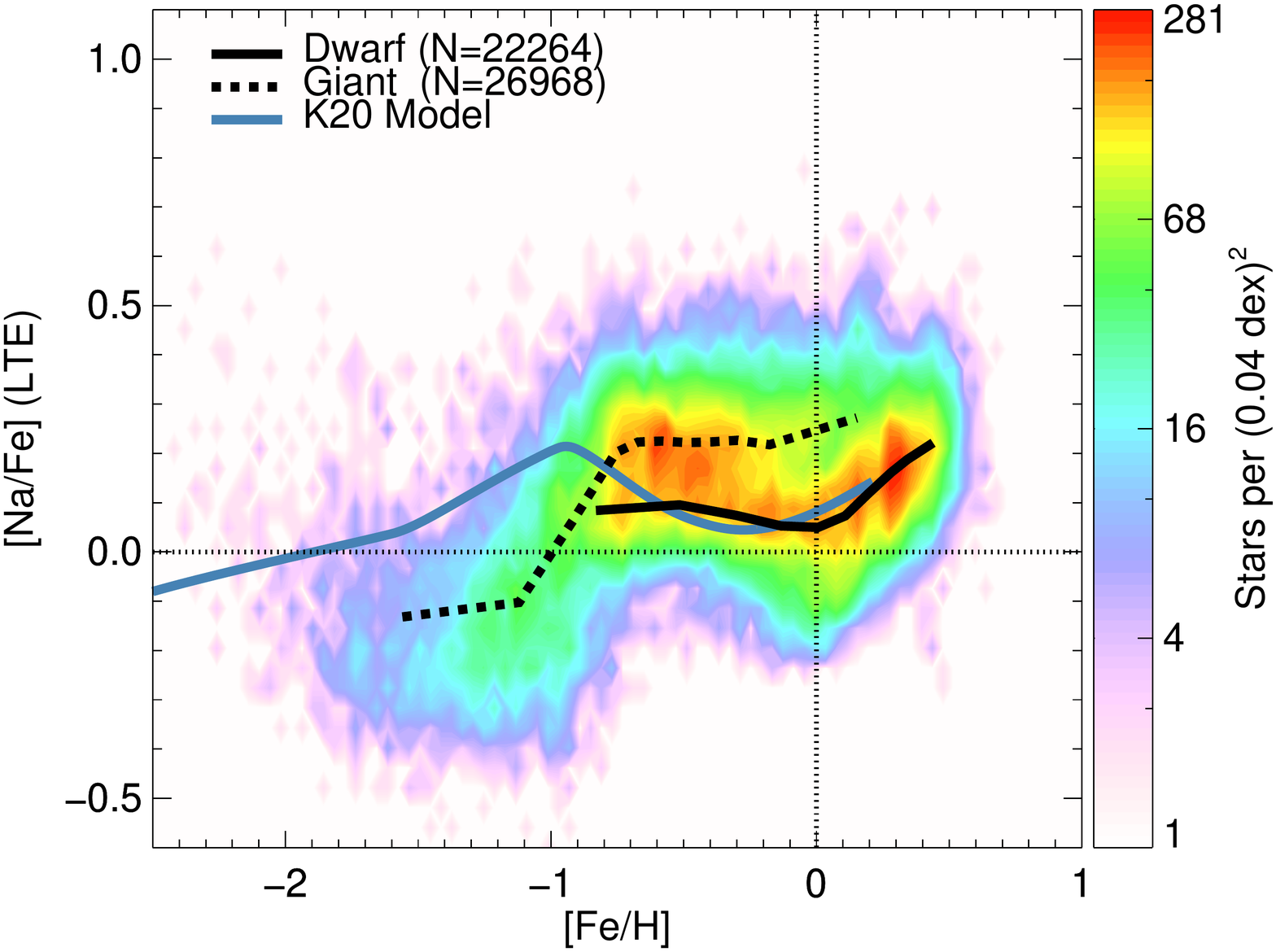}
        \caption{Non-LTE (left) and LTE (right) sodium abundances,
        with non-LTE $\feh$ adopted from GALAH DR3 in both cases.
        Overplotted are binned data for dwarfs ($\lggu>3.5$) and
        giants ($\lggu\leq3.5$), and the GCE model of K20.} 
        \label{fig:na}
    \end{center}
\end{figure*}

\begin{figure*}
    \begin{center}
        \includegraphics[scale=0.31]{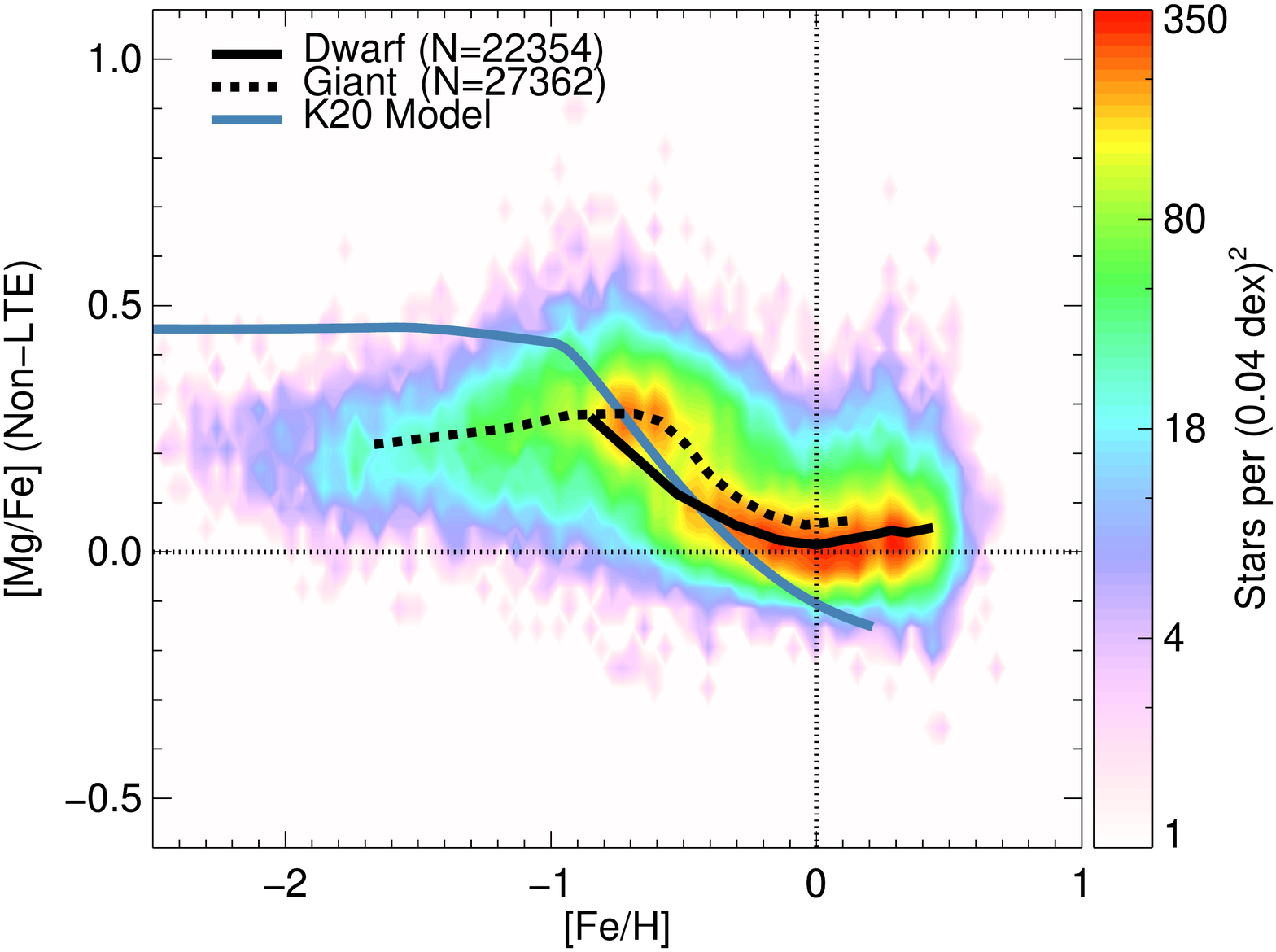}\includegraphics[scale=0.31]{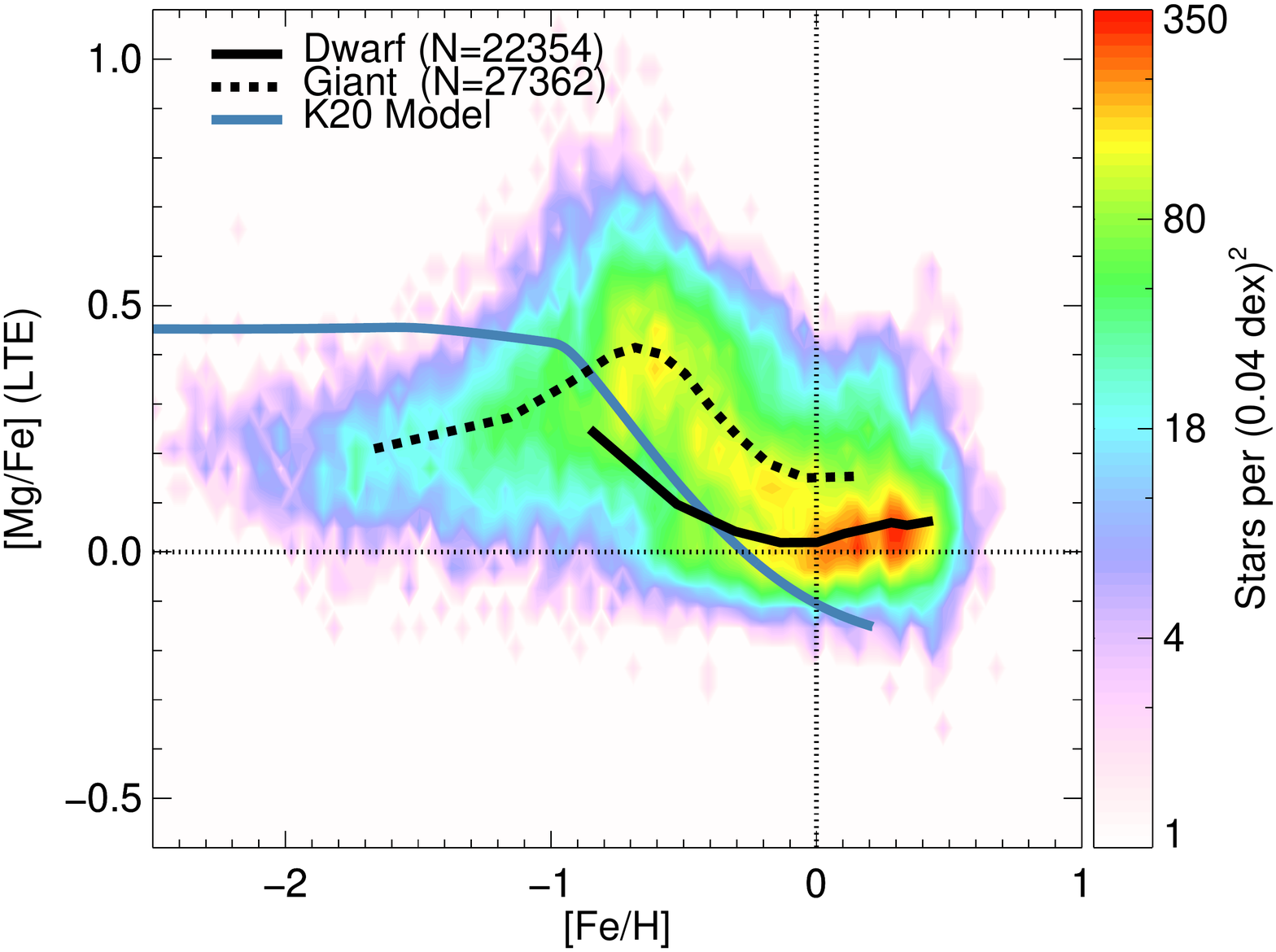}
        \caption{Non-LTE (left) and LTE (right) magnesium abundances,
        with non-LTE $\feh$ adopted from GALAH DR3 in both cases.
        Overplotted are binned data for dwarfs ($\lggu>3.5$) and
        giants ($\lggu\leq3.5$), and the GCE model of K20.} 
        \label{fig:mg}
    \end{center}
\end{figure*}

\begin{figure*}
    \begin{center}
        \includegraphics[scale=0.31]{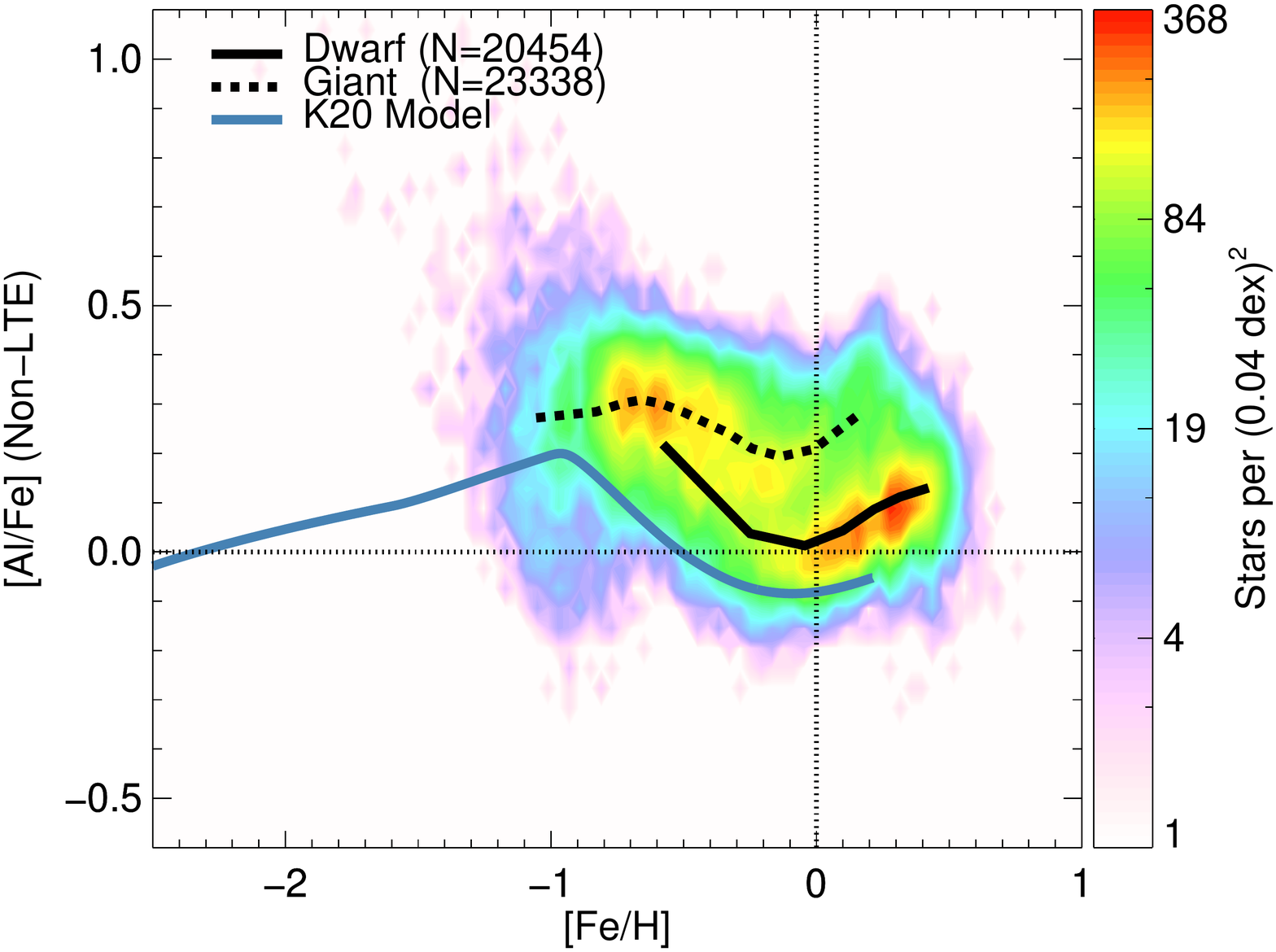}\includegraphics[scale=0.31]{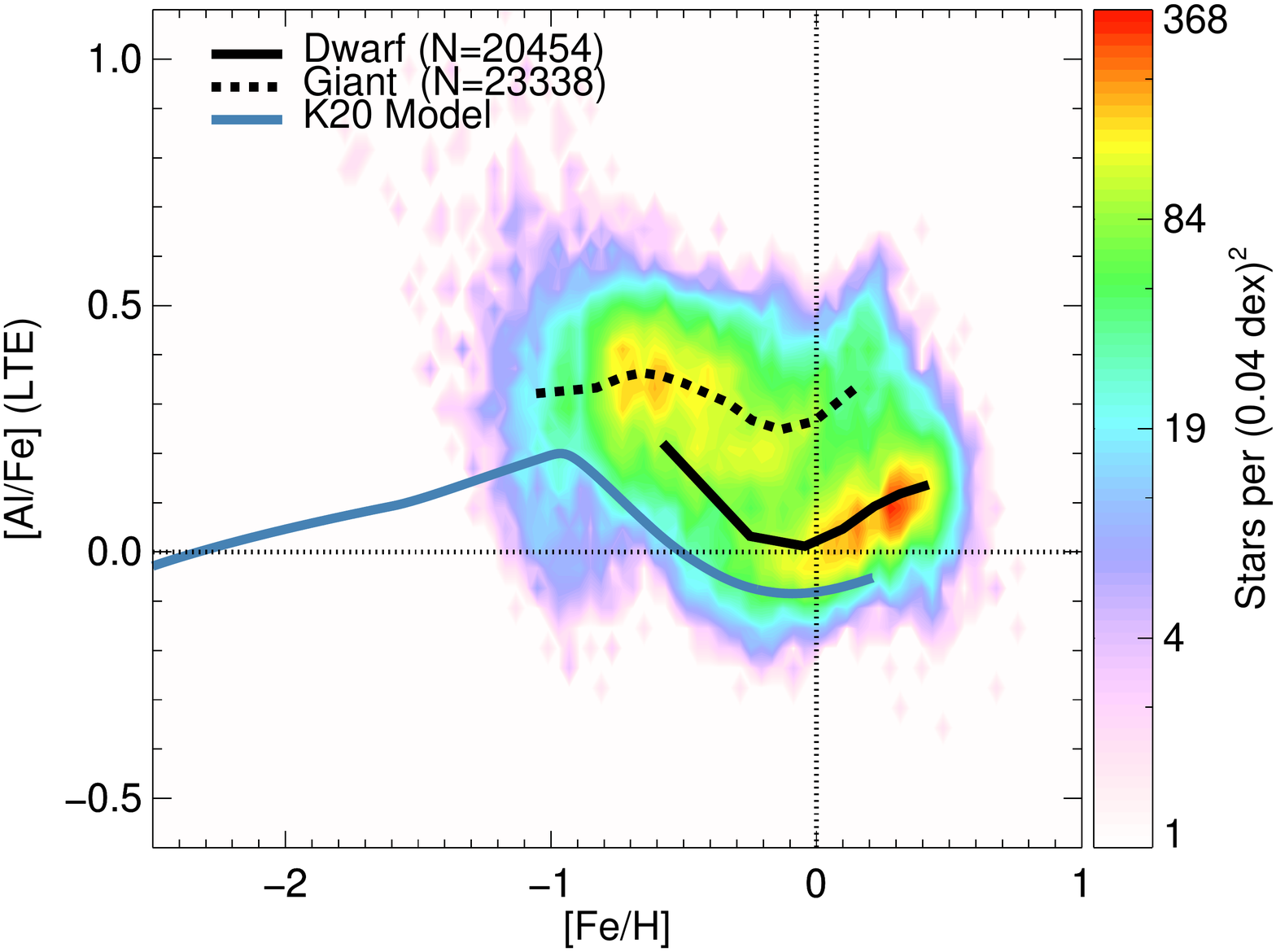}
        \caption{Non-LTE (left) and LTE (right) aluminium abundances,
        with non-LTE $\feh$ adopted from GALAH DR3 in both cases.
        Overplotted are binned data for dwarfs ($\lggu>3.5$) and
        giants ($\lggu\leq3.5$), and the GCE model of K20.} 
        \label{fig:al}
    \end{center}
\end{figure*}

\begin{figure*}
    \begin{center}
        \includegraphics[scale=0.31]{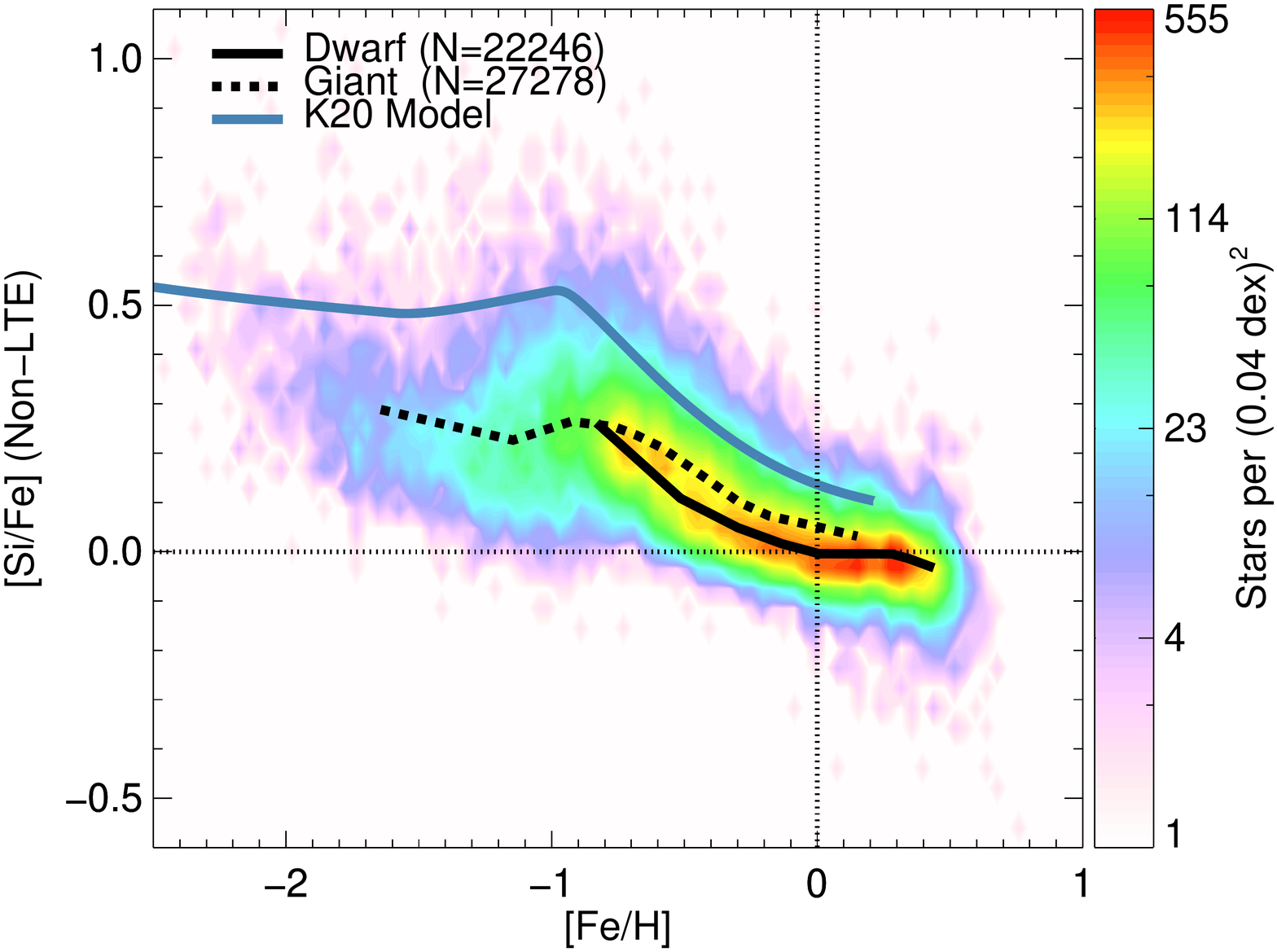}\includegraphics[scale=0.31]{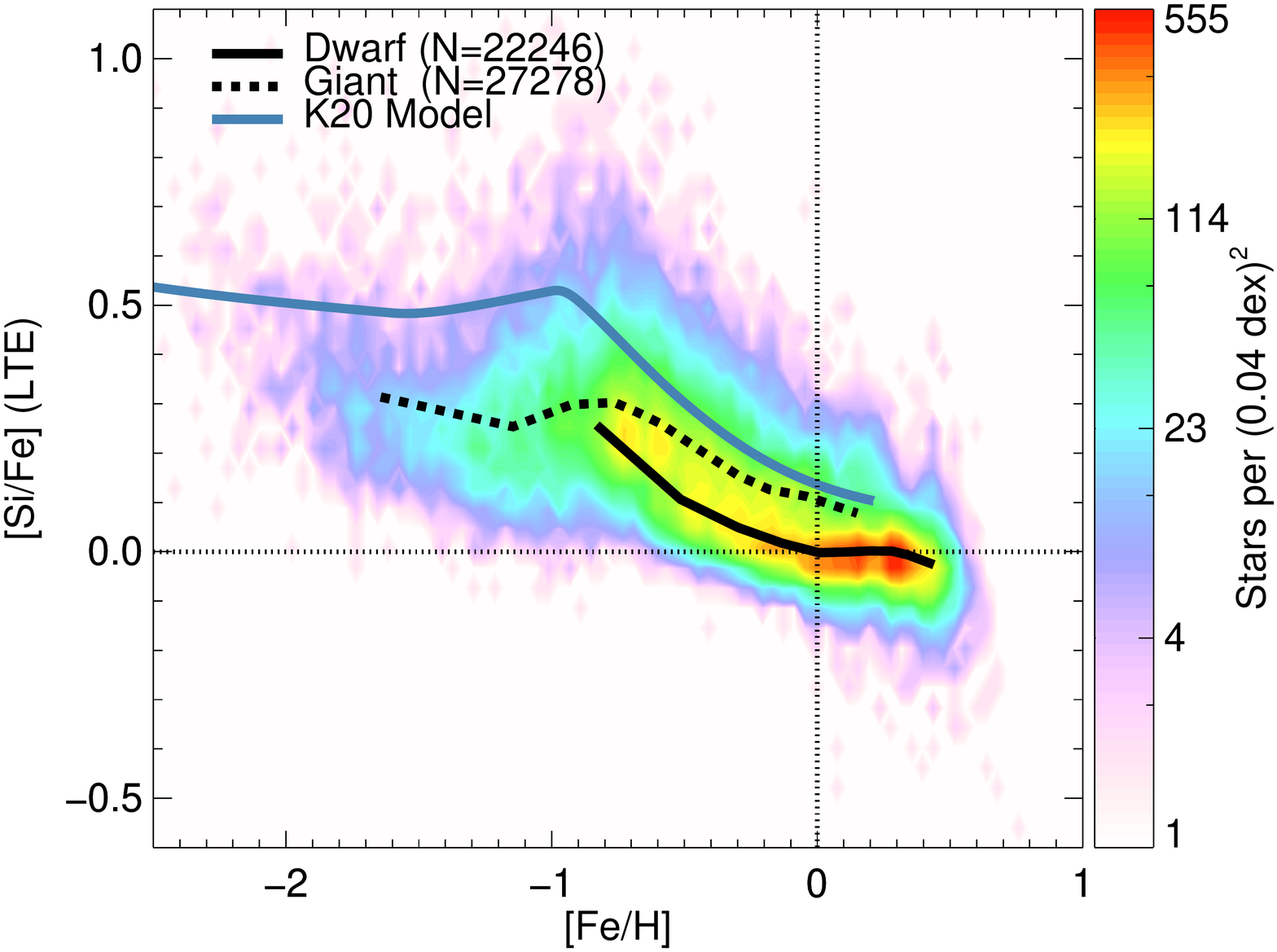}
        \caption{Non-LTE (left) and LTE (right) silicon abundances,
        with non-LTE $\feh$ adopted from GALAH DR3 in both cases.
        Overplotted are binned data for dwarfs ($\lggu>3.5$) and
        giants ($\lggu\leq3.5$), and the GCE model of K20.} 
        \label{fig:si}
    \end{center}
\end{figure*}

\begin{figure*}
    \begin{center}
        \includegraphics[scale=0.31]{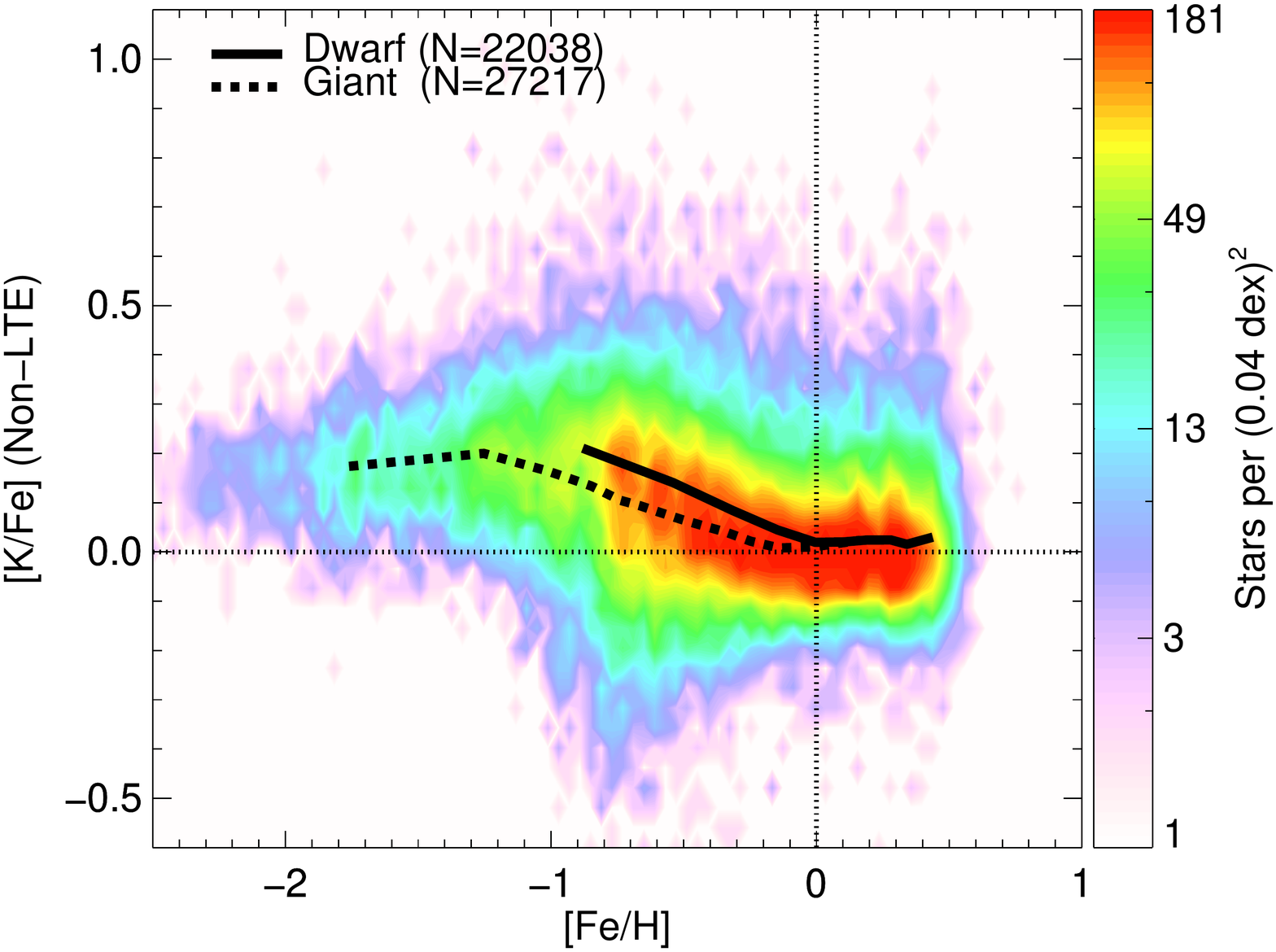}\includegraphics[scale=0.31]{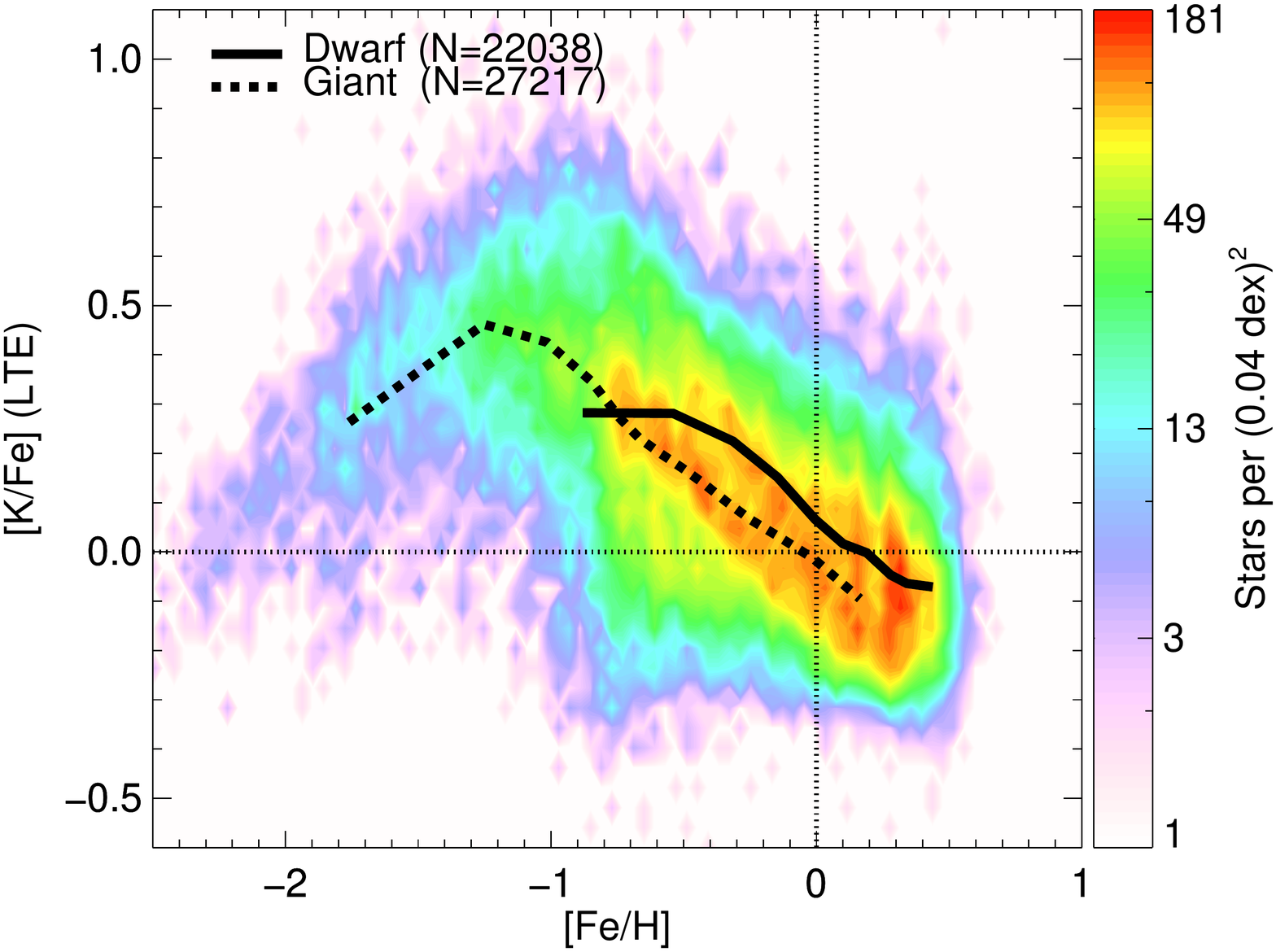}
        \caption{Non-LTE (left) and LTE (right) potassium abundances,
        with non-LTE $\feh$ adopted from GALAH DR3 in both cases.
        Overplotted are binned data for dwarfs ($\lggu>3.5$) and
        giants ($\lggu\leq3.5$). The GCE model of K20 falls below the
        vertical axis of this plot.}
        \label{fig:k}
    \end{center}
\end{figure*}

\begin{figure*}
    \begin{center}
        \includegraphics[scale=0.31]{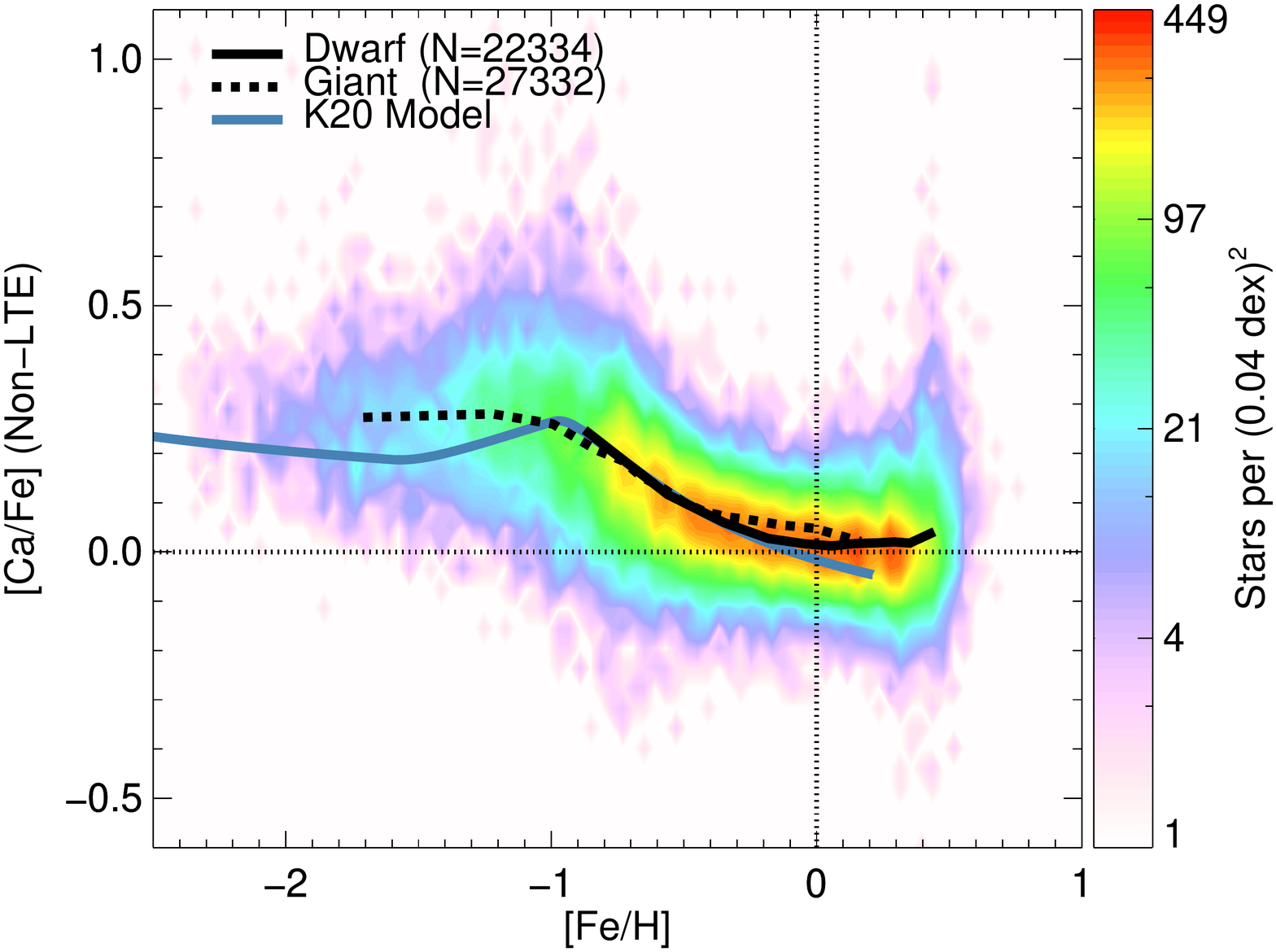}\includegraphics[scale=0.31]{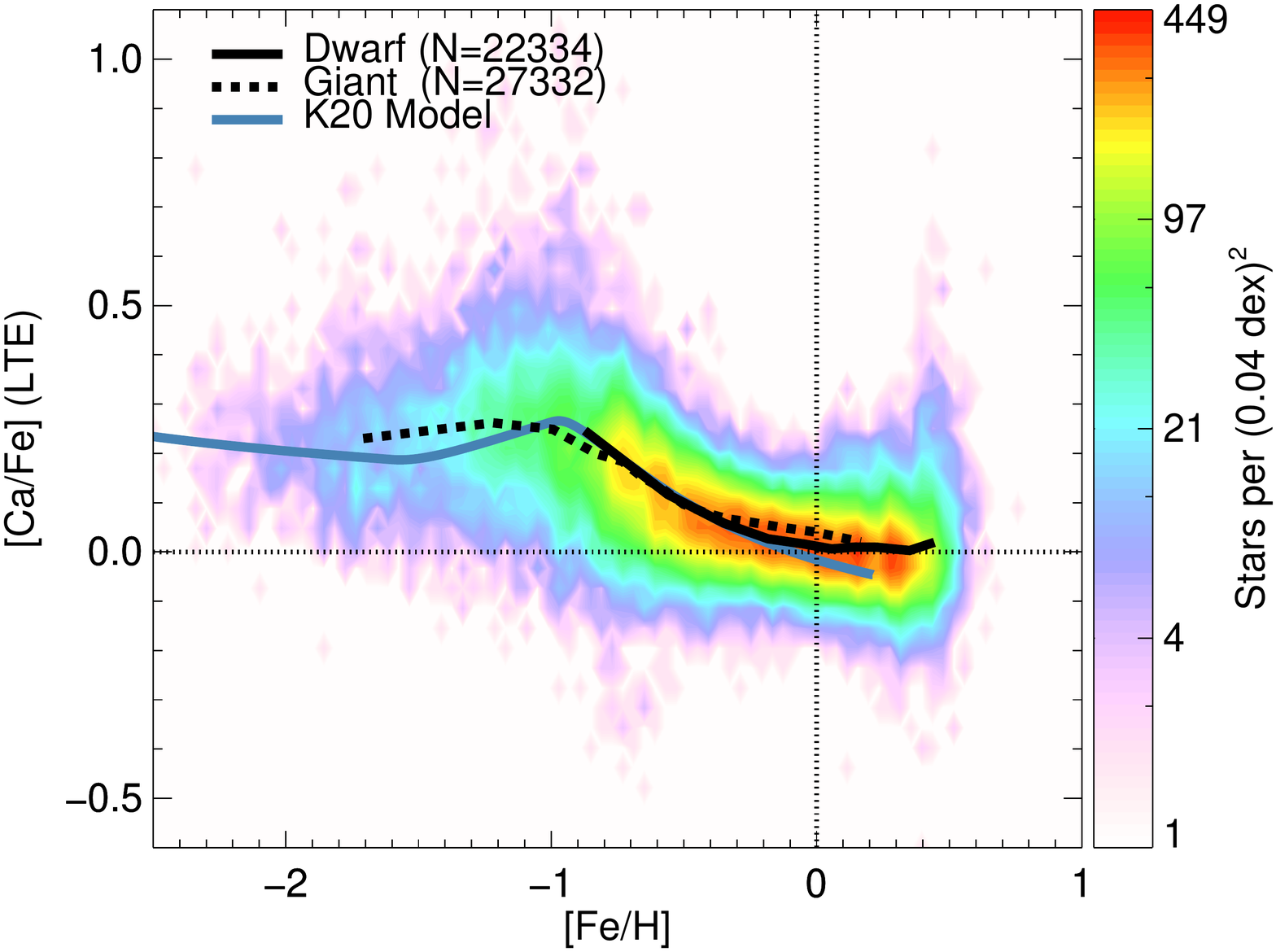}
        \caption{Non-LTE (left) and LTE (right) calcium abundances,
        with non-LTE $\feh$ adopted from GALAH DR3 in both cases.
        Overplotted are binned data for dwarfs ($\lggu>3.5$) and
        giants ($\lggu\leq3.5$), and the GCE model of K20.} 
        \label{fig:ca}
    \end{center}
\end{figure*}

\begin{figure*}
    \begin{center}
        \includegraphics[scale=0.31]{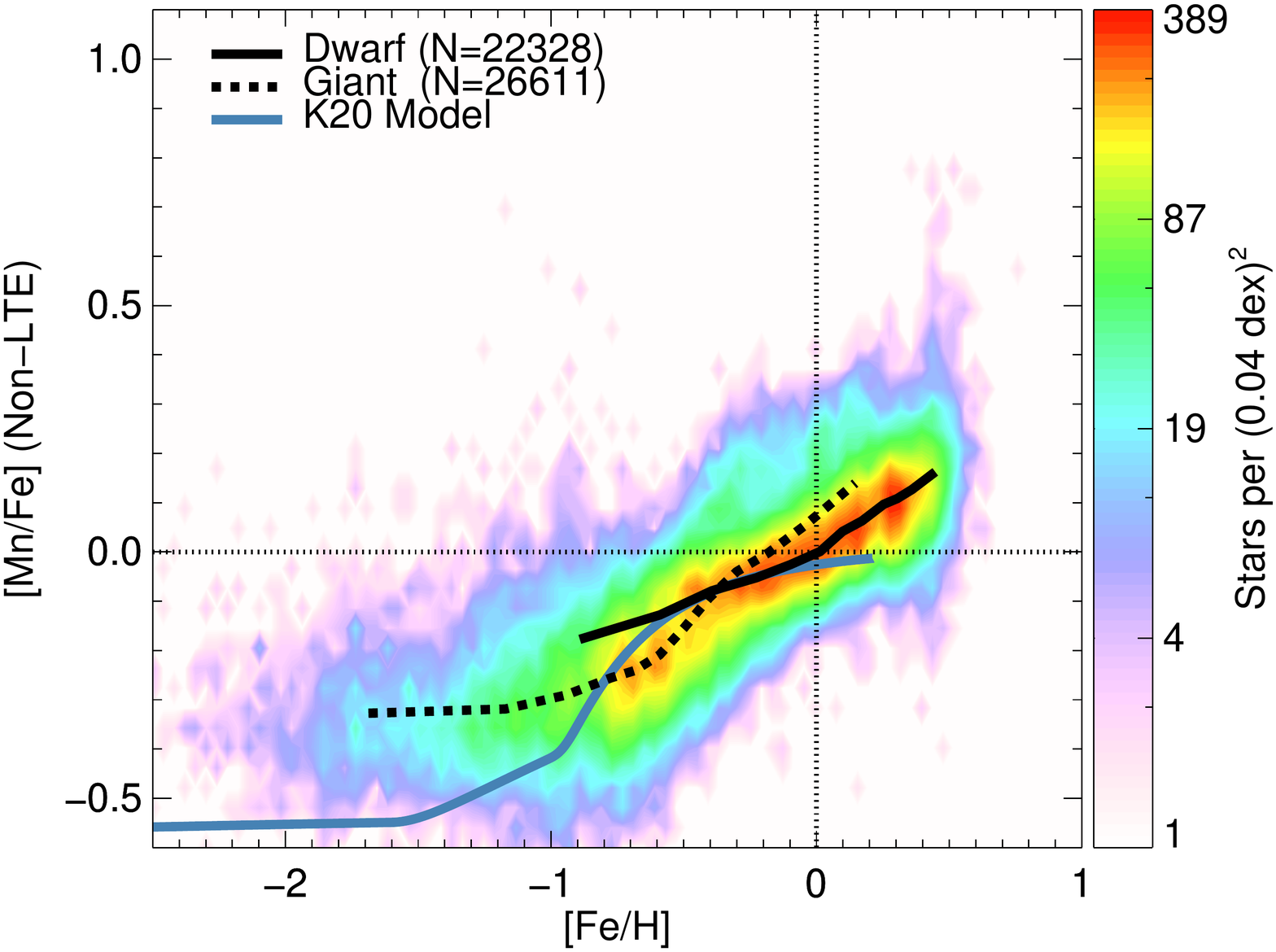}\includegraphics[scale=0.31]{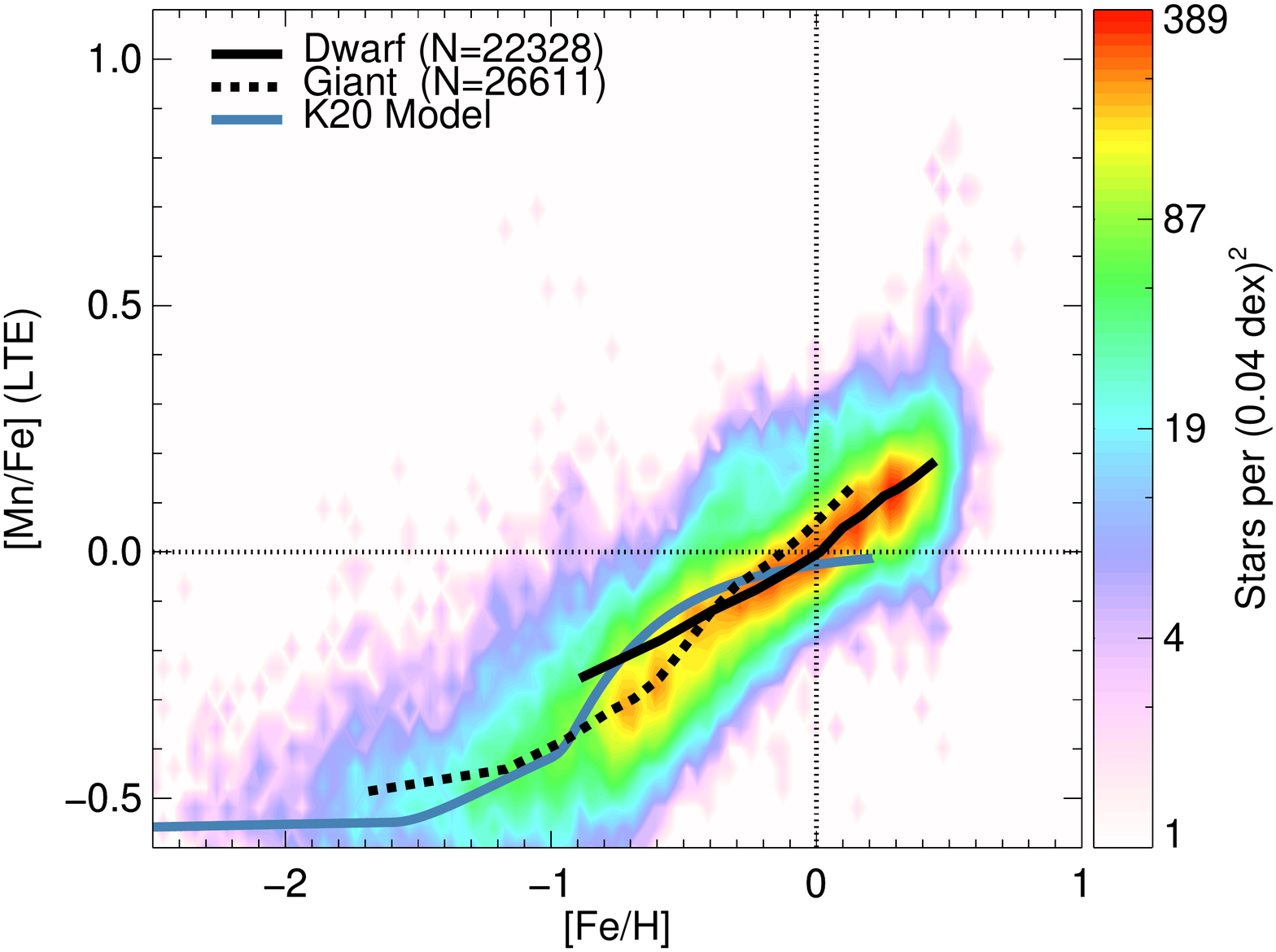}
        \caption{Non-LTE (left) and LTE (right) manganese abundances,
        with non-LTE $\feh$ adopted from GALAH DR3 in both cases.
        Overplotted are binned data for dwarfs ($\lggu>3.5$) and
        giants ($\lggu\leq3.5$), and the GCE model of K20.} 
        \label{fig:mn}
    \end{center}
\end{figure*}

\begin{figure*}
    \begin{center}
        \includegraphics[scale=0.31]{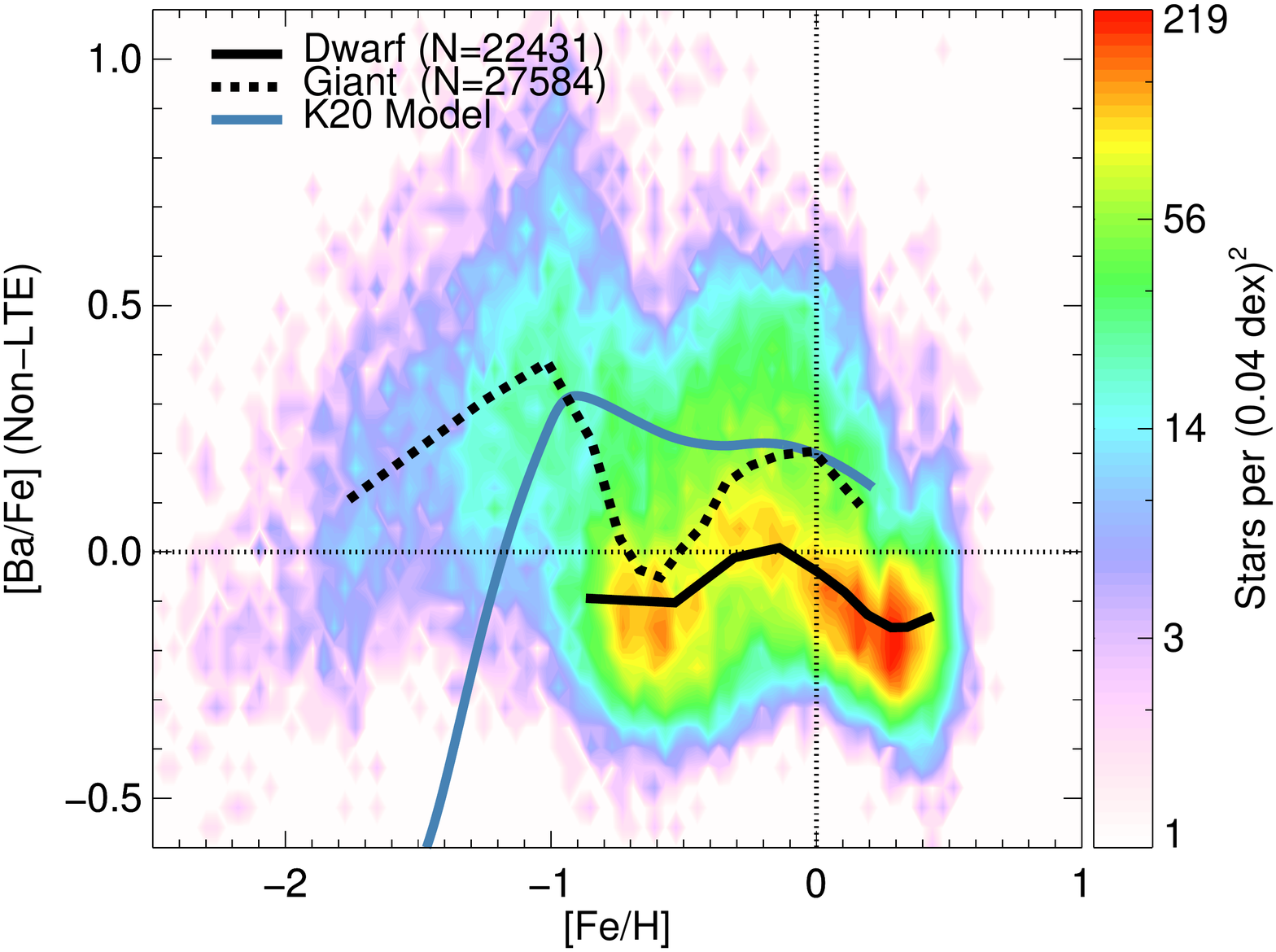}\includegraphics[scale=0.31]{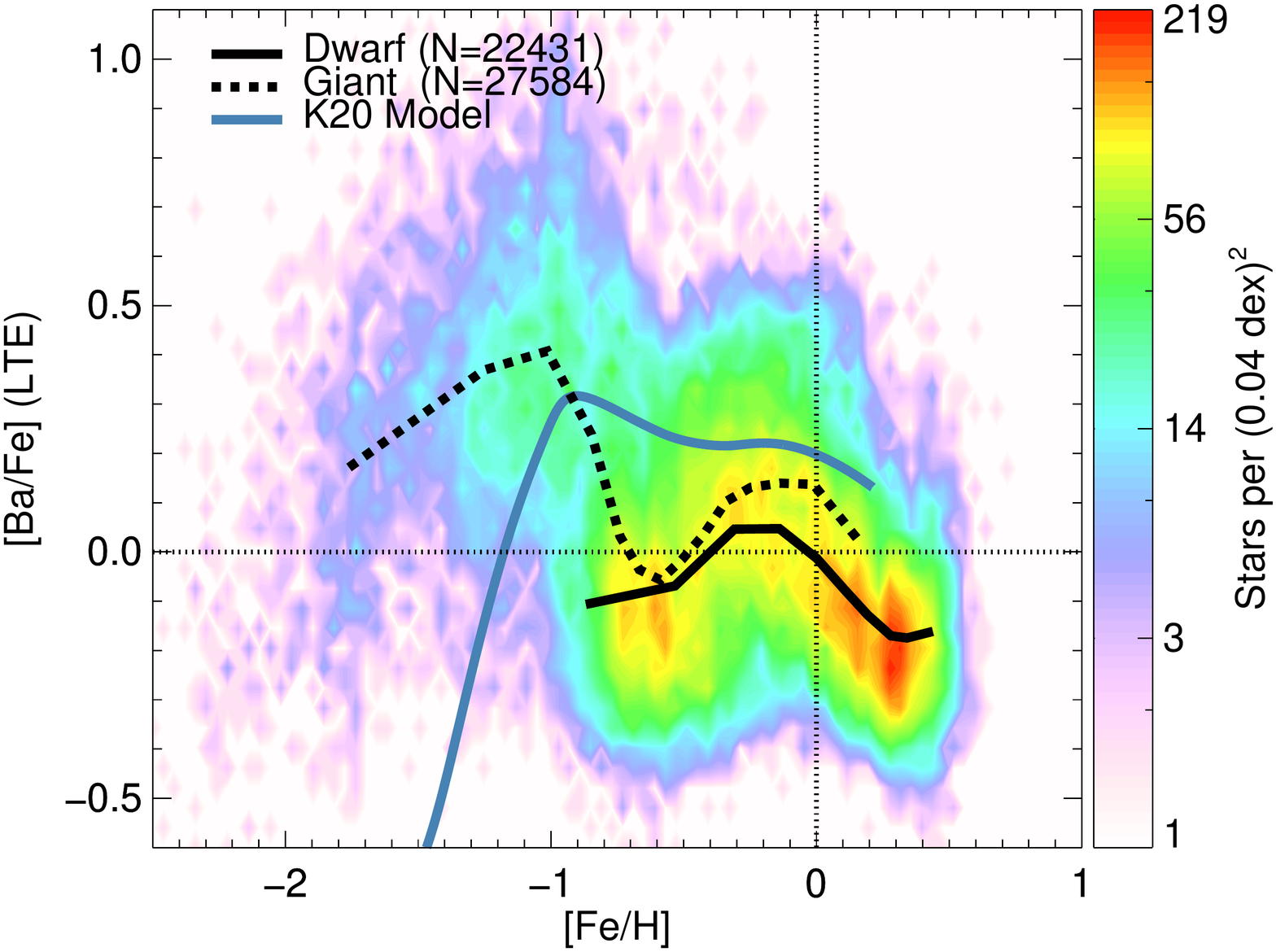}
        \caption{Non-LTE (left) and LTE (right) barium abundances,
        with non-LTE $\feh$ adopted from GALAH DR3 in both cases.
        Overplotted are binned data for dwarfs ($\lggu>3.5$) and
        giants ($\lggu\leq3.5$), and the GCE model of K20.} 
        \label{fig:ba}
    \end{center}
\end{figure*}

\begin{figure*}
    \begin{center}
        \includegraphics[scale=0.31]{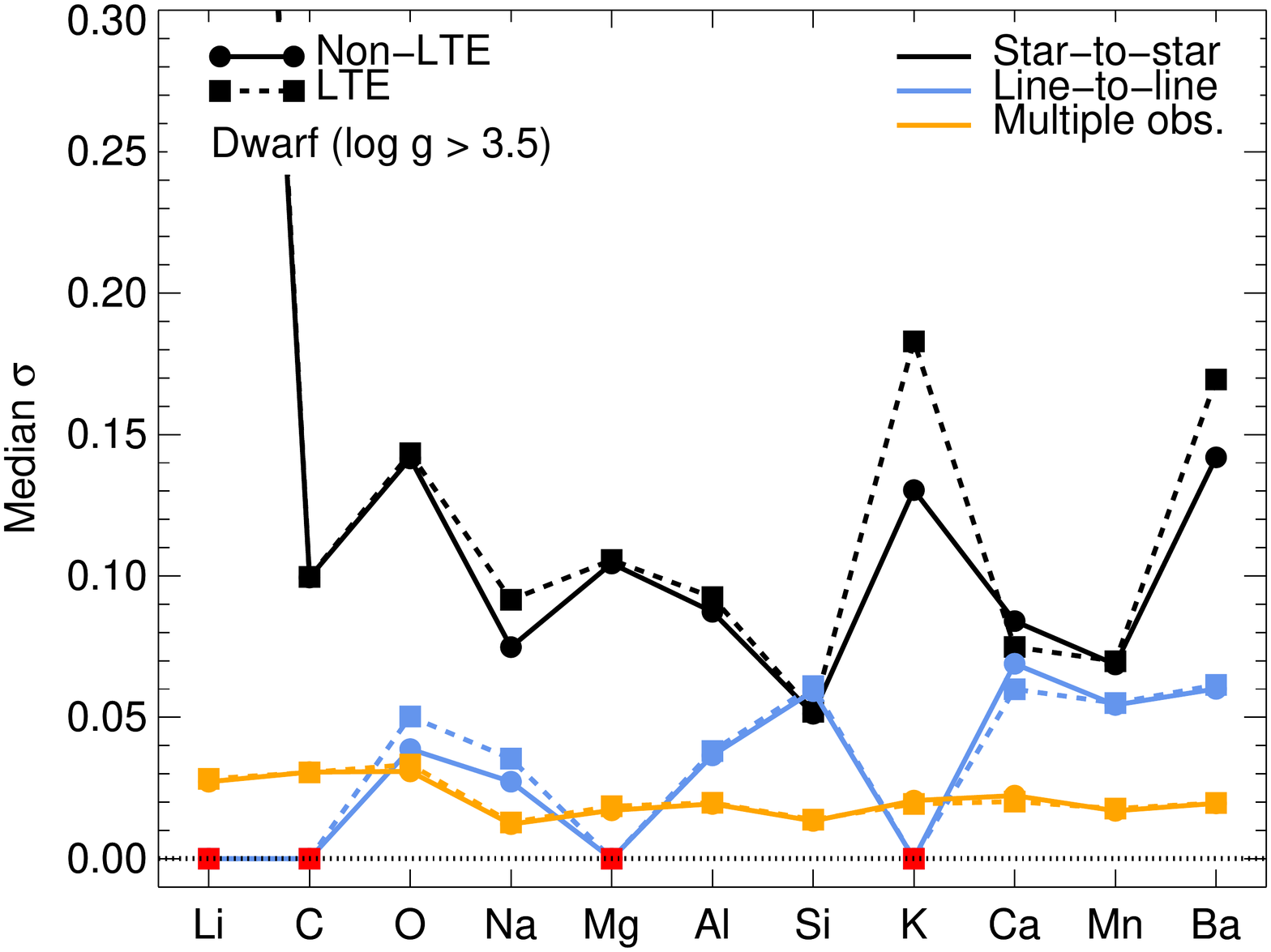}
        \includegraphics[scale=0.31]{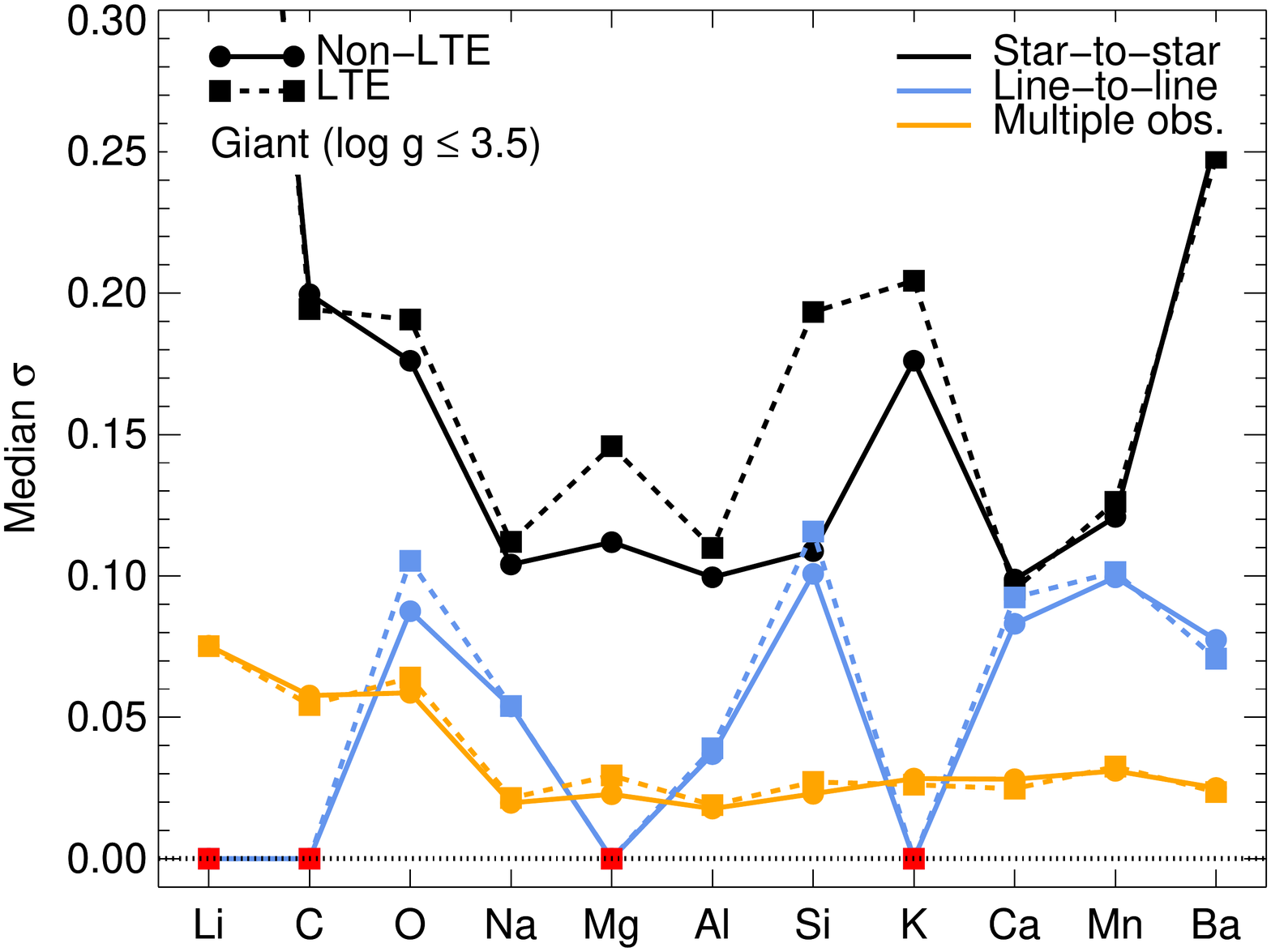}
        \caption{Dispersions in the abundance results,
            in non-LTE (solid) and LTE (dashed)
            for dwarfs (left; $\lggu>3.5$)
            and giants (right; $\lggu\leq3.5$).
            `Multiple obs.' (orange) show median
            standard deviations in $\xfe{A}$ over multiple observations
            of the same stars.
            `Line-to-line' (blue) 
            show the median standard deviations
            in $\xfe{A}$ over multiple spectral lines 
            for the same star;
            lithium, carbon, magnesium, and potassium in red are
            set to zero because they were inferred from single lines. 
            `Star-to-star' (black) 
            show the median standard deviation in 
            $\xfe{A}$ (or in $\lgeps{A}$, in the case of lithium)
            across the different bins shown in
            Figs \ref{fig:li} to \ref{fig:ba}.}
        \label{fig:disp}
    \end{center}
\end{figure*}

\subsection{Stellar sample}
\label{resultsselection}

The full GALAH DR3 stellar sample (Buder et al. in prep.) includes
over $600000$ spectra corresponding to over $500000$ stars. 
The GALAH DR3 data products include $\teff$, $\lgg$, 
a first estimate of the iron abundance $\feh_{\text{atmo}}$,
microturbulence $\vmic$, and a broadening parameter $\vbroad$
that reflects the combined effects of stellar rotation and macroturbulence.
These stellar parameters were determined simultaneously,
with hydrogen modelled in non-LTE
using the departure coefficients presented in this work,
and iron modelled in non-LTE using the departure coefficients
presented in an older, but similar, set of calculations
\citep{2016MNRAS.463.1518A}.
Also provided are
elemental abundance ratios $\xfe{A}$, that were determined in a second step
with the stellar parameters fixed.
In particular, the non-LTE abundances were determined
using the departure coefficients presented in this work
(albeit without background scattering;
\sect{methodscattering}).

To show the impact of the departure coefficients,
a re-analysis of GALAH DR3
was carried out on $55159$ spectra corresponding
to $50126$ stars.
These spectra were selected out of GALAH DR3
such that they correspond to field stars, of spectral type FGK,
and have the lowest uncertainties in the stellar parameters
as stipulated by the GALAH analysis pipeline
\citep{2018MNRAS.478.4513B}.
The stellar parameters ($\teff$, $\lgg$, $\feh$) 
were fixed to the
values provided with GALAH DR3, however
the abundances were rederived using reduced
GALAH spectra \citep{2017MNRAS.464.1259K}
and the GALAH analysis pipeline, modifying the spectral masks 
so that separate abundances were derived from 
the different lines (\sect{resultssolar}).
This re-analysis was necessary
because LTE abundances are not provided with
GALAH DR3, for the elements listed in
\tab{tab:atoms}. Moreover, for the elements considered here,
GALAH DR3 does not include
information about the line-by-line abundance dispersion,
because the elemental abundances were determined from
simultaneous fits of all of the available lines of that element.

\subsection{Line list and solar abundances}
\label{resultssolar}

In \tab{tab:linelist} we list the lines used in the
re-analysis of $50126$ stars from GALAH DR3. All of the elements
in \tab{tab:atoms} were considered, 
except nitrogen, due to the lack of suitable
lines within the GALAH spectral range. 
The line list has a large overlap with that of GALAH DR3
(Buder et al. in prep.).

We also list the adopted line-by-line absolute solar elemental abundances 
$\lgeps{}$ in \tab{tab:linelist}.
These were determined from the mean result
of $281$ solar twins present in the sample:
stars with $\teff$ within $100\,\K$,
$\lgg$ within $0.1\,\dex$, and 
$\feh$ within $0.1\,\dex$ from the Sun
\citep{2018A&ARv..26....6N}.
The number of solar twins is sufficient to average out
the intrinsic dispersions that are due to their different ages
\citep{2015A&amp;A...579A..52N,2016A&A...585A.152S,2018ApJ...865...68B},
so the zero points in \sect{results} should closely
reflect the actual results for the Sun.
For lithium the standard value of
$\lgeps{Li}=1.05$ \citep{2009ARA&amp;A..47..481A}
is listed in \tab{tab:linelist}, however 
the stellar lithium abundances discussed in this
paper are given in the absolute sense, as per convention.

Subsequent sections are based on
the abundance ratios $\xh{A}$ or $\xfe{A}$ relative to the Sun;
these were determined differentially
on a line-by-line basis using 
the solar elemental abundances given in 
\tab{tab:linelist}, prior to averaging over the lines.
Owing to the large number of solar twins in the sample,
this method of normalising the elemental abundances
precisely puts the different spectral lines onto the same 
scale.  It thus leads to a significant cancellation of errors
originating from the adopted oscillator strengths,
systematic offsets in the continuum placements,
and the neglected effects of stellar surface convection.
In the LTE results, this also cancels some of the 
errors arising from the neglected non-LTE effects,
as we show in \sect{discussionabcor}.

\subsection{Stellar elemental abundances}
\label{resultsstars}

We show the difference between the inferred LTE and non-LTE elemental
abundances 
(the differential abundance corrections; \eqn{eq:abcor2})
for the entire
sample of up to $50126$ stars in 
Figs \ref{fig:abcorli} to \ref{fig:abcorba}.
We show the abundances themselves in
Figs \ref{fig:li} to \ref{fig:ba}:
\fig{fig:li} shows the absolute lithium abundances,
$\lgeps{Li}$, whereas and Figs \ref{fig:c} to \ref{fig:ba} show
the run of $\xfe{A}$ from carbon to barium, as functions of 
the non-LTE iron abundance $\feh$
adopted from GALAH DR3.
For a particular element, the plots are restricted to those stars
for which both LTE and non-LTE abundances were successfully measured,
so the same number of stars are included in both cases.
The results were also binned, separately for dwarfs ($\lggu>3.5$) and
for giants ($\lggu\leq3.5$), and these are overplotted to
illustrate the average abundance trends.
The chemical evolution model
of \citet{2020arXiv200804660K},
discussed in \sect{gce}, 
is overplotted and can be used to compare the LTE and non-LTE 
trends. For further insight,
we also plot various measures of the dispersion in the abundances,
in \fig{fig:disp}.

There are several features in these figures that 
help us
to gauge the impact of non-LTE effects on large spectroscopic surveys.
These include:
the severity of the non-LTE abundance corrections
(\sect{discussionabcor});
the appearance of the average abundance trends (\sect{discussiontrends});
the differences between the results for dwarfs and giants
(\sect{discussiondwarfgiant});
and the overall dispersions in the elemental abundances
(\sect{discussiondisp}).
We discuss the implications for the chemical evolution of our Galaxy
in \sect{gce}.

\section{Discussion}
\label{discussion}

\subsection{Typical non-LTE abundance corrections}
\label{discussionabcor}

In \tab{tab:linelist} we show the typical 
line-by-line differences between the absolute abundances
inferred in non-LTE and in LTE:
$\lgeps{A}^{\text{Non-LTE}}-\lgeps{A}^{\text{LTE}}$.
These are based on the abundances inferred from the present sample,
and are shown for different regions of
the HR diagram illustrated in \fig{fig:grid}
and the abundance plane illustrated in \fig{fig:abgrid},
that is for K dwarfs, F dwarfs, and K giants,
at high and low metallicities.

In metal-rich F dwarfs,
the absolute abundance corrections can be as severe as
$-0.6\,\dex$ for the \ion{K}{I} $769.9\,\nm$ resonance line
\citep{2019A&A...627A.177R},
 $-0.5\,\dex$ for the \ion{O}{I} $777\,\nm$ triplet
\citep{2016MNRAS.455.3735A},
and $-0.3\,\dex$ for the
\ion{Ba}{II} $649.7\,\nm$ line 
\citep{2011MNRAS.415.2093K,2020A&A...634A..55G}.
Typically the most severe absolute abundance corrections
are found for metal-rich K-giants;
it is also in this cooler, lower-pressure regime where inelastic
hydrogen collisions (\sect{methodatom}) play a greater role.

\tab{tab:linelist} shows that for a particular species, the absolute 
abundance correction goes in the same direction and is of
a similar magnitude for the different lines of that
species: the non-LTE effects are of a similar nature
for the lines of a given species.
A consequence of this is that,
in the present analysis, non-LTE effects do not average out simply 
by using multiple lines of the same species.
It is beyond the scope of this paper to discuss the
non-LTE effects for each of the spectral lines and
elements considered here,
and instead refer the reader to the papers listed in 
\tab{tab:atoms} for further details.

It is important to note that
with the exception of lithium (\fig{fig:li}),
the abundance results presented in \sect{resultsstars} are differential
in the sense that the LTE and non-LTE
solar abundances presented in \tab{tab:linelist}
were used to convert absolute elemental
abundances $\lgeps{A}_{l}$ into 
abundance ratios $\xh{A}_{l}$ or $\xfe{A}_{l}$ relative to the Sun,
for given lines $l$.
These quantities were then averaged over the different lines $l$,
to obtain the final abundance ratios $\xh{A}$ or $\xfe{A}$.
The internal normalisation results in a more accurate zero point,
and, for the elements with multiple line diagnostics, 
the line-by-line normalisation leads to
lower dispersion, owing to significant cancellation of systematic errors
as we discussed in \sect{resultssolar}.

In Figs \ref{fig:abcorli} to \ref{fig:abcorba}
we plot the difference $\mathrm{\xh{A}^{\text{Non-LTE}}}-
\mathrm{\xh{A}^{\text{LTE}}}$, or in other words
the differences between the non-LTE and LTE panels in
Figs \ref{fig:abcorli} to \ref{fig:abcorba}.
This is a rough estimate of the
differential abundance corrections (\eqn{eq:abcor2}).
The severity of these differential corrections can be
less than of the absolute corrections,
if the non-LTE effects in the Sun are of the same sign
and magnitude as in the star under consideration.
For example for oxygen (\fig{fig:abcoro}), the most negative 
differential corrections reach at most only $-0.3\,\dex$ 
compared to the absolute corrections reaching $-0.5\,\dex$
in metal-rich F-dwarfs, which
reflects that the absolute correction in the Sun is 
$-0.2\,\dex$ (\tab{tab:linelist}).
This cancellation is less effective for stars further away from
the Sun in parameter space: for example in the cooler
giants, as can be seen for both sodium (\fig{fig:abcorna}) 
and magnesium (\fig{fig:abcormg});
or in the metal-poor regime, as can be seen for 
manganese (\fig{fig:abcormn}).

\subsection{Non-LTE effects on abundance trends}
\label{discussiontrends}

We find that the assumption of LTE can
lead to average abundance trends
that are quantitatively and sometimes even qualitatively
different compared to what is found in non-LTE.
This is particularly apparent for a few elements:
sodium, magnesium, potassium, and manganese, where there 
are differences in the mean $\xfe{A}$ at given $\feh$ of up to 
$0.2\,\dex$, resulting in qualitatively
different pictures of the Galactic chemical evolution
as we discuss further in \sect{gce}.

For sodium (\fig{fig:na}), in LTE, there is a jump in $\xfe{Na}$ of
from $-0.2\,\dex$ to $+0.2\,\dex$ as
$\feh$ increases past $-1.0\,\dex$.
In non-LTE, this jump of $0.4\,\dex$ is reduced
to just $0.2\,\dex$, 
with $\xfe{Na}$ increasing from $-0.1\,\dex$ to $+0.1\,\dex$.

For magnesium (\fig{fig:mg}) in LTE there is a clear
local maximum 
in $\xfe{Mg}$ of approximately $0.45\,\dex$ at 
$\feh\approx-0.75$.  
In non-LTE, $\xfe{Mg}$ shows the usual
plateau at low metallicities at $0.25\,\dex$,
and a decreasing trend towards higher metallicities.

For potassium (\fig{fig:k}), 
there is also a local maximum in LTE,
with $\xfe{K}$ peaking at approximately $0.45\,\dex$
at $\feh\approx-1.0$.
There is also a linear decrease in $\xfe{K}$ at
super-solar metallicities.
In non-LTE, potassium instead shows a trend
similar to the $\upalpha$-elements,
plateauing at $\xfe{K}\approx0.25$ for
$\feh\lesssim-1.0$, and 
showing a flat trend at super-solar metallicities.

Finally for manganese (\fig{fig:mn}),
in LTE, $\xfe{Mn}$ increases rather steeply
from $-0.5\,\dex$ at $\feh\approx-2.0$,
up to $0.2\,\dex$ at $\feh\approx0.5$.
In non-LTE instead,
$\xfe{Mn}$ plateaus at low metallicities
at approximately $-0.3\,\dex$ below $\feh\lesssim-1.0$,
and the abundance trend is less steep.

\subsection{Non-LTE effects on offsets between dwarfs and giants}
\label{discussiondwarfgiant}

We find that the assumption of LTE 
can impart large offsets between
dwarfs and giants in abundance space
\citep{2009A&A...501..519B}.
That is, the giants tend to sit either far above or below the dwarfs 
in $\xfe{A}$ versus $\feh$ space,
even if the abundance tracks are parallel to each other,
This offset is as large as $0.3\,\dex$ in the case of 
magnesium (\fig{fig:mg});
smaller offsets can also be seen for 
sodium (\fig{fig:na}),
silicon (\fig{fig:si}),
and potassium (\fig{fig:k}).
These offsets shrink significantly 
for each of these elements when non-LTE methods are used.
For example, for sodium it is reduced by $0.2\,\dex$,
and the tracks are in good agreement between
$-1\lesssim\feh\lesssim-0.5$.

For some elements,
taking non-LTE effects into account does not 
negate entirely the offsets between the dwarfs and giants.
The residual offsets for most elements tend to be only of
the order $0.05\,\dex$.
For aluminium (\fig{fig:al}) and barium (\fig{fig:ba}),
the offsets reach $0.2\,\dex$.
These large offsets may 
reflect other systematic errors 
within the GALAH analysis pipeline, even though
the different \ion{Al}{I} and \ion{Ba}{II} lines are in
reasonable agreement 
(the line-by-line dispersion is not anomalously large
for aluminium and barium in \fig{fig:ba}).
For extremely metal-poor stars
\citet{2008A&A...481..481A} and \citet{2009A&A...501..519B} found that 
taking non-LTE effects into account bring the 
aluminium abundances of dwarfs and giants into agreement,
which might indicate that the offsets seen here at 
higher metallicities are not physical.

For the $\upalpha$-elements oxygen (\fig{fig:o}), 
magnesium (\fig{fig:mg}), and 
silicon (\fig{fig:si}), the residual offsets in non-LTE
amount to approximately $0.1\,\dex$.  
These offsets may in part be due to selection effects:
GALAH is a magnitude limited survey, and
the giants are brighter and observed to larger distances,
and thus probe more of the Galactic thick disc,
that is typically $\upalpha$-rich.
Conversely, the dwarfs probe more 
of the Galactic thin disc, that is 
typically $\upalpha$-poor 
\citep{2015ApJ...808..132H,2019MNRAS.489.1742F}.
However, one would then expect a similar offset for 
calcium (\fig{fig:ca}), whereas instead the mean trends for
the dwarfs and giants lie on top of each other.
This may reflect residual systematics with the non-LTE calcium abundances.
For instance, \citet{2020A&A...637A..80O} showed recently that
non-LTE \ion{Mg}{I} continuous opacities can make a significant impact on the
predicted \ion{Ca}{I} departure coefficients; this effect is not
taken into account in the present calculations.

\subsection{Non-LTE effects on abundance dispersions}
\label{discussiondisp}

We find that the assumption of LTE can
substantially increase the spread of $\xfe{A}$ at given $\feh$.
This is immediately evident by comparing the 
densities of stars in LTE and non-LTE 
in many of the plots presented in
in \sect{resultsstars}, and
is particularly obvious for
oxygen (\fig{fig:o}),
sodium (\fig{fig:na}),
magnesium (\fig{fig:mg}), and 
potassium (\fig{fig:k}).

The dispersions in the stellar elemental abundances can also be quantified.
In \fig{fig:disp} we show the median standard deviation in 
$\xfe{A}$ (or for lithium, $\lgeps{Li}$) 
across the different bins shown in
Figs \ref{fig:li} to \ref{fig:ba}, separately
for dwarfs, and for giants; this is labelled `Star-to-star'.
For comparison, also plotted are the median standard deviations
in $\xfe{A}$ over multiple spectral lines,
as inferred for a given element and
for a given star (after already
averaging over multiple observations, if necessary);
this is labelled `Line-to-line'.
Another comparison can be made, by noting that
there are $3237$ stars in the sample
of $50126$ for which two or more observations exist.
Thus, the median standard deviations in $\xfe{A}$ over multiple observations
of the same stars are also plotted; this
is labelled `Multiple obs.'.
The stipulated signal-to-noise ratios of this sample
is in close agreement to that of the entire sample.

The assumption of LTE does not significantly
impact the dispersions in the results from multiple observations.
This reflects the random noise that are propagated
from the finite signal-to-noise ratios of the observations  
into the inferred abundances, either directly 
or via the inferred stellar parameters.
In \fig{fig:disp} the dispersions are
approximately $0.02\,\dex$ in dwarfs and $0.03\,\dex$ in giants
for most elements, in both LTE and non-LTE; 
they are slightly larger
for lithium, carbon, and oxygen,
namely $0.03\,\dex$ in dwarfs, and $0.06\,\dex$ in giants.
These small dispersions reflect the great potential 
of GALAH data for high-precision studies of 
stellar abundances and Galactic chemical evolution.

The assumption of LTE has a small impact on the line-to-line dispersions.
This dispersion reflects (in addition to the random noise) the
systematic errors arising from the pipeline,
arising from deficiencies in modelling the stellar spectra
including 3D and non-LTE effects,
as well as the impact of blends, and consistent offsets in
the continuum placement.
We caution that the entirety of the 
systematic errors are not represented here,
because the spectral lines being used for a particular 
element belong to the same ionisation stage and also
tend to have similar wavelengths and excitation potentials
(\tab{tab:linelist}), and thus 
tend to have similar sensitivities to systematic errors in
the stellar parameters, and suffer similar 3D and non-LTE effects.
In \fig{fig:disp} the dispersions are, in non-LTE,
approximately $0.05\,\dex$ for dwarfs, and
$0.07\,\dex$ for giants.
In LTE, they are typically only
about $0.01\,\dex$ larger, owing
to the cancellation effects
obtained by using line-by-line solar abundances
(\sect{resultssolar}).

Finally, the assumption of LTE does significantly
impact the star-to-star dispersions.
This reflects
random noise and systematic errors in the pipeline,
as well as intrinsic dispersion arising
from different stars with the same $\feh$ actually having different 
values of $\xfe{A}$ in their atmospheres.
The systematic errors are better represented here 
than in the line-to-line dispersions.
In \fig{fig:disp}, in non-LTE and for most elements,
the dispersions are
approximately $0.05$ to $0.15\,\dex$ for dwarfs and
$0.1$ to $0.2\,\dex$ for giants,
much larger than the dispersions in the results from multiple observations
and also
typically larger than the line-to-line dispersions.
In LTE the star-to-star dispersions can be much larger:
the largest difference is for magnesium and silicon in giants,
where the dispersion is increased by $0.1\,\dex$.

It is difficult to make general conclusions about 
whether or not the non-LTE results
shown in Figs \ref{fig:li} to \ref{fig:ba}
are dominated by intrinsic dispersion, or are dominated by
systematics errors. The small dispersions in 
the results from multiple observations in
\fig{fig:disp} suggests that random noise
are not dominating the dispersions in the abundance relations.
For lithium (\fig{fig:li}), intrinsic dispersion
is clearly dominant; lithium is fragile and easily
destroyed in stellar atmospheres.
Some elements such as 
silicon (\fig{fig:si}) show tight abundance relations in non-LTE,
with star-to-star dispersions that are similar to the
line-to-line dispersions in
\fig{fig:disp}, which suggests that 
systematics could be the limiting factor here.
For other elements with broader spreads in the abundance relations
including aluminium (\fig{fig:al}) and barium (\fig{fig:ba}), 
intrinsic effects could be dominant, although
residual systematic errors still cannot be ruled out
as discussed above.

\section{Galactic chemical evolution and comparison with previous studies}
\label{gce}

In this section we briefly discuss the
non-LTE abundance trends (Figs \ref{fig:li}---\ref{fig:ba}),
and contrast them with results recently presented elsewhere
in the literature. 
For elements heavier than lithium,
we compare the observed trends to the
standard Galactic chemical evolution model
of the solar neighbourhood
presented in \citet{2020arXiv200804660K},
hereafter
referred to as the K20 model.
In brief, the K20 model assumes instantaneous mixing
in the interstellar medium (a one-zone model),
adopting an initial mass function from
\citet{2008ASPC..390....3K}.
Enrichment by intermediate-mass and massive
stars via winds and core collapse supernova,
neutron-star mergers, and Type Ia supernova are considered:
for Type Ia supernovae, the progenitor model and
nucleosynthesis yields from \citet{2020ApJ...895..138K} are adopted.
The K20 model is based 
on theoretical nucleosynthesis
yields and event rates and avoids empirical relations
and calibrations; in particular, the
zero-points of the model have not been adjusted.

\subsection{Lithium}
\label{gceli}

Lithium was one of th few nuclei that formed during the Big Bang 
\citep{2011ARNPS..61...47F},
and so lithium abundances impose constraints on the primordial
nucleosynthesis and baryon density of the early Universe. 
Later in cosmic time
lithium may be produced by spallation reactions in cosmic rays and by various
stellar sources \citep{2012A&A...542A..67P}.
As a fragile element, lithium can be
destroyed by proton capture reactions at relatively low temperatures in the
stellar interior, making it a sensitive tracer of mixing within stars
\citep{1997ARA&amp;A..35..557P}.
However, recent observations have shown that surface
lithium abundances could also be enriched, by multiple mechanisms that act in
different parts of the HR diagram 
\citep{2020MNRAS.497L..30G,2020arXiv200602106M}.

The lithium abundances here confirm
those recently presented in the literature.
\fig{fig:li} shows a large intrinsic spread of abundances,
with the dwarfs and giants following two distinct
trends: in giant stars the convection zones extend deeper
into the stellar interior, to high enough temperatures for lithium
to be burnt.
The most metal-poor dwarfs in the sample
converge to the well-known `Spite-plateau' \citep{1982A&A...115..357S}.
In non-LTE, this is at 
$\lgeps{Li}\approx2.1$.
In LTE this is about $0.1\,\dex$ higher, as the abundance
corrections for the \ion{Li}{I} $670.8\,\nm$ line
is slightly negative and not too sensitive to metallicity
(\fig{fig:abcorli}).
In both LTE and non-LTE, there 
is a mild increase in $\lgeps{Li}$ with increasing $\feh$
indicative of a gradual enrichment of lithium in the Galaxy
as found in earlier studies of dwarf stars
\citep[][]{2018A&A...615A.151B}.

Recent studies have reported a puzzling
drop in $\lgeps{Li}$ at super-solar metallicities
\citep[][]{2015A&A...576A..69D,2020AJ....159...90S}.
The same trend is found in the present study, in both LTE and non-LTE.
It has been proposed that this is a signature of radial migration
\citep{2019A&A...623A..99G}.
According to this picture,
towards higher metallicities the sample
is dominated by stars that were
born closer to the Galactic centre, in a more metal-rich environment.
To have reached the solar neighbourhood, these stars have to be old,
implying that they have had time to deplete significant amounts
of their birth lithium abundance.

\subsection{Carbon and oxygen}
\label{gceco}

Carbon is a light element that traces multiple sources
including Asymptotic Giant Branch (AGB) stars, core-collapse supernova,
and Wolf-Rayet stars
\citep{2014PASA...31...30K,2019MNRAS.490.2838R,
2020A&A...639A..37R}.
Some molecular features are present in the GALAH
spectral range and can be used to infer carbon abundances
in cooler stars to low metallicities
\citep{2019MNRAS.483.3196C}.
However in the present study,
carbon abundances were determined from
the high-excitation \ion{C}{I} $658.8\,\nm$ line
that is usually too weak to be observed in cool giant stars.
Thus the steep trend seen in \fig{fig:c} at low metallicity
is based on a relatively
small number of giant stars, and may be due to the GALAH analysis
pipeline interpreting noise as a detected line; it should
be treated with scepticism.
The Galactic evolution of carbon inferred here
is based on the warmer dwarf stars:
\fig{fig:c} shows a mild linear decrease in $\xfe{C}$ 
from approximately $0.2\,\dex$ at $\feh\approx-0.5$,
to $-0.1\,\dex$ at $\feh\approx+0.5$.
This likely reflects that the cosmic production of iron
from Type Ia supernova outpaces that of carbon
from AGB stars or from massive stars.
The trend is very similar in LTE owing to very
mild abundance corrections
for the \ion{C}{I} $658.8\,\nm$ line (\fig{fig:abcorc}).

Oxygen is an $\upalpha$-element that forms
almost entirely via hydrostatic burning in
massive stars \citep{2006ApJ...653.1145K}.
The dwarfs and giants form parallel tracks 
in $\xfe{O}$ versus $\feh$ 
in \fig{fig:o}, albeit offset by approximately 
$0.05$ to $0.10\,\dex$. 
This offset may reflect selection effects:
the giants probe more of the $\upalpha$-rich
Galactic thick disc (\sect{discussiondwarfgiant}).
There is a steep linear decrease in $\xfe{O}$ 
from $0.5\,\dex$ at $\feh\approx-1.0$,
down to below $-0.2\,\dex$ at $\feh\approx0.5$,
that reflects that the cosmic production of iron
from Type Ia supernova vastly outweighs that of oxygen from massive stars
at high metallicities. 
In LTE, the abundance gradient is similar, however the dwarf trends are offset
to higher abundances; the LTE results
imply that the Sun is underabundant in oxygen.

In general at moderate metallicities 
these results for carbon and oxygen are consistent with those
from previous high-resolution, high signal-to-noise ratio
studies in LTE and in non-LTE \citep{2014A&amp;A...562A..71B,
2016ApJ...833..225Z,2019A&A...630A.104A,2020AJ....159...90S}.
The gradients in the abundance trends are also 
captured well in the K20 model;
for oxygen the non-LTE results are clearly
in better agreement than the LTE results.
However they are not quite consistent
with results for giants from APOGEE;
these tend to find much flatter relationships,
with a plateau in $\xfe{O}$ or even a slight increase 
in $\xfe{C}$ at super-solar metallicities
\citep[][]{2018ApJ...852...49H}.
The APOGEE abundances are inferred in LTE from
molecular lines, prone to different systematics
(in particular from 3D effects; 
\citealt{2007A&amp;A...469..687C}),
and this may be the reason for the discrepancies.

In the more metal-poor regime, 
$\xfe{O}$ still shows a linear decrease with
increasing $\feh$, albeit with a more gentle slope,
as well as a hint of a plateau at $\feh\approx-1.0$, 
The very high values of $\xfe{O}$ at the lowest
metallicities is difficult to reconcile with the
canonical picture of a $\upalpha$-abundance plateau
at low metallicities, as also seen in the K20 model below
$\feh\lesssim-1.0$. However it is
qualitatively consistent with what was found in
earlier non-LTE studies of the 
\ion{O}{I} $777\,\nm$ triplet
\citep{2012ApJ...757..164R,2019A&A...630A.104A}.
Similar to carbon, the metal-poor trend is driven by the
giants and should also be treated with
scepticism because the high-excitation triplet is very weak in 
this regime.  One or two of the
weaker components becoming immeasurable over
the noise, and perhaps spuriously measured,
may also explain the relatively
large star-to-star and line-to-line dispersions
seen for oxygen in the giant stars in \fig{fig:disp}.

\subsection{Sodium, aluminium, and potassium}
\label{gcenaalk}

Sodium, aluminium, and potassium are light elements with odd 
proton numbers, formed primarily in hydrostatic burning
in massive stars with metallicity-dependent yields
\citep{2006ApJ...653.1145K}.
In \fig{fig:na} the non-LTE analysis indicates that 
$\xfe{Na}$ plateaus at
approximately $-0.1\,\dex$ at $\feh\lesssim-1.0$, 
and then jumps by $0.2\,\dex$.
Between $-1.0\lesssim\feh\lesssim0.0$,
sodium abundances show a gentle decrease from
$0.1\,\dex$ down to $0.0\,\dex$,
that can be attributed to the onset of Type Ia supernova.
At super-solar metallicities, $\xfe{Na}$ increases again,
reaching $0.25\,\dex$ at $\feh\approx0.5$.
As discuss in \sect{discussiontrends},
the trend from the LTE analysis is qualitatively similar, however with a much
larger jump at low metallicity.

At sub-solar metallicities, aluminium and potassium behave similarly to the
$\upalpha$-elements 
(\sect{gcemgsica}).
In \fig{fig:al}, in both LTE and non-LTE, there appears to
be an aluminium-poor component with 
$\xfe{Al}\approx0.0$ at 
$\feh\approx0.0$, and an 
aluminium-rich component with 
$\xfe{Al}\approx0.25$ at 
$\feh\approx-1.0$, (traced by
dwarfs and giants respectively; \sect{discussiondwarfgiant}). 
Similarly, in \fig{fig:k} $\xfe{K}$ shows
a plateau at approximately $0.25\,\dex$ below $\feh\lesssim-1.0$,
and a mild decrease down to $\feh\approx0.0$.
This plateau is significantly higher, and the trend
much steeper, in the LTE plot 
(\sect{discussiontrends}).  At super-solar metallicities
aluminium behaves qualitatively similar to sodium:
in the dwarfs, $\xfe{Al}$ increases from
$0.0\,\dex$ at $\feh\approx0.0$,
to $0.2\,\dex$ at $\feh\approx0.5$. In
contrast, at these metallicities potassium shows a flat trend, similar to the
canonical $\upalpha$-elements magnesium, silicon, and calcium 
(\sect{gcemgsica}).

The observed run of $\xfe{Na}$ is in qualitative agreement
with what has been found in other
recent high-resolution, high signal-to-noise ratio non-LTE studies
\citep{2014A&amp;A...562A..71B,2016ApJ...833..225Z}.
The aluminium and potassium abundances agree well 
with LTE results for giants from APOGEE
\citep[][]{2018ApJ...852...49H},
but the increasing trend of $\xfe{Al}$ at super-solar metallicities
is not seen in LTE in \citet{2014A&amp;A...562A..71B}
nor in non-LTE in \citet{2016ApJ...833..225Z}.
The K20 model quantitatively reproduces the sodium abundances, and the shape of
the $\xfe{Al}$ trend for dwarfs. However, as with most
GCE models \citep{2019A&A...627A.177R}
the K20 model significantly underestimates the cosmic abundance
of potassium.

As pointed out already in \sect{discussiondwarfgiant},
there is an offset between the dwarfs and giants
for aluminium of $0.2\,\dex$. 
Since the dwarf and giant trends have similar shapes,
any such offset could be calibrated away in APOGEE results,
which are strictly for giants.
A similar offset between dwarfs and
giants was found in LTE in the Gaia-ESO survey
\citep{2016A&A...589A.115S}; the authors suspected the offset
was due to systematics errors, albeit not related to departures from LTE.
In the present study, we similarly suspect that the offset signals
residual systematic errors in the aluminium abundances,
and thus that our results should be treated with caution.

\subsection{Magnesium, silicon, and calcium}
\label{gcemgsica}

Like oxygen,
these canonical $\upalpha$-elements form primarily
through hydrostatic burning in massive stars,
however the yields are altered during explosive nucleosynthesis
\citep{2006ApJ...653.1145K}.
The non-LTE results for magnesium, silicon, and calcium
each display a plateau at low metallicities,
and a knee at $\feh\approx-1.0$ that signals the
onset of Type Ia supernova
\citep{1997ARA&amp;A..35..503M}.
Here this is visible via the dichotomy 
between the $\upalpha$-poor thin disc and the 
$\upalpha$-rich thick disc 
(traced by dwarfs and giants respectively; \sect{discussiondwarfgiant}).
In Figs \ref{fig:mg}, \ref{fig:si}, and
\ref{fig:ca}, this thick disc forms a plateau
at approximately $\xfe{\upalpha}\approx0.25$ below $\feh\lesssim-1.0$.
At super-solar metallicities these elements tend to show flat trends,
with abundance ratios that are close to solar.
For silicon and calcium, the LTE trends are
qualitatively similar to the non-LTE ones, 
but for magnesium the LTE abundances
show a local maximum rather than a plateau, as discussed in 
\sect{discussiontrends}, that would be difficult to reconcile
with our standard understanding of Galactic chemical evolution.

The non-LTE results are quantitatively consistent with 
recent high-resolution, high signal-to-noise ratio non-LTE studies
\citep{2016ApJ...833..225Z,
2017ApJ...847...15B,2017ApJ...847...16B,2019ARep...63..726M}.
Is interesting to also compare them 
with LTE results inferred from from giants in
APOGEE \citep{2018ApJ...852...49H},
as that study is based on a large sample size.
Below $\feh<-0.9$ the authors separate their stars into those having low
(LMg) and high (HMg) magnesium abundances. The LMg population, possibly
accreted from dwarf galaxies \citep{2010A&amp;A...511L..10N},
shows a linear decrease in $\xfe{\upalpha}$ with
$\feh$,
whilst the HMg population, probably inherent to our Galaxy, shows a plateau
at approximately $0.3\,\dex$. 
These two populations may well be hidden under the systematics of
our analysis (\sect{discussiondisp}), 
and we note that the dispersion in Figs \ref{fig:mg}, \ref{fig:si}, and
\ref{fig:ca} do visibly increase below $\feh\lesssim-1.0$.

The K20 model is in qualitative agreement with the plateaus
seen in the non-LTE results.
Quantitatively, however, there are some important discrepancies.
The model overestimates the amount of magnesium and silicon
in the metal-poor Galaxy, with
plateaus of $\xfe{Mg}\approx0.5$ and $\xfe{Si}\approx0.5$,
about $0.25\,\dex$ larger than what is observed.
In contrast, the K20 model shows good agreement for calcium, 
with a plateau of $\xfe{Ca}\approx0.25$.
Compared with the other elements, it is much harder to change
the relative ratios among $\upalpha$-elements.
It might be reflect the failure of 1D hydrostatic supernova models
\citep{2016PASA...33...48M},
or perhaps uncertainties in the nuclear reaction rates
\citep{2017RvMP...89c5007D,2018ApJS..234...19F}.

\subsection{Manganese}
\label{gcemn}

Manganese is an iron-peak element that 
is important for constraining the physics 
of Type Ia supernova. More manganese than iron is produced
in these events, resulting in a positive trend of $\xfe{Mn}$ 
with $\feh$, that is the opposite to what is seen for 
the $\upalpha$-elements.  While \citet{2011A&A...530A..15N} found 
that the high-$\upalpha$ halo (enriched by Type Ia supernova)
and low-$\upalpha$ (not enriched) do not show different $\xfe{Mn}$,
the amount of manganese depends on the mode of Type Ia supernova 
\citep{2013A&A...559L...5S,2020ApJ...895..138K}, and the dominant mode
may be environment (metallicitiy) dependent 
\citep{2019ApJ...881...45K,2020ApJ...891...85D}.

As expected, therefore, \fig{fig:mn} shows $\xfe{Mn}$ increases linearly from
$-0.3\,\dex$ at $\feh\approx-1.0$,
to approximately $0.2\,\dex$ at $\feh\approx0.5$;
at lower metallicities there is a plateau
with $\xfe{Mn}\approx-0.3$. This increasing trend
is qualitatively consistent with what was found by
\citet{2020A&A...635A..38E},
although their $\xfe{Mn}$ ratios are closer to solar at all
metallicities. The difference may be because of neglected 3D effects,
and we note that the \ion{Mn}{I} $475.4\,\nm$ and $476.1\,\nm$ lines
used in the present study
appear to be sensitive
to such effects (Fig. 17 of \citealt{2019A&A...631A..80B}).
There is a much steeper trend of $\xfe{Mn}$ in LTE
(\sect{discussiontrends}).
This LTE trend is consistent with previous LTE studies
for dwarfs \citep{2015MNRAS.454.1585M},
and for giants from APOGEE \citep[][]{2018ApJ...852...49H}.

The K20 model predicts a steep trend of $\xfe{Mn}$ with
$\feh$ at low metallicities, and an inflection at 
$\feh\approx-0.75$.  The steep trend in the K20 model is in fact in 
better agreement with the LTE results than with the non-LTE results, at least
at low metallicities. However, the inflection in the K20 model is not seen
in either the LTE or non-LTE results, 
and may therefore be indicative of some missing physics.

\subsection{Barium}
\label{gceba}

Barium is a heavy element that is mainly produced in AGB stars, via the slow
neutron capture process \citep{2014PASA...31...30K},
and possibly also the
intermediate neutron capture process
\citep{2016ApJ...831..171H,2019ApJ...887...11H,2020A&A...634A..84S}.
In \fig{fig:ba} the majority of stars sit at approximately
$\xfe{Ba}\approx-0.1$ at $\feh\approx-1.0$,
gradually rise to 
$\xfe{Ba}\approx0.25$ at $\feh\approx-0.25$,
and decrease again to
$\xfe{Ba}\approx0.0$ at $\feh\approx0.25$.
At lower metallicities the barium abundances 
are slightly elevated, at approximately
$\xfe{Ba}\approx0.5$ at $\feh\lesssim-1.0$.

The general behaviour of a peak at slightly sub-solar metallicities is
in fact qualitatively similar
to what was found in the LTE study of
\citet{2017A&A...606A..94D},
and in the non-LTE study of 
\citet{2011MNRAS.415.2093K}.
In contrast, however, other LTE studies tend to find a flat trend in 
$\xfe{Ba}$, or one that very gradually increases
with increasing $\feh$
\citep{2013A&A...552A.128M,2014A&amp;A...562A..71B}.

This decrease of $\xfe{Ba}$ with $\feh$ at the highest metallicities 
is not captured in the K20 model. It
may suggest a lower production of barium from AGB stars
at higher metallicities than what is found in the K20 model,
such that iron pollution of Type Ia supernova dominates.
It may also have a similar explanation to the trend for lithium
(\sect{gceli}), namely that the highest metallicity stars
are older stars that have migrated into the solar neighbourhood,
since barium is sensitive to stellar age
\citep{2016A&A...593A..65N,2018MNRAS.474.2580S,2019A&A...631A.171S}.

There is a large dispersion in 
\fig{fig:ba} and \fig{fig:disp}.
Some of this dispersion may be intrinsic.
Studies of solar twins indicate that the youngest stars are enhanced 
in $\xfe{Ba}$ by approximately $0.2\,\dex$
(Fig.~4 of \citealt{2016A&A...593A..65N}).
However, this interpretation cannot explain why 
the results for the giants
appear to be skewed to higher barium abundances,
as these are on average $1.5\,\mathrm{Gyr}$ older than the dwarfs
according to ages from GALAH DR3.
As mentioned in \sect{discussiondwarfgiant}, this offset may reflect other
systematic errors within the GALAH analysis pipeline, not related to non-LTE
effects (noting that the LTE trend is qualitatively similar to
the non-LTE one). Possibly the
problems are related to the choice of microturbulence, 
to which the \ion{Ba}{II} lines 
are particularly sensitive \citep{2012A&A...540A.128D}.
The barium abundances presented
here should therefore be treated with caution.

\section{Conclusion}
\label{conclusion}

We have presented extensive grids of departure coefficients
for $13$ different elements, calculated on standard
\marcs{} model atmospheres. These data can be found online
\citep{grid_nlte}
or by contacting the lead author directly.
The departure coefficients can be implemented
into existing stellar analyses pipelines, to simply and cheaply 
account for departures from LTE for a large number of spectral lines.

We illustrated this latter point
by implementing the grids into the GALAH 
analysis pipeline, that is based on the 1D spectrum synthesis code \sme{}.
The grids of departure coefficients clearly 
make an impact on large stellar surveys.
The grids lead to differences in the mean $\xfe{A}$ at given $\feh$ of
as much as $0.2\,\dex$.  This results in qualitatively
different pictures of the Galactic chemical evolution,
for example giving rise to metal-poor plateaus in
$\xfe{Mg}$ (\fig{fig:mg}),
$\xfe{K}$ (\fig{fig:k}), and 
$\xfe{Mn}$ (\fig{fig:mn}), whereas a steep increase
with increasing $\feh$ would be seen in LTE.
The grids affect different parts of stellar parameter space differently,
and thus for example remove offsets
in $\xfe{Na}$ of approximately $0.2\,\dex$ between
dwarfs and giants (\fig{fig:na}).
The grids generally reduce the dispersion
in the $\xfe{A}$ versus $\feh$ plane (\fig{fig:disp}),
by as much as $0.10\,\dex$ in the case of $\xfe{Si}$ (\fig{fig:si}).

Future efforts will extend these calculations
to more elements in the periodic table.
This will make it possible to model even
more elements in non-LTE,
in GALAH as well as in other large spectroscopic surveys.
This is a necessary step towards getting elemental
abundances that are accurate to the $0.05\,\dex$ level,
of the order of millions of stars.

We caution that there
are nevertheless other systematics in stellar models that still
need to be addressed. 
After taking departures from LTE into account,
the next step may be to consider the effects of 
stellar convection \citep{2018A&ARv..26....6N}.
Unfortunately 3D non-LTE corrections \citep{2019A&A...630A.104A}
cannot be implemented into existing
stellar analyses pipelines in an analogous way
to these pre-computed grids of 1D non-LTE departure coefficients;
at present, grids of line-by-line 3D non-LTE abundance corrections
are still required.
However, in the more distant future, stellar analyses pipelines
may move towards using 3D spectrum synthesis codes,
and to using pre-computed grids of 3D non-LTE departure coefficients.

It is also unclear to what extent departures from LTE
may impact the atmospheric stratifications themselves
\citep{2005ApJ...618..926S}.
The grids of departure coefficients presented
here offer a way forward towards 
relaxing the LTE assumption in the construction 
of 1D model atmospheres.  The departure coefficients
for the key electron donors could be fed back into the 
model atmosphere code for a final iteration,
to see the impact of departures from LTE on
the atmospheric temperature stratification.
This will be investigated in a future study.

\begin{acknowledgements}
The GALAH survey is based on observations made at the Anglo-Australia
Telescope (AAT).
We acknowledge the traditional owners of the land on which the AAT
stands, the Gamilaraay people, and pay our respects to elders past and present.
We thank Poul Erik Nissen for providing comments on the manuscript.
AMA and PSB acknowledge 
support from the Swedish Research Council (VR 2016-03765),
and the project grant `The New Milky Way'
(KAW 2013.0052) from the Knut and Alice Wallenberg Foundation.
MA gratefully acknowledges funding from
the Australian Research Council
(DP150100250 and FL110100012).
HR acknowledges support from JHU.
CK acknowledges funding from the UK Science and Technology Facility
Council (STFC) through grants ST/M000958/1 and ST/R000905/1, and the
Stromlo Distinguished Visitorship at the ANU.
SLM acknowledges support from the UNSW Scientia Fellowship program,
and from the Australian Research Council 
(DP180101791). JDS and DBZ acknowledge support from
the Australian Research Council (DP180101791).
This work was also supported by
the Australian Research Council Centre of Excellence for All Sky
Astrophysics in 3 Dimensions (ASTRO 3D).
Some of the computations were performed on resources provided 
(through projects SNIC 2018/3-465 and SNIC 2019/3-532)
by the Swedish National Infrastructure for Computing (SNIC) at 
the Multidisciplinary Center for Advanced Computational Science (UPPMAX) 
and at the High Performance Computing Center North (HPC2N),
partially funded by the Swedish Research Council 
through grant agreement no. 2016-07213.
This work was supported by computational resources provided by 
the Australian Government through the 
National Computational Infrastructure (NCI)
under the National Computational Merit Allocation Scheme.
\end{acknowledgements}

\bibliographystyle{aa} 
\bibliography{/Users/ama51/Documents/work/papers/bibl.bib}
\label{lastpage}
\end{document}

%% file: atoms.tex
\begin{table}
\begin{center}
\caption{Number of levels, lines, and continua, and the original references of the adopted model atoms. A non-LTE grid for iron (highlighted) is also adopted in GALAH; details of those calculations can be found in \citet{2016MNRAS.463.1518A}.}
\label{tab:atoms}
\begin{tabular}{l | c c c c | c}
\hline
\multicolumn{1}{l|}{Element} &
\multicolumn{1}{c}{Species} &
\multicolumn{1}{c}{\# levels} &
\multicolumn{1}{c}{\# lines} &
\multicolumn{1}{c|}{\# continua} &
\multicolumn{1}{c}{Ref.} \\
\hline
\hline
\multirow{2}{*}{H} &
\multicolumn{1}{c}{\ion{H}{I}} &
\multicolumn{1}{c}{  20} &
\multicolumn{1}{c}{ 190} &
\multicolumn{1}{c|}{  20} &
\multirow{ 2}{*}{ 1} \\
 & 
\multicolumn{1}{c}{\ion{H}{II}} &
\multicolumn{1}{c}{   1} &
\multicolumn{1}{c}{-} &
\multicolumn{1}{c|}{-} &
 \\ 
\hline
\multirow{2}{*}{Li} &
\multicolumn{1}{c}{\ion{Li}{I}} &
\multicolumn{1}{c}{  20} &
\multicolumn{1}{c}{ 113} &
\multicolumn{1}{c|}{  20} &
\multirow{ 2}{*}{ 2,  3} \\
 & 
\multicolumn{1}{c}{\ion{Li}{II}} &
\multicolumn{1}{c}{   1} &
\multicolumn{1}{c}{-} &
\multicolumn{1}{c|}{-} &
 \\ 
\hline
\multirow{2}{*}{C} &
\multicolumn{1}{c}{\ion{C}{I}} &
\multicolumn{1}{c}{  46} &
\multicolumn{1}{c}{ 343} &
\multicolumn{1}{c|}{  45} &
\multirow{ 2}{*}{ 4} \\
 & 
\multicolumn{1}{c}{\ion{C}{II}} &
\multicolumn{1}{c}{   1} &
\multicolumn{1}{c}{-} &
\multicolumn{1}{c|}{-} &
 \\ 
\hline
\multirow{2}{*}{N} &
\multicolumn{1}{c}{\ion{N}{I}} &
\multicolumn{1}{c}{  31} &
\multicolumn{1}{c}{ 174} &
\multicolumn{1}{c|}{  32} &
\multirow{ 2}{*}{ 5} \\
 & 
\multicolumn{1}{c}{\ion{N}{II}} &
\multicolumn{1}{c}{   2} &
\multicolumn{1}{c}{-} &
\multicolumn{1}{c|}{-} &
 \\ 
\hline
\multirow{2}{*}{O} &
\multicolumn{1}{c}{\ion{O}{I}} &
\multicolumn{1}{c}{  47} &
\multicolumn{1}{c}{ 322} &
\multicolumn{1}{c|}{  47} &
\multirow{ 2}{*}{ 6} \\
 & 
\multicolumn{1}{c}{\ion{O}{II}} &
\multicolumn{1}{c}{   3} &
\multicolumn{1}{c}{-} &
\multicolumn{1}{c|}{-} &
 \\ 
\hline
\multirow{2}{*}{Na} &
\multicolumn{1}{c}{\ion{Na}{I}} &
\multicolumn{1}{c}{  22} &
\multicolumn{1}{c}{ 166} &
\multicolumn{1}{c|}{  22} &
\multirow{ 2}{*}{ 7} \\
 & 
\multicolumn{1}{c}{\ion{Na}{II}} &
\multicolumn{1}{c}{   1} &
\multicolumn{1}{c}{-} &
\multicolumn{1}{c|}{-} &
 \\ 
\hline
\multirow{3}{*}{Mg} &
\multicolumn{1}{c}{\ion{Mg}{I}} &
\multicolumn{1}{c}{  96} &
\multicolumn{1}{c}{ 721} &
\multicolumn{1}{c|}{  96} &
\multirow{ 3}{*}{ 8} \\
 & 
\multicolumn{1}{c}{\ion{Mg}{II}} &
\multicolumn{1}{c}{  29} &
\multicolumn{1}{c}{ 161} &
\multicolumn{1}{c|}{  29} &
 \\ 
 & 
\multicolumn{1}{c}{\ion{Mg}{III}} &
\multicolumn{1}{c}{   1} &
\multicolumn{1}{c}{-} &
\multicolumn{1}{c|}{-} &
 \\ 
\hline
\multirow{2}{*}{Al} &
\multicolumn{1}{c}{\ion{Al}{I}} &
\multicolumn{1}{c}{  42} &
\multicolumn{1}{c}{ 135} &
\multicolumn{1}{c|}{  33} &
\multirow{ 2}{*}{ 9} \\
 & 
\multicolumn{1}{c}{\ion{Al}{II}} &
\multicolumn{1}{c}{   2} &
\multicolumn{1}{c}{-} &
\multicolumn{1}{c|}{-} &
 \\ 
\hline
\multirow{2}{*}{Si} &
\multicolumn{1}{c}{\ion{Si}{I}} &
\multicolumn{1}{c}{  56} &
\multicolumn{1}{c}{ 634} &
\multicolumn{1}{c|}{  56} &
\multirow{ 2}{*}{10} \\
 & 
\multicolumn{1}{c}{\ion{Si}{II}} &
\multicolumn{1}{c}{   1} &
\multicolumn{1}{c}{-} &
\multicolumn{1}{c|}{-} &
 \\ 
\hline
\multirow{2}{*}{K} &
\multicolumn{1}{c}{\ion{K}{I}} &
\multicolumn{1}{c}{ 133} &
\multicolumn{1}{c}{ 250} &
\multicolumn{1}{c|}{ 133} &
\multirow{ 2}{*}{11} \\
 & 
\multicolumn{1}{c}{\ion{K}{II}} &
\multicolumn{1}{c}{   1} &
\multicolumn{1}{c}{-} &
\multicolumn{1}{c|}{-} &
 \\ 
\hline
\multirow{3}{*}{Ca} &
\multicolumn{1}{c}{\ion{Ca}{I}} &
\multicolumn{1}{c}{  67} &
\multicolumn{1}{c}{ 937} &
\multicolumn{1}{c|}{  67} &
\multirow{ 3}{*}{12} \\
 & 
\multicolumn{1}{c}{\ion{Ca}{II}} &
\multicolumn{1}{c}{  24} &
\multicolumn{1}{c}{  89} &
\multicolumn{1}{c|}{  24} &
 \\ 
 & 
\multicolumn{1}{c}{\ion{Ca}{III}} &
\multicolumn{1}{c}{   1} &
\multicolumn{1}{c}{-} &
\multicolumn{1}{c|}{-} &
 \\ 
\hline
\multirow{3}{*}{Mn} &
\multicolumn{1}{c}{\ion{Mn}{I}} &
\multicolumn{1}{c}{ 198} &
\multicolumn{1}{c}{1645} &
\multicolumn{1}{c|}{ 198} &
\multirow{ 3}{*}{13} \\
 & 
\multicolumn{1}{c}{\ion{Mn}{II}} &
\multicolumn{1}{c}{  82} &
\multicolumn{1}{c}{  44} &
\multicolumn{1}{c|}{-} &
 \\ 
 & 
\multicolumn{1}{c}{\ion{Mn}{III}} &
\multicolumn{1}{c}{   1} &
\multicolumn{1}{c}{-} &
\multicolumn{1}{c|}{-} &
 \\ 
\hline
\multirow{3}{*}{{\textcolor{\colsmall}{Fe}}} &
\multicolumn{1}{c}{{\textcolor{\colsmall}{\ion{Fe}{I}}}} &
\multicolumn{1}{c}{{\textcolor{\colsmall}{ 421}}} &
\multicolumn{1}{c}{{\textcolor{\colsmall}{3923}}} &
\multicolumn{1}{c|}{{\textcolor{\colsmall}{  48}}} &
\multirow{ 3}{*}{{\textcolor{\colsmall}{14, 15}}} \\
 & 
\multicolumn{1}{c}{{\textcolor{\colsmall}{\ion{Fe}{II}}}} &
\multicolumn{1}{c}{{\textcolor{\colsmall}{  41}}} &
\multicolumn{1}{c}{{\textcolor{\colsmall}{  77}}} &
\multicolumn{1}{c|}{{\textcolor{\colsmall}{-}}} &
 \\ 
 & 
\multicolumn{1}{c}{{\textcolor{\colsmall}{\ion{Fe}{III}}}} &
\multicolumn{1}{c}{{\textcolor{\colsmall}{   1}}} &
\multicolumn{1}{c}{{\textcolor{\colsmall}{-}}} &
\multicolumn{1}{c|}{{\textcolor{\colsmall}{-}}} &
 \\ 
\hline
\multirow{3}{*}{Ba} &
\multicolumn{1}{c}{\ion{Ba}{I}} &
\multicolumn{1}{c}{   8} &
\multicolumn{1}{c}{-} &
\multicolumn{1}{c|}{   8} &
\multirow{ 3}{*}{16} \\
 & 
\multicolumn{1}{c}{\ion{Ba}{II}} &
\multicolumn{1}{c}{ 102} &
\multicolumn{1}{c}{ 284} &
\multicolumn{1}{c|}{ 102} &
 \\ 
 & 
\multicolumn{1}{c}{\ion{Ba}{III}} &
\multicolumn{1}{c}{   1} &
\multicolumn{1}{c}{-} &
\multicolumn{1}{c|}{-} &
 \\ 
\hline
\end{tabular}
\end{center}
\tablebib{(1) \citet{2018A&A...615A.139A}; (2) \citet{2013A&amp;A...554A..96L}; (3) Wang et al. (in prep.); (4) \citet{2019A&A...624A.111A}; (5) \citet{2020A&A...636A.120A}; (6) \citet{2018A&A...616A..89A}; (7) \citet{2011A&amp;A...528A.103L}; (8) \citet{2015A&amp;A...579A..53O}; (9) \citet{2017A&amp;A...607A..75N}; (10) \citet{2017MNRAS.464..264A}; (11) \citet{2019A&A...627A.177R}; (12) \citet{2019A&A...623A.103O}; (13) \citet{2019A&A...631A..80B}; (14) \citet{2016MNRAS.463.1518A}; (15) \citet{2017MNRAS.468.4311L}; (16) \citet{2020A&A...634A..55G}. }
\end{table}

%% file: linelist.tex
\begin{table*}
\begin{center}
\caption{Lines used in the reanalysis of GALAH DR3 stars. The two components of
    the \ion{Li}{I} doublet were fit simultaneously. Also shown are the solar
    abundances inferred via the solar twins in the sample; the value for the
    \ion{Li}{I} doublet in brackets was adopted from the literature
    (\sect{resultssolar}). The final columns show the typical absolute
    abundance corrections ($\lgeps{}^{\text{Non-LTE}}-\lgeps{}^{\text{LTE}}$)
    for each line: median results are shown in $(\teff/K\pm250,\lggu\pm0.5)$
    regions corresponding to spectral and luminosity classes K V (5000, 4.5), F V (6500, 4.0), and K III (4250, 2.0). The most negative and positive corrections for a given line are highlighted, and no results are shown for where the lines were too weak to be detected.}
\label{tab:linelist}
\begin{tabular}{l c c c c | c c | c c c | c c c}
\hline
\multirow{2}{*}{Spec.} &
\multirow{2}{*}{$\lambda_{\text{air}} / \mathrm{nm}$} &
\multirow{2}{*}{$\chi_{\text{low}} / \mathrm{eV}$} &
\multirow{2}{*}{$\lggf$} &
\multirow{2}{*}{Ref.} &
\multirow{2}{*}{$\lgeps{}^{\text{Non-LTE}}_{\odot}$} &
\multirow{2}{*}{$\lgeps{}^{\text{LTE}}_{\odot}$} &
\multicolumn{3}{c|}{$-2<\feh<-1$} &
\multicolumn{3}{c}{$0<\feh<1$} \\
& & & & & & & 
\multicolumn{1}{c}{K V} &
\multicolumn{1}{c}{F V} &
\multicolumn{1}{c|}{K III} &
\multicolumn{1}{c}{K V} &
\multicolumn{1}{c}{F V} &
\multicolumn{1}{c}{K III} \\
\hline
\hline
\multirow{2}{*}{\ion{Li}{I}} &
\multicolumn{1}{c}{$   670.776$} &
\multicolumn{1}{c}{$     0.000$} &
\multicolumn{1}{c}{$    -0.002$} &
\multirow{2}{*}{ 1} &
\multirow{2}{*}{($      1.05$)} &
\multirow{2}{*}{($      1.05$)} &
\multirow{2}{*}{$     -0.05$} & 
\multirow{2}{*}{\textcolor{\colsmall}{{$     -0.10$}}} & 
\multirow{2}{*}{$     -0.02$} & 
\multirow{2}{*}{$     -0.02$} & 
\multirow{2}{*}{$     -0.07$} & 
\multirow{2}{*}{\textcolor{\colbig}{{$+      0.02$}}} \\
 & 
\multicolumn{1}{c}{$   670.791$} &
\multicolumn{1}{c}{$     0.000$} &
\multicolumn{1}{c}{$    -0.303$} &
 & 
 & 
 & 
 & & & & & \\
\hline
\multirow{1}{*}{\ion{C}{I}} &
\multicolumn{1}{c}{$   658.761$} &
\multicolumn{1}{c}{$     8.537$} &
\multicolumn{1}{c}{$    -1.021$} &
\multirow{1}{*}{ 2} &
\multirow{1}{*}{$      8.43$} &
\multirow{1}{*}{$      8.43$} &
\multirow{1}{*}{} & 
\multirow{1}{*}{\textcolor{\colsmall}{{$     -0.02$}}} & 
\multirow{1}{*}{$+      0.00$} & 
\multirow{1}{*}{$+      0.00$} & 
\multirow{1}{*}{$     -0.01$} & 
\multirow{1}{*}{\textcolor{\colbig}{{$+      0.00$}}} \\
\hline
\multirow{1}{*}{\ion{O}{I}} &
\multicolumn{1}{c}{$   777.194$} &
\multicolumn{1}{c}{$     9.146$} &
\multicolumn{1}{c}{$+     0.369$} &
\multirow{1}{*}{ 3} &
\multirow{1}{*}{$      8.79$} &
\multirow{1}{*}{$      8.99$} &
\multirow{1}{*}{\textcolor{\colbig}{{$     -0.06$}}} & 
\multirow{1}{*}{$     -0.26$} & 
\multirow{1}{*}{$     -0.12$} & 
\multirow{1}{*}{$     -0.08$} & 
\multirow{1}{*}{\textcolor{\colsmall}{{$     -0.52$}}} & 
\multirow{1}{*}{$     -0.12$} \\
\multirow{1}{*}{\ion{O}{I}} &
\multicolumn{1}{c}{$   777.417$} &
\multicolumn{1}{c}{$     9.146$} &
\multicolumn{1}{c}{$+     0.223$} &
\multirow{1}{*}{ 3} &
\multirow{1}{*}{$      8.80$} &
\multirow{1}{*}{$      8.98$} &
\multirow{1}{*}{\textcolor{\colbig}{{$     -0.04$}}} & 
\multirow{1}{*}{$     -0.22$} & 
\multirow{1}{*}{$     -0.12$} & 
\multirow{1}{*}{$     -0.07$} & 
\multirow{1}{*}{\textcolor{\colsmall}{{$     -0.48$}}} & 
\multirow{1}{*}{$     -0.13$} \\
\multirow{1}{*}{\ion{O}{I}} &
\multicolumn{1}{c}{$   777.539$} &
\multicolumn{1}{c}{$     9.146$} &
\multicolumn{1}{c}{$+     0.002$} &
\multirow{1}{*}{ 3} &
\multirow{1}{*}{$      8.79$} &
\multirow{1}{*}{$      8.94$} &
\multirow{1}{*}{\textcolor{\colbig}{{$     -0.05$}}} & 
\multirow{1}{*}{$     -0.19$} & 
\multirow{1}{*}{$     -0.12$} & 
\multirow{1}{*}{$     -0.08$} & 
\multirow{1}{*}{\textcolor{\colsmall}{{$     -0.42$}}} & 
\multirow{1}{*}{$     -0.07$} \\
\hline
\multirow{1}{*}{\ion{Na}{I}} &
\multicolumn{1}{c}{$   568.263$} &
\multicolumn{1}{c}{$     2.102$} &
\multicolumn{1}{c}{$    -0.706$} &
\multirow{1}{*}{ 4} &
\multirow{1}{*}{$      6.04$} &
\multirow{1}{*}{$      6.19$} &
\multirow{1}{*}{\textcolor{\colbig}{{$     -0.10$}}} & 
\multirow{1}{*}{$     -0.12$} & 
\multirow{1}{*}{$     -0.12$} & 
\multirow{1}{*}{$     -0.13$} & 
\multirow{1}{*}{$     -0.17$} & 
\multirow{1}{*}{\textcolor{\colsmall}{{$     -0.23$}}} \\
\multirow{1}{*}{\ion{Na}{I}} &
\multicolumn{1}{c}{$   568.821$} &
\multicolumn{1}{c}{$     2.104$} &
\multicolumn{1}{c}{$    -0.404$} &
\multirow{1}{*}{ 4} &
\multirow{1}{*}{$      6.05$} &
\multirow{1}{*}{$      6.24$} &
\multirow{1}{*}{\textcolor{\colbig}{{$     -0.10$}}} & 
\multirow{1}{*}{$     -0.14$} & 
\multirow{1}{*}{$     -0.17$} & 
\multirow{1}{*}{$     -0.14$} & 
\multirow{1}{*}{\textcolor{\colsmall}{{$     -0.23$}}} & 
\multirow{1}{*}{$     -0.19$} \\
\hline
\multirow{1}{*}{\ion{Mg}{I}} &
\multicolumn{1}{c}{$   571.109$} &
\multicolumn{1}{c}{$     4.346$} &
\multicolumn{1}{c}{$    -1.724$} &
\multirow{1}{*}{ 5} &
\multirow{1}{*}{$      7.59$} &
\multirow{1}{*}{$      7.63$} &
\multirow{1}{*}{$     -0.03$} & 
\multirow{1}{*}{\textcolor{\colbig}{{$+      0.03$}}} & 
\multirow{1}{*}{$     -0.11$} & 
\multirow{1}{*}{$     -0.06$} & 
\multirow{1}{*}{$     -0.05$} & 
\multirow{1}{*}{\textcolor{\colsmall}{{$     -0.18$}}} \\
\hline
\multirow{1}{*}{\ion{Al}{I}} &
\multicolumn{1}{c}{$   669.867$} &
\multicolumn{1}{c}{$     3.143$} &
\multicolumn{1}{c}{$    -1.870$} &
\multirow{1}{*}{ 6} &
\multirow{1}{*}{$      6.39$} &
\multirow{1}{*}{$      6.41$} &
\multirow{1}{*}{$     -0.03$} & 
\multirow{1}{*}{} & 
\multirow{1}{*}{$     -0.09$} & 
\multirow{1}{*}{$     -0.03$} & 
\multirow{1}{*}{\textcolor{\colbig}{{$     -0.01$}}} & 
\multirow{1}{*}{\textcolor{\colsmall}{{$     -0.10$}}} \\
\multirow{1}{*}{\ion{Al}{I}} &
\multicolumn{1}{c}{$   783.531$} &
\multicolumn{1}{c}{$     4.022$} &
\multicolumn{1}{c}{$    -0.689$} &
\multirow{1}{*}{ 6} &
\multirow{1}{*}{$      6.41$} &
\multirow{1}{*}{$      6.43$} &
\multirow{1}{*}{$     -0.03$} & 
\multirow{1}{*}{} & 
\multirow{1}{*}{$     -0.07$} & 
\multirow{1}{*}{$     -0.03$} & 
\multirow{1}{*}{\textcolor{\colbig}{{$     -0.02$}}} & 
\multirow{1}{*}{\textcolor{\colsmall}{{$     -0.13$}}} \\
\hline
\multirow{1}{*}{\ion{Si}{I}} &
\multicolumn{1}{c}{$   568.448$} &
\multicolumn{1}{c}{$     4.954$} &
\multicolumn{1}{c}{$    -1.553$} &
\multirow{1}{*}{ 7,  8} &
\multirow{1}{*}{$      7.39$} &
\multirow{1}{*}{$      7.42$} &
\multirow{1}{*}{$     -0.01$} & 
\multirow{1}{*}{\textcolor{\colbig}{{$+      0.00$}}} & 
\multirow{1}{*}{$     -0.06$} & 
\multirow{1}{*}{$     -0.02$} & 
\multirow{1}{*}{$     -0.03$} & 
\multirow{1}{*}{\textcolor{\colsmall}{{$     -0.08$}}} \\
\multirow{1}{*}{\ion{Si}{I}} &
\multicolumn{1}{c}{$   569.043$} &
\multicolumn{1}{c}{$     4.930$} &
\multicolumn{1}{c}{$    -1.773$} &
\multirow{1}{*}{ 7,  8} &
\multirow{1}{*}{$      7.43$} &
\multirow{1}{*}{$      7.44$} &
\multirow{1}{*}{$     -0.01$} & 
\multirow{1}{*}{\textcolor{\colbig}{{$+      0.00$}}} & 
\multirow{1}{*}{$     -0.05$} & 
\multirow{1}{*}{$     -0.01$} & 
\multirow{1}{*}{$     -0.02$} & 
\multirow{1}{*}{\textcolor{\colsmall}{{$     -0.06$}}} \\
\multirow{1}{*}{\ion{Si}{I}} &
\multicolumn{1}{c}{$   570.110$} &
\multicolumn{1}{c}{$     4.930$} &
\multicolumn{1}{c}{$    -1.953$} &
\multirow{1}{*}{ 7,  8} &
\multirow{1}{*}{$      7.49$} &
\multirow{1}{*}{$      7.50$} &
\multirow{1}{*}{$     -0.01$} & 
\multirow{1}{*}{} & 
\multirow{1}{*}{\textcolor{\colsmall}{{$     -0.04$}}} & 
\multirow{1}{*}{\textcolor{\colbig}{{$     -0.01$}}} & 
\multirow{1}{*}{$     -0.02$} & 
\multirow{1}{*}{$     -0.03$} \\
\multirow{1}{*}{\ion{Si}{I}} &
\multicolumn{1}{c}{$   577.215$} &
\multicolumn{1}{c}{$     5.082$} &
\multicolumn{1}{c}{$    -1.653$} &
\multirow{1}{*}{ 7,  8} &
\multirow{1}{*}{$      7.50$} &
\multirow{1}{*}{$      7.52$} &
\multirow{1}{*}{$     -0.01$} & 
\multirow{1}{*}{\textcolor{\colbig}{{$     -0.01$}}} & 
\multirow{1}{*}{$     -0.04$} & 
\multirow{1}{*}{$     -0.01$} & 
\multirow{1}{*}{$     -0.02$} & 
\multirow{1}{*}{\textcolor{\colsmall}{{$     -0.07$}}} \\
\multirow{1}{*}{\ion{Si}{I}} &
\multicolumn{1}{c}{$   579.307$} &
\multicolumn{1}{c}{$     4.930$} &
\multicolumn{1}{c}{$    -1.963$} &
\multirow{1}{*}{ 7,  8} &
\multirow{1}{*}{$      7.48$} &
\multirow{1}{*}{$      7.50$} &
\multirow{1}{*}{$     -0.01$} & 
\multirow{1}{*}{\textcolor{\colbig}{{$     -0.01$}}} & 
\multirow{1}{*}{$     -0.04$} & 
\multirow{1}{*}{$     -0.01$} & 
\multirow{1}{*}{$     -0.02$} & 
\multirow{1}{*}{\textcolor{\colsmall}{{$     -0.11$}}} \\
\multirow{1}{*}{\ion{Si}{I}} &
\multicolumn{1}{c}{$   768.027$} &
\multicolumn{1}{c}{$     5.863$} &
\multicolumn{1}{c}{$    -0.590$} &
\multirow{1}{*}{ 7,  8} &
\multirow{1}{*}{$      7.57$} &
\multirow{1}{*}{$      7.61$} &
\multirow{1}{*}{$     -0.02$} & 
\multirow{1}{*}{\textcolor{\colbig}{{$     -0.01$}}} & 
\multirow{1}{*}{$     -0.11$} & 
\multirow{1}{*}{$     -0.04$} & 
\multirow{1}{*}{$     -0.06$} & 
\multirow{1}{*}{\textcolor{\colsmall}{{$     -0.25$}}} \\
\hline
\multirow{1}{*}{\ion{K}{I}} &
\multicolumn{1}{c}{$   769.896$} &
\multicolumn{1}{c}{$     0.000$} &
\multicolumn{1}{c}{$    -0.178$} &
\multirow{1}{*}{ 9} &
\multirow{1}{*}{$      5.05$} &
\multirow{1}{*}{$      5.49$} &
\multirow{1}{*}{$     -0.28$} & 
\multirow{1}{*}{$     -0.28$} & 
\multirow{1}{*}{\textcolor{\colsmall}{{$     -0.66$}}} & 
\multirow{1}{*}{\textcolor{\colbig}{{$     -0.21$}}} & 
\multirow{1}{*}{$     -0.62$} & 
\multirow{1}{*}{$     -0.30$} \\
\hline
\multirow{1}{*}{\ion{Ca}{I}} &
\multicolumn{1}{c}{$   586.756$} &
\multicolumn{1}{c}{$     2.933$} &
\multicolumn{1}{c}{$    -1.570$} &
\multirow{1}{*}{10} &
\multirow{1}{*}{$      6.32$} &
\multirow{1}{*}{$      6.32$} &
\multirow{1}{*}{$     -0.01$} & 
\multirow{1}{*}{} & 
\multirow{1}{*}{$     -0.01$} & 
\multirow{1}{*}{$     -0.01$} & 
\multirow{1}{*}{\textcolor{\colbig}{{$     -0.01$}}} & 
\multirow{1}{*}{\textcolor{\colsmall}{{$     -0.03$}}} \\
\multirow{1}{*}{\ion{Ca}{I}} &
\multicolumn{1}{c}{$   649.965$} &
\multicolumn{1}{c}{$     2.523$} &
\multicolumn{1}{c}{$    -0.818$} &
\multirow{1}{*}{11} &
\multirow{1}{*}{$      6.27$} &
\multirow{1}{*}{$      6.31$} &
\multirow{1}{*}{\textcolor{\colsmall}{{$     -0.06$}}} & 
\multirow{1}{*}{$+      0.00$} & 
\multirow{1}{*}{$     -0.01$} & 
\multirow{1}{*}{$     -0.05$} & 
\multirow{1}{*}{$+      0.01$} & 
\multirow{1}{*}{\textcolor{\colbig}{{$+      0.02$}}} \\
\hline
\multirow{1}{*}{\ion{Mn}{I}} &
\multicolumn{1}{c}{$   475.404$} &
\multicolumn{1}{c}{$     2.282$} &
\multicolumn{1}{c}{$    -0.080$} &
\multirow{1}{*}{12} &
\multirow{1}{*}{$      5.29$} &
\multirow{1}{*}{$      5.28$} &
\multirow{1}{*}{$+      0.06$} & 
\multirow{1}{*}{\textcolor{\colbig}{{$+      0.15$}}} & 
\multirow{1}{*}{$+      0.10$} & 
\multirow{1}{*}{\textcolor{\colsmall}{{$+      0.00$}}} & 
\multirow{1}{*}{$+      0.00$} & 
\multirow{1}{*}{$+      0.02$} \\
\multirow{1}{*}{\ion{Mn}{I}} &
\multicolumn{1}{c}{$   476.151$} &
\multicolumn{1}{c}{$     2.953$} &
\multicolumn{1}{c}{$    -0.274$} &
\multirow{1}{*}{12} &
\multirow{1}{*}{$      5.36$} &
\multirow{1}{*}{$      5.34$} &
\multirow{1}{*}{$+      0.05$} & 
\multirow{1}{*}{\textcolor{\colbig}{{$+      0.12$}}} & 
\multirow{1}{*}{$+      0.12$} & 
\multirow{1}{*}{\textcolor{\colsmall}{{$+      0.01$}}} & 
\multirow{1}{*}{$+      0.03$} & 
\multirow{1}{*}{$+      0.02$} \\
\hline
\multirow{1}{*}{\ion{Ba}{II}} &
\multicolumn{1}{c}{$   585.367$} &
\multicolumn{1}{c}{$     0.604$} &
\multicolumn{1}{c}{$    -0.907$} &
\multirow{1}{*}{13} &
\multirow{1}{*}{$      2.13$} &
\multirow{1}{*}{$      2.22$} &
\multirow{1}{*}{\textcolor{\colbig}{{$     -0.01$}}} & 
\multirow{1}{*}{$     -0.04$} & 
\multirow{1}{*}{$     -0.12$} & 
\multirow{1}{*}{$     -0.04$} & 
\multirow{1}{*}{\textcolor{\colsmall}{{$     -0.14$}}} & 
\multirow{1}{*}{$     -0.03$} \\
\multirow{1}{*}{\ion{Ba}{II}} &
\multicolumn{1}{c}{$   649.689$} &
\multicolumn{1}{c}{$     0.604$} &
\multicolumn{1}{c}{$    -0.407$} &
\multirow{1}{*}{13} &
\multirow{1}{*}{$      2.23$} &
\multirow{1}{*}{$      2.42$} &
\multirow{1}{*}{$     -0.07$} & 
\multirow{1}{*}{$     -0.12$} & 
\multirow{1}{*}{$     -0.18$} & 
\multirow{1}{*}{$     -0.10$} & 
\multirow{1}{*}{\textcolor{\colsmall}{{$     -0.29$}}} & 
\multirow{1}{*}{\textcolor{\colbig}{{$     -0.07$}}} \\
\hline
\end{tabular}
\end{center}
\tablebib{(1) \citet{1998PhRvA..57.1652Y}; (2) \citet{1993A&AS...99..179H}; (3) \citet{1991JPhB...24.3943H}; (4) \citet{2006ADNDT..92..607F}; (5) \citet{1990JQSRT..43..207C}; (6) \citet{2008JPCRD..37..709K}; (7) \citet{1973A&A....26..471G}; (8) \citet{1991PhRvA..44.7134O}; (9) \citet{2017PhRvA..95e2507T}; (10) \citet{1988JPhB...21.2827S}; (11) \citet{1981JPhB...14.4015S}; (12) \citet{2011ApJS..194...35D}; (13) \citet{1992A&A...255..457D}. }
\end{table*}

%% file: paper.bbl
\begin{thebibliography}{151}
\expandafter\ifx\csname natexlab\endcsname\relax\def\natexlab#1{#1}\fi

\bibitem[{{Ahumada} {et~al.}(2020){Ahumada}, {Allende Prieto}, {Almeida},
  {Anders}, {Anderson}, {Andrews}, {Anguiano}, {Arcodia}, {Armengaud},
  {Aubert}, {Avila}, {Avila-Reese}, {Badenes}, {Balland }, {Barger},
  {Barrera-Ballesteros}, {Basu}, {Bautista}, {Beaton}, {Beers}, {Benavides},
  {Bender}, {Bernardi}, {Bershady}, {Beutler}, {Bidin}, {Bird}, {Bizyaev},
  {Blanc}, {Blanton}, {Boquien}, {Borissova}, {Bovy}, {Brand t}, {Brinkmann},
  {Brownstein}, {Bundy}, {Bureau}, {Burgasser}, {Burtin}, {Cano-D{\'\i}az},
  {Capasso}, {Cappellari}, {Carrera}, {Chabanier}, {Chaplin}, {Chapman},
  {Cherinka}, {Chiappini}, {Doohyun Choi}, {Chojnowski}, {Chung}, {Clerc},
  {Coffey}, {Comerford}, {Comparat}, {da Costa}, {Cousinou}, {Covey}, {Crane},
  {Cunha}, {da Silva Ilha}, {Dai}, {Damsted}, {Darling}, {Davidson}, {Davies},
  {Dawson}, {De}, {de la Macorra}, {De Lee}, {de Andrade Queiroz}, {Deconto
  Machado}, {de la Torre}, {Dell'Agli}, {du Mas des Bourboux},
  {Diamond-Stanic}, {Dillon}, {Donor}, {Drory}, {Duckworth}, {Dwelly},
  {Ebelke}, {Eftekharzadeh}, {Eigenbrot}, {Elsworth}, {Eracleous},
  {Erfanianfar}, {Escoffier}, {Fan}, {Farr}, {Fern{\'a}ndez-Trincado},
  {Feuillet}, {Finoguenov}, {Fofie}, {Fraser-McKelvie}, {Frinchaboy},
  {Fromenteau}, {Fu}, {Galbany}, {Garcia}, {Garc{\'\i}a-Hern{\'a}ndez}, {Garma
  Oehmichen}, {Ge}, {Geimba Maia}, {Geisler}, {Gelfand }, {Goddy},
  {Gonzalez-Perez}, {Grabowski}, {Green}, {Grier}, {Guo}, {Guy}, {Harding},
  {Hasselquist}, {Hawken}, {Hayes}, {Hearty}, {Hekker}, {Hogg}, {Holtzman},
  {Horta}, {Hou}, {Hsieh}, {Huber}, {Hunt}, {Ider Chitham}, {Imig}, {Jaber},
  {Jimenez Angel}, {Johnson}, {Jones}, {J{\"o}nsson}, {Jullo}, {Kim},
  {Kinemuchi}, {Kirkpatrick}, {Kite}, {Klaene}, {Kneib}, {Kollmeier}, {Kong},
  {Kounkel}, {Krishnarao}, {Lacerna}, {Lan}, {Lane}, {Law}, {Le Goff}, {Leung},
  {Lewis}, {Li}, {Lian}, {Lin}, {Long}, {Longa-Pe{\~n}a}, {Lundgren}, {Lyke},
  {Ted Mackereth}, {MacLeod}, {Majewski}, {Manchado}, {Maraston}, {Martini},
  {Masseron}, {Masters}, {Mathur}, {McDermid}, {Merloni}, {Merrifield},
  {M{\'e}sz{\'a}ros}, {Miglio}, {Minniti}, {Minsley}, {Miyaji}, {Mohammad},
  {Mosser}, {Mueller}, {Muna}, {Mu{\~n}oz-Guti{\'e}rrez}, {Myers}, {Nadathur},
  {Nair}, {Nandra}, {do Nascimento}, {Nevin}, {Newman}, {Nidever}, {Nitschelm},
  {Noterdaeme}, {O'Connell}, {Olmstead}, {Oravetz}, {Oravetz}, {Osorio},
  {Pace}, {Padilla}, {Palanque-Delabrouille}, {Palicio}, {Pan}, {Pan},
  {Parker}, {Paviot}, {Peirani}, {Pe{\~n}a Ram{\'r}ez}, {Penny}, {Percival},
  {Perez-Fournon}, {P{\'e}rez-R{\`a}fols}, {Petitjean}, {Pieri},
  {Pinsonneault}, {Poovelil}, {Povick}, {Prakash}, {Price-Whelan}, {Raddick},
  {Raichoor}, {Ray}, {Rembold}, {Rezaie}, {Riffel}, {Riffel}, {Rix}, {Robin},
  {Roman-Lopes}, {Rom{\'a}n-Z{\'u}{\~n}iga}, {Rose}, {Ross}, {Rossi}, {Rowland
  s}, {Rubin}, {Salvato}, {S{\'a}nchez}, {S{\'a}nchez-Menguiano},
  {S{\'a}nchez-Gallego}, {Sayres}, {Schaefer}, {Schiavon}, {Schimoia},
  {Schlafly}, {Schlegel}, {Schneider}, {Schultheis}, {Schwope}, {Seo},
  {Serenelli}, {Shafieloo}, {Shamsi}, {Shao}, {Shen}, {Shetrone}, {Shirley},
  {Silva Aguirre}, {Simon}, {Skrutskie}, {Slosar}, {Smethurst}, {Sobeck},
  {Sodi}, {Souto}, {Stark}, {Stassun}, {Steinmetz}, {Stello}, {Stermer},
  {Storchi-Bergmann}, {Streblyanska}, {Stringfellow}, {Stutz}, {Su{\'a}rez},
  {Sun}, {Taghizadeh-Popp}, {Talbot}, {Tayar}, {Thakar}, {Theriault}, {Thomas},
  {Thomas}, {Tinker}, {Tojeiro}, {Toledo}, {Tremonti}, {Troup}, {Tuttle},
  {Unda-Sanzana}, {Valentini}, {Vargas-Gonz{\'a}lez}, {Vargas-Maga{\~n}a},
  {V{\'a}zquez-Mata}, {Vivek}, {Wake}, {Wang}, {Weaver}, {Weijmans}, {Wild},
  {Wilson}, {Wilson}, {Wolthuis}, {Wood-Vasey}, {Yan}, {Yang}, {Y{\`e}che},
  {Zamora}, {Zarrouk}, {Zasowski}, {Zhang}, {Zhao}, {Zhao}, {Zheng}, {Zheng},
  {Zhu}, \& {Zou}}]{2020ApJS..249....3A}
{Ahumada}, R., {Allende Prieto}, C., {Almeida}, A., {et~al.} 2020, \apjs, 249,
  3

\bibitem[{{Allende Prieto} {et~al.}(2004){Allende Prieto}, {Asplund}, \&
  {Fabiani Bendicho}}]{2004A&amp;A...423.1109A}
{Allende Prieto}, C., {Asplund}, M., \& {Fabiani Bendicho}, P. 2004, \aap, 423,
  1109

\bibitem[{{Amarsi}(2020)}]{grid_nlte}
{Amarsi}, A.~M. 2020, Grid/NLTE [Data set], Zenodo,
  \url{http://doi.org/10.5281/zenodo.3888393}

\bibitem[{{Amarsi} \& {Asplund}(2017)}]{2017MNRAS.464..264A}
{Amarsi}, A.~M. \& {Asplund}, M. 2017, \mnras, 464, 264

\bibitem[{{Amarsi} {et~al.}(2016{\natexlab{a}}){Amarsi}, {Asplund}, {Collet},
  \& {Leenaarts}}]{2016MNRAS.455.3735A}
{Amarsi}, A.~M., {Asplund}, M., {Collet}, R., \& {Leenaarts}, J.
  2016{\natexlab{a}}, \mnras, 455, 3735

\bibitem[{{Amarsi} {et~al.}(2018{\natexlab{a}}){Amarsi}, {Barklem}, {Asplund},
  {Collet}, \& {Zatsarinny}}]{2018A&A...616A..89A}
{Amarsi}, A.~M., {Barklem}, P.~S., {Asplund}, M., {Collet}, R., \&
  {Zatsarinny}, O. 2018{\natexlab{a}}, \aap, 616, A89

\bibitem[{{Amarsi} {et~al.}(2019{\natexlab{a}}){Amarsi}, {Barklem}, {Collet},
  {Grevesse}, \& {Asplund}}]{2019A&A...624A.111A}
{Amarsi}, A.~M., {Barklem}, P.~S., {Collet}, R., {Grevesse}, N., \& {Asplund},
  M. 2019{\natexlab{a}}, \aap, 624, A111

\bibitem[{{Amarsi} {et~al.}(2020){Amarsi}, {Grevesse}, {Grumer}, {Asplund},
  {Barklem}, \& {Collet}}]{2020A&A...636A.120A}
{Amarsi}, A.~M., {Grevesse}, N., {Grumer}, J., {et~al.} 2020, \aap, 636, A120

\bibitem[{{Amarsi} {et~al.}(2016{\natexlab{b}}){Amarsi}, {Lind}, {Asplund},
  {Barklem}, \& {Collet}}]{2016MNRAS.463.1518A}
{Amarsi}, A.~M., {Lind}, K., {Asplund}, M., {Barklem}, P.~S., \& {Collet}, R.
  2016{\natexlab{b}}, \mnras, 463, 1518

\bibitem[{{Amarsi} {et~al.}(2019{\natexlab{b}}){Amarsi}, {Nissen}, \&
  {Sk{\'u}lad{\'o}ttir}}]{2019A&A...630A.104A}
{Amarsi}, A.~M., {Nissen}, P.~E., \& {Sk{\'u}lad{\'o}ttir}, {\'A}.
  2019{\natexlab{b}}, \aap, 630, A104

\bibitem[{{Amarsi} {et~al.}(2018{\natexlab{b}}){Amarsi}, {Nordlander},
  {Barklem}, {Asplund}, {Collet}, \& {Lind}}]{2018A&A...615A.139A}
{Amarsi}, A.~M., {Nordlander}, T., {Barklem}, P.~S., {et~al.}
  2018{\natexlab{b}}, \aap, 615, A139

\bibitem[{{Andrievsky} {et~al.}(2008){Andrievsky}, {Spite}, {Korotin}, {Spite},
  {Bonifacio}, {Cayrel}, {Hill}, \& {Fran{\c{c}}ois}}]{2008A&A...481..481A}
{Andrievsky}, S.~M., {Spite}, M., {Korotin}, S.~A., {et~al.} 2008, \aap, 481,
  481

\bibitem[{{Asplund}(2005)}]{2005ARA&amp;A..43..481A}
{Asplund}, M. 2005, \araa, 43, 481

\bibitem[{{Asplund} {et~al.}(2009){Asplund}, {Grevesse}, {Sauval}, \&
  {Scott}}]{2009ARA&amp;A..47..481A}
{Asplund}, M., {Grevesse}, N., {Sauval}, A.~J., \& {Scott}, P. 2009, \araa, 47,
  481

\bibitem[{{Barklem}(2016{\natexlab{a}})}]{2016A&amp;ARv..24....9B}
{Barklem}, P.~S. 2016{\natexlab{a}}, \aapr, 24, 9

\bibitem[{{Barklem}(2016{\natexlab{b}})}]{2016PhRvA..93d2705B}
{Barklem}, P.~S. 2016{\natexlab{b}}, \pra, 93, 042705

\bibitem[{{Barklem} {et~al.}(2003){Barklem}, {Belyaev}, \&
  {Asplund}}]{2003A&amp;A...409L...1B}
{Barklem}, P.~S., {Belyaev}, A.~K., \& {Asplund}, M. 2003, \aap, 409, L1

\bibitem[{{Barklem} {et~al.}(2010){Barklem}, {Belyaev}, {Dickinson}, \&
  {Gad{\'e}a}}]{2010A&amp;A...519A..20B}
{Barklem}, P.~S., {Belyaev}, A.~K., {Dickinson}, A.~S., \& {Gad{\'e}a}, F.~X.
  2010, \aap, 519, A20

\bibitem[{{Barklem} {et~al.}(2011){Barklem}, {Belyaev}, {Guitou}, {Feautrier},
  {Gad{\'e}a}, \& {Spielfiedel}}]{2011A&amp;A...530A..94B}
{Barklem}, P.~S., {Belyaev}, A.~K., {Guitou}, M., {et~al.} 2011, \aap, 530, A94

\bibitem[{{Barklem} {et~al.}(2012){Barklem}, {Belyaev}, {Spielfiedel},
  {Guitou}, \& {Feautrier}}]{2012A&amp;A...541A..80B}
{Barklem}, P.~S., {Belyaev}, A.~K., {Spielfiedel}, A., {Guitou}, M., \&
  {Feautrier}, N. 2012, \aap, 541, A80

\bibitem[{{Barklem} \& {Collet}(2016)}]{2016A&amp;A...588A..96B}
{Barklem}, P.~S. \& {Collet}, R. 2016, \aap, 588, A96

\bibitem[{{Bedell} {et~al.}(2018){Bedell}, {Bean}, {Mel{\'e}ndez}, {Spina},
  {Ram{\'{\i}}rez}, {Asplund}, {Alves-Brito}, {dos Santos}, {Dreizler}, {Yong},
  {Monroe}, \& {Casagrande}}]{2018ApJ...865...68B}
{Bedell}, M., {Bean}, J.~L., {Mel{\'e}ndez}, J., {et~al.} 2018, \apj, 865, 68

\bibitem[{{Belyaev}(2013)}]{2013PhRvA..88e2704B}
{Belyaev}, A.~K. 2013, \pra, 88, 052704

\bibitem[{{Belyaev} \& {Barklem}(2003)}]{2003PhRvA..68f2703B}
{Belyaev}, A.~K. \& {Barklem}, P.~S. 2003, \pra, 68, 062703

\bibitem[{{Belyaev} {et~al.}(2010){Belyaev}, {Barklem}, {Dickinson}, \&
  {Gad{\'e}a}}]{2010PhRvA..81c2706B}
{Belyaev}, A.~K., {Barklem}, P.~S., {Dickinson}, A.~S., \& {Gad{\'e}a}, F.~X.
  2010, \pra, 81, 032706

\bibitem[{{Belyaev} {et~al.}(2012){Belyaev}, {Barklem}, {Spielfiedel},
  {Guitou}, {Feautrier}, {Rodionov}, \& {Vlasov}}]{2012PhRvA..85c2704B}
{Belyaev}, A.~K., {Barklem}, P.~S., {Spielfiedel}, A., {et~al.} 2012, \pra, 85,
  032704

\bibitem[{{Belyaev} \& {Yakovleva}(2017)}]{2017A&A...606A.147B}
{Belyaev}, A.~K. \& {Yakovleva}, S.~A. 2017, \aap, 606, A147

\bibitem[{{Bensby} {et~al.}(2014){Bensby}, {Feltzing}, \&
  {Oey}}]{2014A&amp;A...562A..71B}
{Bensby}, T., {Feltzing}, S., \& {Oey}, M.~S. 2014, \aap, 562, A71

\bibitem[{{Bensby} \& {Lind}(2018)}]{2018A&A...615A.151B}
{Bensby}, T. \& {Lind}, K. 2018, \aap, 615, A151

\bibitem[{{Bergemann} {et~al.}(2017{\natexlab{a}}){Bergemann}, {Collet},
  {Amarsi}, {Kovalev}, {Ruchti}, \& {Magic}}]{2017ApJ...847...15B}
{Bergemann}, M., {Collet}, R., {Amarsi}, A.~M., {et~al.} 2017{\natexlab{a}},
  \apj, 847, 15

\bibitem[{{Bergemann} {et~al.}(2017{\natexlab{b}}){Bergemann}, {Collet},
  {Sch{\"o}nrich}, {Andrae}, {Kovalev}, {Ruchti}, {Hansen}, \&
  {Magic}}]{2017ApJ...847...16B}
{Bergemann}, M., {Collet}, R., {Sch{\"o}nrich}, R., {et~al.}
  2017{\natexlab{b}}, \apj, 847, 16

\bibitem[{{Bergemann} {et~al.}(2019){Bergemann}, {Gallagher}, {Eitner},
  {Bautista}, {Collet}, {Yakovleva}, {Mayriedl}, {Plez}, {Carlsson},
  {Leenaarts}, {Belyaev}, \& {Hansen}}]{2019A&A...631A..80B}
{Bergemann}, M., {Gallagher}, A.~J., {Eitner}, P., {et~al.} 2019, \aap, 631,
  A80

\bibitem[{{Bergemann} {et~al.}(2012){Bergemann}, {Hansen}, {Bautista}, \&
  {Ruchti}}]{2012A&A...546A..90B}
{Bergemann}, M., {Hansen}, C.~J., {Bautista}, M., \& {Ruchti}, G. 2012, \aap,
  546, A90

\bibitem[{{Bonifacio} {et~al.}(2009){Bonifacio}, {Spite}, {Cayrel}, {Hill},
  {Spite}, {Fran{\c{c}}ois}, {Plez}, {Ludwig}, {Caffau}, {Molaro}, {Depagne},
  {Andersen}, {Barbuy}, {Beers}, {Nordstr{\"o}m}, \&
  {Primas}}]{2009A&A...501..519B}
{Bonifacio}, P., {Spite}, M., {Cayrel}, R., {et~al.} 2009, \aap, 501, 519

\bibitem[{{Buder} {et~al.}(2018){Buder}, {Asplund}, {Duong}, {Kos}, {Lind},
  {Ness}, {Sharma}, {Bland-Hawthorn}, {Casey}, {De Silva}, {D'Orazi},
  {Freeman}, {Lewis}, {Lin}, {Martell}, {Schlesinger}, {Simpson}, {Zucker},
  {Zwitter}, {Amarsi}, {Anguiano}, {Carollo}, {Casagrande}, {{\v C}otar},
  {Cottrell}, {Da Costa}, {Gao}, {Hayden}, {Horner}, {Ireland}, {Kafle},
  {Munari}, {Nataf}, {Nordlander}, {Stello}, {Ting}, {Traven}, {Watson},
  {Wittenmyer}, {Wyse}, {Yong}, {Zinn}, \& {{\v Z}erjal}}]{2018MNRAS.478.4513B}
{Buder}, S., {Asplund}, M., {Duong}, L., {et~al.} 2018, \mnras, 478, 4513

\bibitem[{{Chang} \& {Tang}(1990)}]{1990JQSRT..43..207C}
{Chang}, T.~N. \& {Tang}, X. 1990, \jqsrt, 43, 207

\bibitem[{{Collet} {et~al.}(2007){Collet}, {Asplund}, \&
  {Trampedach}}]{2007A&amp;A...469..687C}
{Collet}, R., {Asplund}, M., \& {Trampedach}, R. 2007, \aap, 469, 687

\bibitem[{{Dalton} {et~al.}(2016){Dalton}, {Trager}, {Abrams}, {Bonifacio},
  {Aguerri}, {Middleton}, {Benn}, {Dee}, {Say{\`e}de}, {Lewis}, {Pragt},
  {Pico}, {Walton}, {Rey}, {Allende Prieto}, {Pe{\~n}ate}, {Lhome},
  {Ag{\'o}cs}, {Alonso}, {Terrett}, {Brock}, {Gilbert}, {Schallig}, {Ridings},
  {Guinouard}, {Verheijen}, {Tosh}, {Rogers}, {Lee}, {Steele}, {Stuik},
  {Tromp}, {Jask{\'o}}, {Carrasco}, {Farcas}, {Kragt}, {Lesman}, {Kroes},
  {Mottram}, {Bates}, {Rodriguez}, {Gribbin}, {Delgado}, {Herreros}, {Martin},
  {Cano}, {Navarro}, {Irwin}, {Lewis}, {Gonzalez Solares}, {Murphy}, {Worley},
  {Bassom}, {O'Mahoney}, {Bianco}, {Zurita}, {ter Horst}, {Molinari}, {Lodi},
  {Guerra}, {Martin}, {Vallenari}, {Salasnich}, {Baruffolo}, {Jin}, {Hill},
  {Smith}, {Drew}, {Poggianti}, {Pieri}, {Dominquez Palmero}, \&
  {Farina}}]{2016SPIE.9908E..1GD}
{Dalton}, G., {Trager}, S., {Abrams}, D.~C., {et~al.} 2016, in \procspie, Vol.
  9908, Ground-based and Airborne Instrumentation for Astronomy VI, 99081G

\bibitem[{{Davidson} {et~al.}(1992){Davidson}, {Snoek}, {Volten}, \&
  {Doenszelmann}}]{1992A&A...255..457D}
{Davidson}, M.~D., {Snoek}, L.~C., {Volten}, H., \& {Doenszelmann}, A. 1992,
  \aap, 255, 457

\bibitem[{{de Jong} {et~al.}(2019){de Jong}, {Agertz}, {Berbel}, {Aird},
  {Alexander}, {Amarsi}, {Anders}, {Andrae}, {Ansarinejad}, {Ansorge}, \&
  et~al.}]{2019Msngr.175....3D}
{de Jong}, R.~S., {Agertz}, O., {Berbel}, A.~A., {et~al.} 2019, The Messenger,
  175, 3

\bibitem[{{de los Reyes} {et~al.}(2020){de los Reyes}, {Kirby}, {Seitenzahl},
  \& {Shen}}]{2020ApJ...891...85D}
{de los Reyes}, M. A.~C., {Kirby}, E.~N., {Seitenzahl}, I.~R., \& {Shen}, K.~J.
  2020, \apj, 891, 85

\bibitem[{{De Silva} {et~al.}(2015){De Silva}, {Freeman}, {Bland-Hawthorn},
  {Martell}, {de Boer}, {Asplund}, {Keller}, {Sharma}, {Zucker}, {Zwitter},
  {Anguiano}, {Bacigalupo}, {Bayliss}, {Beavis}, {Bergemann}, {Campbell},
  {Cannon}, {Carollo}, {Casagrande}, {Casey}, {Da Costa}, {D'Orazi}, {Dotter},
  {Duong}, {Heger}, {Ireland}, {Kafle}, {Kos}, {Lattanzio}, {Lewis}, {Lin},
  {Lind}, {Munari}, {Nataf}, {O'Toole}, {Parker}, {Reid}, {Schlesinger},
  {Sheinis}, {Simpson}, {Stello}, {Ting}, {Traven}, {Watson}, {Wittenmyer},
  {Yong}, \& {{\v Z}erjal}}]{2015MNRAS.449.2604D}
{De Silva}, G.~M., {Freeman}, K.~C., {Bland-Hawthorn}, J., {et~al.} 2015,
  \mnras, 449, 2604

\bibitem[{{deBoer} {et~al.}(2017){deBoer}, {G{\"o}rres}, {Wiescher}, {Azuma},
  {Best}, {Brune}, {Fields}, {Jones}, {Pignatari}, {Sayre}, {Smith}, {Timmes},
  \& {Uberseder}}]{2017RvMP...89c5007D}
{deBoer}, R.~J., {G{\"o}rres}, J., {Wiescher}, M., {et~al.} 2017, Reviews of
  Modern Physics, 89, 035007

\bibitem[{{Delgado Mena} {et~al.}(2015){Delgado Mena}, {Bertr{\'a}n de Lis},
  {Adibekyan}, {Sousa}, {Figueira}, {Mortier}, {Gonz{\'a}lez Hern{\'a}ndez},
  {Tsantaki}, {Israelian}, \& {Santos}}]{2015A&A...576A..69D}
{Delgado Mena}, E., {Bertr{\'a}n de Lis}, S., {Adibekyan}, V.~Z., {et~al.}
  2015, \aap, 576, A69

\bibitem[{{Delgado Mena} {et~al.}(2017){Delgado Mena}, {Tsantaki}, {Adibekyan},
  {Sousa}, {Santos}, {Gonz{\'a}lez Hern{\'a}ndez}, \&
  {Israelian}}]{2017A&A...606A..94D}
{Delgado Mena}, E., {Tsantaki}, M., {Adibekyan}, V.~Z., {et~al.} 2017, \aap,
  606, A94

\bibitem[{{Den Hartog} {et~al.}(2011){Den Hartog}, {Lawler}, {Sobeck},
  {Sneden}, \& {Cowan}}]{2011ApJS..194...35D}
{Den Hartog}, E.~A., {Lawler}, J.~E., {Sobeck}, J.~S., {Sneden}, C., \&
  {Cowan}, J.~J. 2011, \apjs, 194, 35

\bibitem[{{Dobrovolskas} {et~al.}(2012){Dobrovolskas}, {Ku{\v{c}}inskas},
  {Andrievsky}, {Korotin}, {Mishenina}, {Bonifacio}, {Ludwig}, \&
  {Caffau}}]{2012A&A...540A.128D}
{Dobrovolskas}, V., {Ku{\v{c}}inskas}, A., {Andrievsky}, S.~M., {et~al.} 2012,
  \aap, 540, A128

\bibitem[{{Drawin}(1968)}]{1968ZPhy..211..404D}
{Drawin}, H.-W. 1968, \zphys, 211, 404

\bibitem[{{Drawin}(1969)}]{1969ZPhy..225..483D}
{Drawin}, H.~W. 1969, \zphys, 225, 483

\bibitem[{{Eitner} {et~al.}(2020){Eitner}, {Bergemann}, {Hansen}, {Cescutti},
  {Seitenzahl}, {Larsen}, \& {Plez}}]{2020A&A...635A..38E}
{Eitner}, P., {Bergemann}, M., {Hansen}, C.~J., {et~al.} 2020, \aap, 635, A38

\bibitem[{{Feuillet} {et~al.}(2019){Feuillet}, {Frankel}, {Lind}, {Frinchaboy},
  {Garc{\'\i}a-Hern{\'a}ndez}, {Lane}, {Nitschelm}, \&
  {Roman-Lopes}}]{2019MNRAS.489.1742F}
{Feuillet}, D.~K., {Frankel}, N., {Lind}, K., {et~al.} 2019, \mnras, 489, 1742

\bibitem[{{Fields}(2011)}]{2011ARNPS..61...47F}
{Fields}, B.~D. 2011, Annual Review of Nuclear and Particle Science, 61, 47

\bibitem[{{Fields} {et~al.}(2018){Fields}, {Timmes}, {Farmer}, {Petermann},
  {Wolf}, \& {Couch}}]{2018ApJS..234...19F}
{Fields}, C.~E., {Timmes}, F.~X., {Farmer}, R., {et~al.} 2018, \apjs, 234, 19

\bibitem[{{Froese Fischer} {et~al.}(2006){Froese Fischer}, {Tachiev}, \&
  {Irimia}}]{2006ADNDT..92..607F}
{Froese Fischer}, C., {Tachiev}, G., \& {Irimia}, A. 2006, \adndt, 92, 607

\bibitem[{{Gaia Collaboration} {et~al.}(2018){Gaia Collaboration}, {Brown},
  {Vallenari}, {Prusti}, {de Bruijne}, {Babusiaux}, {Bailer-Jones}, {Biermann},
  {Evans}, {Eyer}, \& et~al.}]{2018A&A...616A...1G}
{Gaia Collaboration}, {Brown}, A.~G.~A., {Vallenari}, A., {et~al.} 2018, \aap,
  616, A1

\bibitem[{{Gallagher} {et~al.}(2020){Gallagher}, {Bergemann}, {Collet}, {Plez},
  {Leenaarts}, {Carlsson}, {Yakovleva}, \& {Belyaev}}]{2020A&A...634A..55G}
{Gallagher}, A.~J., {Bergemann}, M., {Collet}, R., {et~al.} 2020, \aap, 634,
  A55

\bibitem[{{Gao} {et~al.}(2020){Gao}, {Lind}, {Amarsi}, {Buder},
  {Bland-Hawthorn}, {Campbell}, {Asplund}, {Casey}, {de Silva}, {Freeman},
  {Hayden}, {Lewis}, {Martell}, {Simpson}, {Sharma}, {Zucker}, {Zwitter},
  {Horner}, {Munari}, {Nordlander}, {Stello}, {Ting}, {Traven}, {Wittenmyer},
  \& {GALAH Collaboration}}]{2020MNRAS.497L..30G}
{Gao}, X., {Lind}, K., {Amarsi}, A.~M., {et~al.} 2020, \mnras, 497, L30

\bibitem[{{Garc{\'\i}a P{\'e}rez} {et~al.}(2016){Garc{\'\i}a P{\'e}rez},
  {Allende Prieto}, {Holtzman}, {Shetrone}, {M{\'e}sz{\'a}ros}, {Bizyaev},
  {Carrera}, {Cunha}, {Garc{\'\i}a-Hern{\'a}ndez}, {Johnson}, {Majewski},
  {Nidever}, {Schiavon}, {Shane}, {Smith}, {Sobeck}, {Troup}, {Zamora},
  {Weinberg}, {Bovy}, {Eisenstein}, {Feuillet}, {Frinchaboy}, {Hayden},
  {Hearty}, {Nguyen}, {O'Connell}, {Pinsonneault}, {Wilson}, \&
  {Zasowski}}]{2016AJ....151..144G}
{Garc{\'\i}a P{\'e}rez}, A.~E., {Allende Prieto}, C., {Holtzman}, J.~A.,
  {et~al.} 2016, \aj, 151, 144

\bibitem[{{Garz}(1973)}]{1973A&A....26..471G}
{Garz}, T. 1973, \aap, 26, 471

\bibitem[{{Grevesse} {et~al.}(2007){Grevesse}, {Asplund}, \&
  {Sauval}}]{2007coma.book..105G}
{Grevesse}, N., {Asplund}, M., \& {Sauval}, A.~J. 2007, {The Solar Chemical
  Composition}, ed. R.~{von Steiger}, G.~{Gloeckler}, \& G.~M. {Mason}
  (Springer Science+Business Media), 105

\bibitem[{{Guiglion} {et~al.}(2019){Guiglion}, {Chiappini}, {Romano},
  {Matteucci}, {Anders}, {Steinmetz}, {Minchev}, {de Laverny}, \&
  {Recio-Blanco}}]{2019A&A...623A..99G}
{Guiglion}, G., {Chiappini}, C., {Romano}, D., {et~al.} 2019, \aap, 623, A99

\bibitem[{{Gustafsson} {et~al.}(2008){Gustafsson}, {Edvardsson}, {Eriksson},
  {J{\o}rgensen}, {Nordlund}, \& {Plez}}]{2008A&amp;A...486..951G}
{Gustafsson}, B., {Edvardsson}, B., {Eriksson}, K., {et~al.} 2008, \aap, 486,
  951

\bibitem[{{Hampel} {et~al.}(2019){Hampel}, {Karakas}, {Stancliffe}, {Meyer}, \&
  {Lugaro}}]{2019ApJ...887...11H}
{Hampel}, M., {Karakas}, A.~I., {Stancliffe}, R.~J., {Meyer}, B.~S., \&
  {Lugaro}, M. 2019, \apj, 887, 11

\bibitem[{{Hampel} {et~al.}(2016){Hampel}, {Stancliffe}, {Lugaro}, \&
  {Meyer}}]{2016ApJ...831..171H}
{Hampel}, M., {Stancliffe}, R.~J., {Lugaro}, M., \& {Meyer}, B.~S. 2016, \apj,
  831, 171

\bibitem[{{Hayden} {et~al.}(2015){Hayden}, {Bovy}, {Holtzman}, {Nidever},
  {Bird}, {Weinberg}, {Andrews}, {Majewski}, {Allende Prieto}, {Anders},
  {Beers}, {Bizyaev}, {Chiappini}, {Cunha}, {Frinchaboy},
  {Garc{\'{\i}}a-Her{\'n}andez}, {Garc{\'{\i}}a P{\'e}rez}, {Girardi},
  {Harding}, {Hearty}, {Johnson}, {M{\'e}sz{\'a}ros}, {Minchev}, {O'Connell},
  {Pan}, {Robin}, {Schiavon}, {Schneider}, {Schultheis}, {Shetrone},
  {Skrutskie}, {Steinmetz}, {Smith}, {Wilson}, {Zamora}, \&
  {Zasowski}}]{2015ApJ...808..132H}
{Hayden}, M.~R., {Bovy}, J., {Holtzman}, J.~A., {et~al.} 2015, \apj, 808, 132

\bibitem[{{Hayes} {et~al.}(2018){Hayes}, {Majewski}, {Shetrone},
  {Fern{\'a}ndez-Alvar}, {Allende Prieto}, {Schuster}, {Carigi}, {Cunha},
  {Smith}, {Sobeck}, {Almeida}, {Beers}, {Carrera}, {Fern{\'a}ndez-Trincado},
  {Garc{\'{\i}}a-Hern{\'a}ndez}, {Geisler}, {Lane}, {Lucatello}, {Matthews},
  {Minniti}, {Nitschelm}, {Tang}, {Tissera}, \& {Zamora}}]{2018ApJ...852...49H}
{Hayes}, C.~R., {Majewski}, S.~R., {Shetrone}, M., {et~al.} 2018, \apj, 852, 49

\bibitem[{{Hibbert} {et~al.}(1991){Hibbert}, {Biemont}, {Godefroid}, \&
  {Vaeck}}]{1991JPhB...24.3943H}
{Hibbert}, A., {Biemont}, E., {Godefroid}, M., \& {Vaeck}, N. 1991, \jpb, 24,
  3943

\bibitem[{{Hibbert} {et~al.}(1993){Hibbert}, {Biemont}, {Godefroid}, \&
  {Vaeck}}]{1993A&AS...99..179H}
{Hibbert}, A., {Biemont}, E., {Godefroid}, M., \& {Vaeck}, N. 1993, \aaps, 99,
  179

\bibitem[{{Hubeny} \& {Mihalas}(2014)}]{2014tsa..book.....H}
{Hubeny}, I. \& {Mihalas}, D. 2014, {Theory of Stellar Atmospheres} (Princeton
  Univ. Press, Princeton, NJ)

\bibitem[{{Ibgui} {et~al.}(2013){Ibgui}, {Hubeny}, {Lanz}, \&
  {Stehl{\'e}}}]{2013A&amp;A...549A.126I}
{Ibgui}, L., {Hubeny}, I., {Lanz}, T., \& {Stehl{\'e}}, C. 2013, \aap, 549,
  A126

\bibitem[{{Kao} \& {Auer}(1990)}]{1990MWRv..118.1551K}
{Kao}, C.-Y.~J. \& {Auer}, L.~H. 1990, Monthly Weather Review, 118, 1551

\bibitem[{{Karakas} \& {Lattanzio}(2014)}]{2014PASA...31...30K}
{Karakas}, A.~I. \& {Lattanzio}, J.~C. 2014, \pasa, 31, e030

\bibitem[{{Karovicova} {et~al.}(2020){Karovicova}, {White}, {Nordlander},
  {Casagrand e}, {Ireland}, {Huber}, \& {Jofr{\'e}}}]{2020A&A...640A..25K}
{Karovicova}, I., {White}, T.~R., {Nordlander}, T., {et~al.} 2020, \aap, 640,
  A25

\bibitem[{{Karovicova} {et~al.}(2018){Karovicova}, {White}, {Nordlander},
  {Lind}, {Casagrande}, {Ireland}, {Huber}, {Creevey}, {Mourard}, {Schaefer},
  {Gilmore}, {Chiavassa}, {Wittkowski}, {Jofr{\'e}}, {Heiter}, {Th{\'e}venin},
  \& {Asplund}}]{2018MNRAS.475L..81K}
{Karovicova}, I., {White}, T.~R., {Nordlander}, T., {et~al.} 2018, \mnras, 475,
  L81

\bibitem[{{Kaulakys}(1985)}]{1985JPhB...18L.167K}
{Kaulakys}, B.~P. 1985, \jpb, 18, L167

\bibitem[{{Kaulakys}(1986)}]{kaulakys1986free}
{Kaulakys}, B.~P. 1986, JETP, 91, 391

\bibitem[{{Kaulakys}(1991)}]{1991JPhB...24L.127K}
{Kaulakys}, B.~P. 1991, \jpb, 24, L127

\bibitem[{{Kelleher} \& {Podobedova}(2008)}]{2008JPCRD..37..709K}
{Kelleher}, D.~E. \& {Podobedova}, L.~I. 2008, \jpcrd, 37, 709

\bibitem[{{Kirby} {et~al.}(2019){Kirby}, {Xie}, {Guo}, {de los Reyes},
  {Bergemann}, {Kovalev}, {Shen}, {Piro}, \& {McWilliam}}]{2019ApJ...881...45K}
{Kirby}, E.~N., {Xie}, J.~L., {Guo}, R., {et~al.} 2019, \apj, 881, 45

\bibitem[{{Kobayashi} {et~al.}(2020{\natexlab{a}}){Kobayashi}, {Karakas}, \&
  {Lugaro}}]{2020arXiv200804660K}
{Kobayashi}, C., {Karakas}, A.~I., \& {Lugaro}, M. 2020{\natexlab{a}}, arXiv
  e-prints, arXiv:2008.04660

\bibitem[{{Kobayashi} {et~al.}(2020{\natexlab{b}}){Kobayashi}, {Leung}, \&
  {Nomoto}}]{2020ApJ...895..138K}
{Kobayashi}, C., {Leung}, S.-C., \& {Nomoto}, K. 2020{\natexlab{b}}, \apj, 895,
  138

\bibitem[{{Kobayashi} {et~al.}(2006){Kobayashi}, {Umeda}, {Nomoto}, {Tominaga},
  \& {Ohkubo}}]{2006ApJ...653.1145K}
{Kobayashi}, C., {Umeda}, H., {Nomoto}, K., {Tominaga}, N., \& {Ohkubo}, T.
  2006, \apj, 653, 1145

\bibitem[{{Korotin} {et~al.}(2011){Korotin}, {Mishenina}, {Gorbaneva}, \&
  {Soubiran}}]{2011MNRAS.415.2093K}
{Korotin}, S., {Mishenina}, T., {Gorbaneva}, T., \& {Soubiran}, C. 2011,
  \mnras, 415, 2093

\bibitem[{{Korotin} {et~al.}(2015){Korotin}, {Andrievsky}, {Hansen}, {Caffau},
  {Bonifacio}, {Spite}, {Spite}, \& {Fran{\c{c}}ois}}]{2015A&A...581A..70K}
{Korotin}, S.~A., {Andrievsky}, S.~M., {Hansen}, C.~J., {et~al.} 2015, \aap,
  581, A70

\bibitem[{{Kos} {et~al.}(2017){Kos}, {Lin}, {Zwitter}, {{\v{Z}}erjal},
  {Sharma}, {Bland -Hawthorn}, {Asplund}, {Casey}, {De Silva}, {Freeman},
  {Martell}, {Simpson}, {Schlesinger}, {Zucker}, {Anguiano}, {Bacigalupo},
  {Bedding}, {Betters}, {Da Costa}, {Duong}, {Hyde}, {Ireland}, {Kafle},
  {Leon-Saval}, {Lewis}, {Munari}, {Nataf}, {Stello}, {Tinney}, {Traven},
  {Watson}, \& {Wittenmyer}}]{2017MNRAS.464.1259K}
{Kos}, J., {Lin}, J., {Zwitter}, T., {et~al.} 2017, \mnras, 464, 1259

\bibitem[{Kramida {et~al.}(2012)Kramida, Ralchenko, Reader,
  {et~al.}}]{kramida2012nist}
Kramida, A., Ralchenko, Y., Reader, J., {et~al.} 2012, NIST atomic spectra
  database (version 5)

\bibitem[{{Kroupa}(2008)}]{2008ASPC..390....3K}
{Kroupa}, P. 2008, in Astronomical Society of the Pacific Conference Series,
  Vol. 390, Pathways Through an Eclectic Universe, ed. J.~H. {Knapen}, T.~J.
  {Mahoney}, \& A.~{Vazdekis}, 3

\bibitem[{{Lambert}(1993)}]{1993PhST...47..186L}
{Lambert}, D.~L. 1993, Physica Scripta Volume T, 47, 186

\bibitem[{{Lee} \& {Kim}(2004)}]{2004MNRAS.347..802L}
{Lee}, H.-W. \& {Kim}, H.~I. 2004, \mnras, 347, 802

\bibitem[{{Leenaarts} \& {Carlsson}(2009)}]{2009ASPC..415...87L_short}
{Leenaarts}, J. \& {Carlsson}, M. 2009, in Astronomical Society of the Pacific
  Conference Series, Vol. 415, The Second Hinode Science Meeting, ed.
  B.~{Lites}, M.~{Cheung}, T.~{Magara}, J.~{Mariska}, \& K.~{Reeves}, 87

\bibitem[{{Lind} {et~al.}(2017){Lind}, {Amarsi}, {Asplund}, {Barklem},
  {Bautista}, {Bergemann}, {Collet}, {Kiselman}, {Leenaarts}, \&
  {Pereira}}]{2017MNRAS.468.4311L}
{Lind}, K., {Amarsi}, A.~M., {Asplund}, M., {et~al.} 2017, \mnras, 468, 4311

\bibitem[{{Lind} {et~al.}(2011){Lind}, {Asplund}, {Barklem}, \&
  {Belyaev}}]{2011A&amp;A...528A.103L}
{Lind}, K., {Asplund}, M., {Barklem}, P.~S., \& {Belyaev}, A.~K. 2011, \aap,
  528, A103

\bibitem[{{Lind} {et~al.}(2012){Lind}, {Bergemann}, \&
  {Asplund}}]{2012MNRAS.427...50L}
{Lind}, K., {Bergemann}, M., \& {Asplund}, M. 2012, \mnras, 427, 50

\bibitem[{{Lind} {et~al.}(2013){Lind}, {Melendez}, {Asplund}, {Collet}, \&
  {Magic}}]{2013A&amp;A...554A..96L}
{Lind}, K., {Melendez}, J., {Asplund}, M., {Collet}, R., \& {Magic}, Z. 2013,
  \aap, 554, A96

\bibitem[{{Liu} {et~al.}(2020){Liu}, {Fu}, {Shi}, {Wu}, {Han}, {Chen}, {Dong},
  {Zhao}, {Chen}, {Zhang}, {Bai}, {Chen}, {Cui}, {Du}, {Hsia}, {Jiang}, {Hou},
  {Hou}, {Li}, {Li}, {Li}, {Liu}, {Liu}, {Luo}, {Ren}, {Tian}, {Tian}, {Wang},
  {Wu}, {Xie}, {Yan}, {Yang}, {Yu}, {Zhang}, {Zhang}, {Zhang}, {Zhang}, {Zhao},
  {Zhong}, {Zong}, \& {Zuo}}]{2020arXiv200507210L}
{Liu}, C., {Fu}, J., {Shi}, J., {et~al.} 2020, arXiv e-prints, arXiv:2005.07210

\bibitem[{{Martell} {et~al.}(2020){Martell}, {Simpson}, {Balasubramaniam},
  {Buder}, {Sharma}, {Hon}, {Stello}, {Ting}, {Asplund}, {Bland -Hawthorn}, {De
  Silva}, {Freeman}, {Hayden}, {Kos}, {Lewis}, {Lind}, {Zucker}, {Zwitter},
  {Campbell}, {Cotar}, {Horner}, {Montet}, \&
  {Wittenmyer}}]{2020arXiv200602106M}
{Martell}, S., {Simpson}, J., {Balasubramaniam}, A., {et~al.} 2020, arXiv
  e-prints, arXiv:2006.02106

\bibitem[{{Mashonkina} {et~al.}(2019){Mashonkina}, {Neretina}, {Sitnova}, \&
  {Pakhomov}}]{2019ARep...63..726M}
{Mashonkina}, L.~I., {Neretina}, M.~D., {Sitnova}, T.~M., \& {Pakhomov}, Y.~V.
  2019, Astronomy Reports, 63, 726

\bibitem[{{Mashonkina} {et~al.}(2016){Mashonkina}, {Sitnova}, \&
  {Pakhomov}}]{2016AstL...42..606M}
{Mashonkina}, L.~I., {Sitnova}, T.~N., \& {Pakhomov}, Y.~V. 2016, Astronomy
  Letters, 42, 606

\bibitem[{{McWilliam}(1997)}]{1997ARA&amp;A..35..503M}
{McWilliam}, A. 1997, \araa, 35, 503

\bibitem[{{Mihalas} \& {Athay}(1973)}]{1973ARA&amp;A..11..187M}
{Mihalas}, D. \& {Athay}, R.~G. 1973, \araa, 11, 187

\bibitem[{{Mishenina} {et~al.}(2015){Mishenina}, {Gorbaneva}, {Pignatari},
  {Thielemann}, \& {Korotin}}]{2015MNRAS.454.1585M}
{Mishenina}, T., {Gorbaneva}, T., {Pignatari}, M., {Thielemann}, F.~K., \&
  {Korotin}, S.~A. 2015, \mnras, 454, 1585

\bibitem[{{Mishenina} {et~al.}(2013){Mishenina}, {Pignatari}, {Korotin},
  {Soubiran}, {Charbonnel}, {Thielemann}, {Gorbaneva}, \&
  {Basak}}]{2013A&A...552A.128M}
{Mishenina}, T.~V., {Pignatari}, M., {Korotin}, S.~A., {et~al.} 2013, \aap,
  552, A128

\bibitem[{{M{\"u}ller}(2016)}]{2016PASA...33...48M}
{M{\"u}ller}, B. 2016, \pasa, 33, e048

\bibitem[{{Ness} {et~al.}(2015){Ness}, {Hogg}, {Rix}, {Ho}, \&
  {Zasowski}}]{2015ApJ...808...16N}
{Ness}, M., {Hogg}, D.~W., {Rix}, H.-W., {Ho}, A.~Y.~Q., \& {Zasowski}, G.
  2015, \apj, 808, 16

\bibitem[{{Nissen}(2015)}]{2015A&amp;A...579A..52N}
{Nissen}, P.~E. 2015, \aap, 579, A52

\bibitem[{{Nissen}(2016)}]{2016A&A...593A..65N}
{Nissen}, P.~E. 2016, \aap, 593, A65

\bibitem[{{Nissen} \& {Gustafsson}(2018)}]{2018A&ARv..26....6N}
{Nissen}, P.~E. \& {Gustafsson}, B. 2018, \aapr, 26, 6

\bibitem[{{Nissen} \& {Schuster}(2010)}]{2010A&amp;A...511L..10N}
{Nissen}, P.~E. \& {Schuster}, W.~J. 2010, \aap, 511, L10

\bibitem[{{Nissen} \& {Schuster}(2011)}]{2011A&A...530A..15N}
{Nissen}, P.~E. \& {Schuster}, W.~J. 2011, \aap, 530, A15

\bibitem[{{Nordlander} \& {Lind}(2017)}]{2017A&amp;A...607A..75N}
{Nordlander}, T. \& {Lind}, K. 2017, \aap, 607, A75

\bibitem[{{O'Brian} \& {Lawler}(1991)}]{1991PhRvA..44.7134O}
{O'Brian}, T.~R. \& {Lawler}, J.~E. 1991, \pra, 44, 7134

\bibitem[{{Osorio} {et~al.}(2020){Osorio}, {Allende Prieto}, {Hubeny},
  {M{\'e}sz{\'a}ros}, \& {Shetrone}}]{2020A&A...637A..80O}
{Osorio}, Y., {Allende Prieto}, C., {Hubeny}, I., {M{\'e}sz{\'a}ros}, S., \&
  {Shetrone}, M. 2020, \aap, 637, A80

\bibitem[{{Osorio} \& {Barklem}(2016)}]{2016A&A...586A.120O}
{Osorio}, Y. \& {Barklem}, P.~S. 2016, \aap, 586, A120

\bibitem[{{Osorio} {et~al.}(2015){Osorio}, {Barklem}, {Lind}, {Belyaev},
  {Spielfiedel}, {Guitou}, \& {Feautrier}}]{2015A&amp;A...579A..53O}
{Osorio}, Y., {Barklem}, P.~S., {Lind}, K., {et~al.} 2015, \aap, 579, A53

\bibitem[{{Osorio} {et~al.}(2019){Osorio}, {Lind}, {Barklem}, {Allende Prieto},
  \& {Zatsarinny}}]{2019A&A...623A.103O}
{Osorio}, Y., {Lind}, K., {Barklem}, P.~S., {Allende Prieto}, C., \&
  {Zatsarinny}, O. 2019, \aap, 623, A103

\bibitem[{{Pinsonneault}(1997)}]{1997ARA&amp;A..35..557P}
{Pinsonneault}, M. 1997, \araa, 35, 557

\bibitem[{{Piskunov} \& {Valenti}(2017)}]{2017A&amp;A...597A..16P}
{Piskunov}, N. \& {Valenti}, J.~A. 2017, \aap, 597, A16

\bibitem[{{Prantzos}(2012)}]{2012A&A...542A..67P}
{Prantzos}, N. 2012, \aap, 542, A67

\bibitem[{{Ram{\'{\i}}rez} {et~al.}(2012){Ram{\'{\i}}rez}, {Mel{\'e}ndez}, \&
  {Chanam{\'e}}}]{2012ApJ...757..164R}
{Ram{\'{\i}}rez}, I., {Mel{\'e}ndez}, J., \& {Chanam{\'e}}, J. 2012, \apj, 757,
  164

\bibitem[{{Randich} {et~al.}(2013){Randich}, {Gilmore}, \& {Gaia-ESO
  Consortium}}]{2013Msngr.154...47R}
{Randich}, S., {Gilmore}, G., \& {Gaia-ESO Consortium}. 2013, The Messenger,
  154, 47

\bibitem[{{Recio-Blanco} {et~al.}(2016){Recio-Blanco}, {de Laverny}, {Allende
  Prieto}, {Fustes}, {Manteiga}, {Arcay}, {Bijaoui}, {Dafonte}, {Ordenovic}, \&
  {Ordo{\~n}ez Blanco}}]{2016A&A...585A..93R}
{Recio-Blanco}, A., {de Laverny}, P., {Allende Prieto}, C., {et~al.} 2016,
  \aap, 585, A93

\bibitem[{{Reggiani} {et~al.}(2019){Reggiani}, {Amarsi}, {Lind}, {Barklem},
  {Zatsarinny}, {Bartschat}, {Fursa}, {Bray}, {Spina}, \&
  {Mel{\'e}ndez}}]{2019A&A...627A.177R}
{Reggiani}, H., {Amarsi}, A.~M., {Lind}, K., {et~al.} 2019, \aap, 627, A177

\bibitem[{{Romano} {et~al.}(2020){Romano}, {Franchini}, {Grisoni}, {Spitoni},
  {Matteucci}, \& {Morossi}}]{2020A&A...639A..37R}
{Romano}, D., {Franchini}, M., {Grisoni}, V., {et~al.} 2020, \aap, 639, A37

\bibitem[{{Romano} {et~al.}(2019){Romano}, {Matteucci}, {Zhang}, {Ivison}, \&
  {Ventura}}]{2019MNRAS.490.2838R}
{Romano}, D., {Matteucci}, F., {Zhang}, Z.-Y., {Ivison}, R.~J., \& {Ventura},
  P. 2019, \mnras, 490, 2838

\bibitem[{{Rutten}(2003)}]{2003rtsa.book.....R}
{Rutten}, R.~J. 2003, {Radiative Transfer in Stellar Atmospheres, 8th edn.}
  (Utrecht University)

\bibitem[{{Rybicki} \& {Hummer}(1992)}]{1992A&amp;A...262..209R}
{Rybicki}, G.~B. \& {Hummer}, D.~G. 1992, \aap, 262, 209

\bibitem[{Saad(2003)}]{saad2003iterative}
Saad, Y. 2003, Iterative methods for sparse linear systems, Vol.~82 (siam)

\bibitem[{{Seitenzahl} {et~al.}(2013){Seitenzahl}, {Cescutti}, {R{\"o}pke},
  {Ruiter}, \& {Pakmor}}]{2013A&A...559L...5S}
{Seitenzahl}, I.~R., {Cescutti}, G., {R{\"o}pke}, F.~K., {Ruiter}, A.~J., \&
  {Pakmor}, R. 2013, \aap, 559, L5

\bibitem[{{Short} \& {Hauschildt}(2005)}]{2005ApJ...618..926S}
{Short}, C.~I. \& {Hauschildt}, P.~H. 2005, \apj, 618, 926

\bibitem[{{Silva Aguirre} {et~al.}(2012){Silva Aguirre}, {Casagrande}, {Basu},
  {Campante}, {Chaplin}, {Huber}, {Miglio}, {Serenelli}, {Ballot}, {Bedding},
  {Christensen-Dalsgaard}, {Creevey}, {Elsworth}, {Garc{\'\i}a}, {Gilliland},
  {Hekker}, {Kjeldsen}, {Mathur}, {Metcalfe}, {Monteiro}, {Mosser},
  {Pinsonneault}, {Stello}, {Weiss}, {Tenenbaum}, {Twicken}, \&
  {Uddin}}]{2012ApJ...757...99S}
{Silva Aguirre}, V., {Casagrande}, L., {Basu}, S., {et~al.} 2012, \apj, 757, 99

\bibitem[{{Sk{\'u}lad{\'o}ttir} {et~al.}(2020){Sk{\'u}lad{\'o}ttir}, {Hansen},
  {Choplin}, {Salvadori}, {Hampel}, \& {Campbell}}]{2020A&A...634A..84S}
{Sk{\'u}lad{\'o}ttir}, {\'A}., {Hansen}, C.~J., {Choplin}, A., {et~al.} 2020,
  \aap, 634, A84

\bibitem[{{Sk{\'u}lad{\'o}ttir} {et~al.}(2019){Sk{\'u}lad{\'o}ttir}, {Hansen},
  {Salvadori}, \& {Choplin}}]{2019A&A...631A.171S}
{Sk{\'u}lad{\'o}ttir}, {\'A}., {Hansen}, C.~J., {Salvadori}, S., \& {Choplin},
  A. 2019, \aap, 631, A171

\bibitem[{{Smiljanic} {et~al.}(2016){Smiljanic}, {Romano}, {Bragaglia},
  {Donati}, {Magrini}, {Friel}, {Jacobson}, {Randich}, {Ventura}, {Lind},
  {Bergemann}, {Nordlander}, {Morel}, {Pancino}, {Tautvai{\v{s}}ien{\.{e}}},
  {Adibekyan}, {Tosi}, {Vallenari}, {Gilmore}, {Bensby}, {Fran{\c{c}}ois},
  {Koposov}, {Lanzafame}, {Recio-Blanco}, {Bayo}, {Carraro}, {Casey},
  {Costado}, {Franciosini}, {Heiter}, {Hill}, {Hourihane}, {Jofr{\'e}},
  {Lardo}, {de Laverny}, {Lewis}, {Monaco}, {Morbidelli}, {Sacco}, {Sbordone},
  {Sousa}, {Worley}, \& {Zaggia}}]{2016A&A...589A.115S}
{Smiljanic}, R., {Romano}, D., {Bragaglia}, A., {et~al.} 2016, \aap, 589, A115

\bibitem[{{Smith}(1988)}]{1988JPhB...21.2827S}
{Smith}, G. 1988, \jpb, 21, 2827

\bibitem[{{Smith} \& {Raggett}(1981)}]{1981JPhB...14.4015S}
{Smith}, G. \& {Raggett}, D. S.~J. 1981, \jpb, 14, 4015

\bibitem[{{Spina} {et~al.}(2018){Spina}, {Mel{\'e}ndez}, {Karakas}, {dos
  Santos}, {Bedell}, {Asplund}, {Ram{\'\i}rez}, {Yong}, {Alves-Brito}, {Bean},
  \& {Dreizler}}]{2018MNRAS.474.2580S}
{Spina}, L., {Mel{\'e}ndez}, J., {Karakas}, A.~I., {et~al.} 2018, \mnras, 474,
  2580

\bibitem[{{Spina} {et~al.}(2016){Spina}, {Mel{\'e}ndez}, \&
  {Ram{\'\i}rez}}]{2016A&A...585A.152S}
{Spina}, L., {Mel{\'e}ndez}, J., \& {Ram{\'\i}rez}, I. 2016, \aap, 585, A152

\bibitem[{{Spite} \& {Spite}(1982)}]{1982A&A...115..357S}
{Spite}, F. \& {Spite}, M. 1982, \aap, 115, 357

\bibitem[{{Steenbock} \& {Holweger}(1984)}]{1984A&amp;A...130..319S}
{Steenbock}, W. \& {Holweger}, H. 1984, \aap, 130, 319

\bibitem[{{Steffen} {et~al.}(2015){Steffen}, {Prakapavi{\v c}ius}, {Caffau},
  {Ludwig}, {Bonifacio}, {Cayrel}, {Ku{\v c}inskas}, \&
  {Livingston}}]{2015A&amp;A...583A..57S}
{Steffen}, M., {Prakapavi{\v c}ius}, D., {Caffau}, E., {et~al.} 2015, \aap,
  583, A57

\bibitem[{{Steinmetz} {et~al.}(2020){Steinmetz}, {Guiglion}, {McMillan},
  {Matijevi{\v{c}}}, {Enke}, {Kordopatis}, {Zwitter}, {Valentini}, {Chiappini},
  {Casagrande}, {Wojno}, {Anguiano}, {Bienaym{\'e}}, {Bijaoui}, {Binney},
  {Burton}, {Cass}, {de Laverny}, {Fiegert}, {Freeman}, {Fulbright}, {Gibson},
  {Gilmore}, {Grebel}, {Helmi}, {Kunder}, {Munari}, {Navarro}, {Parker},
  {Ruchti}, {Recio-Blanco}, {Reid}, {Seabroke}, {Siviero}, {Siebert}, {Stupar},
  {Watson}, {Williams}, {Wyse}, {Anders}, {Antoja}, {Birko}, {Bland-Hawthorn},
  {Bossini}, {Garc{\'\i}a}, {Carrillo}, {Chaplin}, {Elsworth}, {Famaey},
  {Gerhard}, {Jofre}, {Just}, {Mathur}, {Miglio}, {Minchev}, {Monari},
  {Mosser}, {Ritter}, {Rodrigues}, {Scholz}, {Sharma}, {Sysoliatina}, \& {RAVE
  Collaboration}}]{2020AJ....160...83S}
{Steinmetz}, M., {Guiglion}, G., {McMillan}, P.~J., {et~al.} 2020, \aj, 160, 83

\bibitem[{{Stonkut{\.{e}}} {et~al.}(2020){Stonkut{\.{e}}}, {Chorniy},
  {Tautvai{\v{s}}ien{\.{e}}}, {Drazdauskas}, {Minkevi{\v{c}}i{\={u}}t{\.{e}}},
  {Mikolaitis}, {Kjeldsen}, {Essen}, {Pak{\v{s}}tien{\.{e}}}, \&
  {Bagdonas}}]{2020AJ....159...90S}
{Stonkut{\.{e}}}, E., {Chorniy}, Y., {Tautvai{\v{s}}ien{\.{e}}}, G., {et~al.}
  2020, \aj, 159, 90

\bibitem[{Thomson(1912)}]{thomson1912xlii}
Thomson, J.~J. 1912, The London, Edinburgh, and Dublin Philosophical Magazine
  and Journal of Science, 23, 449

\bibitem[{{Ting} {et~al.}(2019){Ting}, {Conroy}, {Rix}, \&
  {Cargile}}]{2019ApJ...879...69T}
{Ting}, Y.-S., {Conroy}, C., {Rix}, H.-W., \& {Cargile}, P. 2019, \apj, 879, 69

\bibitem[{{Trubko} {et~al.}(2017){Trubko}, {Gregoire}, {Holmgren}, \&
  {Cronin}}]{2017PhRvA..95e2507T}
{Trubko}, R., {Gregoire}, M.~D., {Holmgren}, W.~F., \& {Cronin}, A.~D. 2017,
  \pra, 95, 052507

\bibitem[{{Valenti} \& {Piskunov}(1996)}]{1996A&amp;AS..118..595V}
{Valenti}, J.~A. \& {Piskunov}, N. 1996, \aaps, 118, 595

\bibitem[{{van Regemorter}(1962)}]{1962ApJ...136..906V}
{van Regemorter}, H. 1962, \apj, 136, 906

\bibitem[{{{\v{C}}otar} {et~al.}(2019){{\v{C}}otar}, {Zwitter}, {Kos},
  {Munari}, {Martell}, {Asplund}, {Bland -Hawthorn}, {Buder}, {de Silva},
  {Freeman}, {Sharma}, {Anguiano}, {Carollo}, {Horner}, {Lewis}, {Nataf},
  {Nordlander}, {Stello}, {Ting}, {Tinney}, {Traven}, {Wittenmyer}, \& {Galah
  Collaboration}}]{2019MNRAS.483.3196C}
{{\v{C}}otar}, K., {Zwitter}, T., {Kos}, J., {et~al.} 2019, \mnras, 483, 3196

\bibitem[{{Xiang} {et~al.}(2019){Xiang}, {Ting}, {Rix}, {Sand ford}, {Buder},
  {Lind}, {Liu}, {Shi}, \& {Zhang}}]{2019ApJS..245...34X}
{Xiang}, M., {Ting}, Y.-S., {Rix}, H.-W., {et~al.} 2019, \apjs, 245, 34

\bibitem[{{Yan} {et~al.}(1998){Yan}, {Tambasco}, \&
  {Drake}}]{1998PhRvA..57.1652Y}
{Yan}, Z.-C., {Tambasco}, M., \& {Drake}, G.~W.~F. 1998, \pra, 57, 1652

\bibitem[{{Zhao} {et~al.}(2016){Zhao}, {Mashonkina}, {Yan}, {Alexeeva},
  {Kobayashi}, {Pakhomov}, {Shi}, {Sitnova}, {Tan}, {Zhang}, {Zhang}, {Zhou},
  {Bolte}, {Chen}, {Li}, {Liu}, \& {Zhai}}]{2016ApJ...833..225Z}
{Zhao}, G., {Mashonkina}, L., {Yan}, H.~L., {et~al.} 2016, \apj, 833, 225

\end{thebibliography}
